\documentclass[galaxies,review,accept,moreauthors,pdftex]{Definitions/mdpi} 
\usepackage[utf8]{inputenc}
\usepackage{bm} 
\usepackage{amsmath, physics, amsfonts,amssymb} 
\usepackage{relsize} 
\usepackage{hyperref} 
\usepackage{caption}

\extrafloats{100}

\def\apj{\rm{ApJ}}
\def\apjl{\rm{ApJ}}

\def\aapr{\rm{A\&A~Rev.}}

\def\jcap{\rm{J. Cosmology Astropart. Phys.}}
\def\mnras{\rm{MNRAS}}
\def\prd{\rm{PRD}}
\def\prl{\rm{PRL}}

\def\pasa{\rm{PASA}}

\def\sovast{\rm{Sov.~Astron.}}

\usepackage{stackengine}
\stackMath

\newcommand{\timeresids}[0]{\delta\boldsymbol{t}}   
 
\Title{Stochastic Gravitational-Wave Backgrounds: Current Detection Efforts and Future Prospects}
\TitleCitation{Stochastic Gravitational-Wave Backgrounds: Current Detection Efforts and Future Prospects}
\Author{Arianna I. Renzini
 $^{1,2,}$*, Boris Goncharov $^{3,4}$, Alexander C. Jenkins $^{5,6}$ and Patrick M. Meyers $^{7,8,9}$} 
\AuthorNames{Arianna I. Renzini, Boris Goncharov, Alexander Jenkins, Patrick M. Meyers}
\address{%
$^1$ \quad LIGO  Laboratory,  California  Institute  of  Technology,  Pasadena,  CA  91125,  USA\\
$^2$ \quad Department of Physics, California Institute of Technology, Pasadena, CA 91125, USA\\
$^3$ \quad Gran Sasso Science Institute (GSSI), I-67100 L'Aquila, Italy; boris.goncharov@gssi.it
 \\
$^4$ \quad INFN, Laboratori Nazionali del Gran Sasso, I-67100 Assergi, Italy\\
$^5$ \quad Department of Physics and Astronomy, University College London, Gower Street, London WC1E 6BT, UK; alex.jenkins@ucl.ac.uk\\
$^6$ \quad Theoretical Particle Physics and Cosmology Group, Physics Department, King’s College London, Strand, London WC2R 2LS, UK\\
$^7$ \quad Theoretical Astrophysics Group, California Institute of Technology, Pasadena, CA 91125, USA; pmeyers@caltech.edu\\
$^8$ \quad OzGrav, University of Melbourne, Parkville, VIC 3010, Australia\\
$^9$ \quad School of Physics, University of Melbourne, Parkville, VIC 3010, Australia}
\corres{Correspondence: arenzini@caltech.edu}

\abstract{The collection of individually resolvable gravitational wave (GW) events makes up a tiny fraction of all GW signals that reach our detectors, while most lie below the confusion limit and are undetected. Similarly to voices in a crowded room, the collection of unresolved signals gives rise to a background that is well-described via stochastic variables and, hence, referred to as the stochastic GW background (SGWB). In this review, we provide an overview of stochastic GW signals and characterise them based on features of interest such as generation processes and observational properties. We then review the current detection strategies for stochastic backgrounds, offering a ready-to-use manual for stochastic GW searches in real data. In the process, we distinguish between interferometric measurements of GWs, either by ground-based or space-based laser interferometers, and timing-residuals analyses with pulsar timing arrays (PTAs). These detection methods have been applied to real data both by  large GW collaborations and smaller research groups, and the most recent and instructive results are reported here. We close this review with an outlook on future observations with third generation detectors, space-based interferometers, and potential noninterferometric detection methods proposed in the literature.}

\firstpage{1} 
\pubvolume{10}
\issuenum{1}
\articlenumber{0}
\pubyear{2022}
\copyrightyear{2020}
\datereceived{16 December 2021} 
\dateaccepted{05 February 2022} 
\datepublished{14 February 2022} 
\pdfoutput=1

\begin{document}

\section{Introduction}

Gravitational waves (GWs) are perturbations of the spacetime metric caused by extremely energetic events throughout the Universe. Up until now, direct detections of GWs have been coherent measurements of resolved waveforms in detector datastreams which may be traced back to single, point-like sources. %
These make up a tiny fraction of the gravitational-wave sky: The vast collection of unresolved signals corresponding to multiple point sources or extended sources adds up incoherently, giving rise to gravitational-wave backgrounds (GWBs). %
A variety of different backgrounds is expected given the range of GW sources in the Universe; however, regardless of their origin, most of these are treated as \emph{stochastic}, as they may be described by a non-deterministic strain signal and are, hence, referred to as stochastic gravitational-wave backgrounds (SGWBs).

Some backgrounds are stochastic by generation processes (e.g., inflationary tensor modes), whereas others are stochastic due to the characteristics or limitations of the specific detector used to observe them (e.g., the cumulative signal from many binary black hole coalescences). An SGWB of this latter nature is by definition at the threshold of detection, making it effectively a detector-dependent observable. %
Determining whether a signal is a background and also whether it inherently behaves as a stochastic field is then an iterative process, where first all signals present in the detector timestream undergo signal-to-noise ratio (SNR) estimation, which determines their resolvability, and then the background is estimated as the cumulative sum of only the unresolved and/or sub-threshold signals. %
In practice, each GW detector will be able to access qualitatively different backgrounds, depending on the noise levels and frequency ranges probed. 

With this review, we intend to provide a clear and complete yet concise reference for the broader community interested in stochastic search efforts while surveying our field of research. In laying out stochastic background detection techniques, we focus on three categories of detectors: ground-based detector networks, pulsar timing arrays (PTAs) monitored by radio telescopes, and spaceborne detectors comprised of sets of satellites. We do not discuss in detail the impressive work on data acquisition and processing necessary to put the data in the form required for the implementation of the searches we report (see Sections~\ref{sec:approaches} and~\ref{sec:efforts}); we refer the reader to several other papers that delineate these efforts (see, e.g.,~\cite{Davis:2022dnd, Sun:2020wke,Sun:2021qcg,VIRGO:2021kfv,KAGRA:2020agh} for recent discussions of ground-based laser interferometer detector characterization and calibration and~\cite{VerbiestOslowski2020,Tiburzi2018,HobbsDai2017} for reviews of PTA experiments).

We consider the present \emph{second generation} (2G) ground-based detector network to be made up of three detectors that are fully operational: the NSF-funded Laser Interferometer Gravitational-wave Observatory (LIGO) pair of detectors in the United States~\cite{TheLIGOScientific:2016agk} and the Virgo detector in Italy~\cite{2011CQGra..28k4002A}. To these, we can add the GEO600~\cite{Affeldt:2014rza} detector in Germany, which, however, does not have comparable sensitivity to the others given its smaller size, and the Kamioka Gravitational-wave Detector (KAGRA) in Japan~\cite{KAGRA}, which has not reached initial design sensitivity yet. The individual collaborations behind these detectors have joined efforts under the LIGO-Virgo-Kagra (LVK) collaboration. Future detectors will include the Indian Initiative in Gravitational-wave Observations~\cite{LIGOIndia}. Beyond these, we discuss the potential of the upcoming \emph{third generation} (3G) detector network, which will include the Einstein Telscope (ET)~\cite{Punturo:2010zz,Sathyaprakash:2011bh} and Cosmic Explorer (CE)~\cite{LIGOScientific:2011yag,Reitze:2019iox,Hall:2020dps}. All these monitor frequencies around $10^2$ Hz, spanning roughly two orders of magnitude in total, $\sim$10--10$^3$ Hz, depending on specific characteristics.

Regarding nanohertz-frequency stochastic backgrounds, $f \sim 10^{-9}$ Hz, we outline the methodology of pulsar timing array (PTA) experiments and present results from the North American Nanohertz Observatory for Gravitational Waves~(NANOGrav,~\cite{McLaughlin2013}), the Parkes Pulsar Timing Array~(PPTA,~\cite{ManchesterHobbs2013}), the European Pulsar Timing Array~([EPTA,~\cite{KramerChampion2013,Desvignes2016}), and the consortium of these collaborations known as the International Pulsar Timing Array~(\cite[IPTA, ][]{HobbsArchibald2010,VerbiestLentati2016}), of which the Indian PTA~(\cite[InPTA, ][]{JoshiArumugasamy2018}) is also now a member.
We comment on the common-spectrum noise process reported by the above collaborations that may be related to stochastic GWs or similar noise spectra in timing array pulsars~\cite{goncharov2021cpgwb}. 
We describe standard search methods for an isotropic GWB, and we also briefly discuss methods for resolving anisotropies in the GWB.
We further comment on the role of future observation efforts such as the Square Kilometer Array~(SKA,~\cite{DewdneyHall2009}) and the five-hundered-meter aperture spherical radio telescope~(FAST,~\cite{NanLi2011}).

As for spaceborne interferometers, we primarily discuss the potential detection capabilities of the Laser Interferometer Space Antenna (LISA), which is an ESA-led mission in collaboration with NASA planned to launch in the mid 2030s. LISA will scan a broad range of frequencies, $f \sim 10^{-5}$--$10^{-1}$ Hz, with a maximum sensitivity around a millihertz, allowing us to tune into a new, vast range of the GW sky at unprecedented depth and volume~\cite{Amaro2017}.

We will start by introducing the theoretical underpinnings of GWB science in Section~\ref{sec:theory}. In Section~\ref{sec:sources}, we review expected sources of GWBs typically considered in the literature. We outline the different properties of the backgrounds in relation to the detectors that measure GWs today and those that will probe them in the foreseeable future. We offer a detailed explanation of stochastic search methods in Section~\ref{sec:approaches}, differentiating between different observing strategies and detection regimes. In Section~\ref{sec:efforts}, we summarise current detection results, reporting the most stringent upper limits yet on SGWBs while highlighting how close we are to detection. Finally, we close this review in Section~\ref{sec:conclusions} with an overview of future prospects with the upcoming GW detectors. 

\section{Theory of Stochastic Backgrounds}\label{sec:theory}

 In this section, we provide the theoretical foundations of GW science essential to the description of generation and propagation of GWs in the Universe. First, we review the mathematical description of stochastic backgrounds in a common framework, regardless of their astrophysical or cosmological origin, which lies at the basis of (almost) all stochastic search methods reviewed in Section~\ref{sec:approaches}. Then, we introduce the fractional energy density of GWs, which is the main observable targeted by stochastic searches.

\subsection{Gravitational-Wave Strain and Stokes Parameters}
We work in the linearised regime of general relativity, such that the spacetime metric is close to that of flat Minkowski space, with small perturbations that encode  GWs, $g_{\mu\nu}=\eta_{\mu\nu}+h_{\mu\nu}$. Neglecting terms of order $h^2$, the vacuum Einstein field equation can then be written as a wave equation in the De Donder gauge~\cite{Wald1984},
\begin{equation}
	\Box \qty( h_{\mu\nu} - \frac{1}{2} \eta_{\mu\nu} h^\alpha_{\,\,\,\alpha} ) = 0,
	\label{eq:Einvac}
\end{equation} 
where the box operator is the D'Alembert operator, defined as $\Box = \partial_\mu \partial^\mu$.
Solutions to this equation may be constructed as linear superpositions of plane waves,
\begin{equation}
	h_{\mu\nu} = A e_{\mu\nu}\,e^{i\qty(k_\alpha x^\alpha)}\,,
	\label{eq:planewave}
\end{equation}
where $e_{\mu\nu}$ is the normalised polarisation tensor, and $k^\alpha$ is the wave four-vector. By inserting this solution in the equation above, one recovers the dispersion relation $k_\alpha k^\alpha=0$. This implies that the waves travel at the speed of light in the linearised regime. We may then pick the transverse-traceless (TT) gauge and reduce $h_{\mu\nu}$ to a purely spatial perturbation, $h_{ij}$, which carries two degrees of freedom or equivalently two independent, spin-2, polarisation states. 

Generalising Equation~(\ref{eq:planewave}) as in~\cite{Cornish2001}, for example, the GW strain at time $t$ and position vector $\bm x$ given by an \emph{infinite superposition} of plane waves incoming from all directions on the sky $\hat{\bm n}$,
\begin{equation}
    \hat{\bm n} = (\sin\theta\cos\phi,\sin\theta\sin\phi,\cos\theta)\,,
\end{equation}
in standard angular coordinates $(\theta,\,\phi)$ on the 2-sphere may be written as
\begin{equation}
h_{ij}\,(t,\bm{x})=\int_{-\infty}^{+\infty} \!df \int_{S^2} \!d\hat{\bm n}\!\!\sum_{P=+,\,\times}\!\!h_P\,(f,\,\hat{\bm n})\,\epsilon_{ij}^P(\hat{\bm n})\,e^{i2\pi f(\hat{\bm n}\cdot \bm{x}-t)}\,.
\label{expan}
\end{equation}
The spatial wave vector is written explicitly as $\bm{k} = 2\pi f \hat{\bm n}$ (we will use $c=1$ throughout unless otherwise noted), $P$ labels the polarisation states, and $h_P$ are the respective amplitudes assigned to each state. Choosing the linear polarisation basis $\{+,\,\times\}$, the orthogonal polarisation base tensors $\epsilon^P$ may be written as
\begin{align}
{\bm \epsilon}^+ &={\bm e}_\theta\otimes {\bm e}_\theta - {\bm e}_\phi\otimes {\bm e}_\phi \,,\\
{\bm \epsilon}^\times &= {\bm e}_\theta\otimes {\bm e}_\phi+{\bm e}_\phi\otimes {\bm e}_\theta\,,
\end{align}
with
\begin{align}
{\bm e}_\theta &= (\cos\theta\cos\phi,\cos\theta\sin\phi,-\sin\theta) \,,\\
{\bm e}_\phi &= (-\sin\phi,\cos\phi,0)\,,
\end{align}
as defined in~\cite{Allen1996}.
For a stochastic signal, $h_P$ represents a random complex number. Most often, these are assumed to be drawn from a Gaussian probability distribution. This is likely a good approximation for a cosmological GWB, as we will see in Section~\ref{sec:sources}. For a stochastic GWB of astrophysical sources, this assumption may break down but the central limit theorem will guarantee that the statistics approach with respect to a Gaussian random field if the signal is sourced by a sufficiently large number of independent events and that any high signal-to-noise outliers have been subtracted from the detector's timestreams. Under the Gaussian assumption, the statistical properties of the amplitudes are then characterised solely by the second moments of the distribution $\langle h_P^{}(f,\,\hat{\bm n}) h^\star_{P'}(f',\,\hat{\bm n}')\rangle$, which, assuming statistical homogeneity, correspond to ensemble averages as~\cite{Romano2017}
\begin{equation}
\begin{split}
 \begin{pmatrix} \langle h^{}_+(f,\,\hat{\bm n})\,h^\star_+(f',\,\hat{\bm n}')\rangle  & \langle h^{}_+(f,\,\hat{\bm n})\,h^\star_\times(f',\,\hat{\bm n}')\rangle  \\ \langle h^{}_\times(f,\,\hat{\bm n})\,h^\star_+(f',\,\hat{\bm n}')\rangle  & \langle h^{}_\times(f,\,\hat{\bm n})\,h^\star_\times(f',\,\hat{\bm n}')\rangle  \end{pmatrix} = \,\delta^{(2)}(\bm{n},\bm{n'})\,\delta(f-f')\,\times&\\
 \begin{pmatrix} I(f,\,\hat{\bm n})+Q(f,\,\hat{\bm n}) & U(f,\,\hat{\bm n})-iV(f,\,\hat{\bm n}) \\ U(f,\,\hat{\bm n})+iV(f,\,\hat{\bm n}) & I(f,\,\hat{\bm n})-Q(f,\,\hat{\bm n}) \end{pmatrix}&\,,
\end{split}
\label{stokey}
\end{equation}
where the Stokes parameters $I$, the intensity, $Q$ and $U$, giving the linear polarisation, and $V$, the circular polarisation, have been introduced. The four Stokes parameters completely describe the polarisation of the observed signal in analogy with electromagnetic Stokes parameters for the photon~\cite{jackson_classical_1999}. The difference here is that whilst the electromagnetic $Q$ and $U$ Stokes parameters transform as spin-2 quantities with respect to rotations, their GW strain counterparts transform as spin-4. In both cases,  intensity $I$ behaves as a scalar under rotations, while $V$ transforms as a pseudo-scalar. More details on these and their spin-weighted behaviour may be found in~\cite{AIR_thesis}.

\subsection{The Energy Density of Gravitational Waves}
The evolution of our Universe is described in terms of its expansion rate, also referred to as the Hubble rate,
    \begin{equation}
        H(t)\equiv\dv{t}\ln a,
    \end{equation}
where $a(t)$ is the cosmological scale factor.
Using general relativity, we can write this rate in terms of the different forms of energy density present in the Universe (this is most conveniently performed as a function of redshift $z$),
\begin{equation}
	H(z) = H_0 \qty( \Omega_{\rm R}\,(1+z)^4 + \Omega_{\rm M}\,(1+z)^3 + \Omega_k (1+z)^2 + \Omega_\Lambda) ^{1/2}\,,
	\label{eq:friedmannH}
\end{equation}
where $H_0 \approx 70$ km s$^{-1}$ Mpc$^{-1}$ is the Hubble rate today.
In Equation~(\ref{eq:friedmannH}), $\Omega_i = \rho_i/\rho_\mathrm{c}$ are the density parameters of each energy component of the Universe: radiation $\mathrm{R}$, which includes photons and relativistic neutrinos; matter $\mathrm{M}$, which includes baryons and cold dark matter; curvature $k$; and the cosmological constant $\Lambda$. The critical energy density $\rho_\mathrm{c}$ is a normalisation that can be interpreted as the energy density required to close our (flat) Universe today, which is explicitly
\begin{equation}
	\rho_\mathrm{c} = \frac{3H_0^2}{8\pi G}\,,
\end{equation}
where $G = 6.67 \times 10^{-11}$ m$^3$ kg$^{-1}$ s$^{-1}$ is the universal gravitational constant.

At this stage, one may ask the following: where do GWs fit in? In fact, GWs feature in Equation~\eqref{eq:friedmannH}, as a part of the radiation energy density, as they are considered to be carried by massless particles propagating at the speed of light\footnote{There are several theories that postulate modifications to this statement, but these modifications are strongly constrained by multimessenger observations of the binary neutron star merger GW170817~\cite{LIGOScientific:2017zic}.}. This allows us to write
    \begin{equation}
        \Omega_\mathrm{R}=\Omega_\gamma+\Omega_\nu+\Omega_\mathrm{GW}+\cdots,
    \end{equation}
    with the terms in the sum representing the energy density in photons, neutrinos, and GWs, respectively.
While the present day value of the radiation density is very small ($\Omega_\mathrm{R}\sim10^{-4}$), we see from Equation~\eqref{eq:friedmannH} that at very high redshift, $z\gtrsim\order{10^3}$, it dominated the Universe's expansion.
As a result, any additional contributions to the radiation density, including those from primordial GWs, will leave an imprint on cosmological observables generated at these early epochs: principally, the cosmic microwave background (CMB) and the light element abundances predicted by Big Bang nucleosynthesis (BBN)~\cite{Kolb:1990vq}. These imprints are identical for all forms of additional radiation density; thus, by convention they are parameterised in terms of the effective number of neutrino species, $N_\mathrm{eff}$. Constraints on $N_\mathrm{eff}$ can straightforwardly be converted into constraints on  GWB energy density.
The most up-to-date analysis combines CMB and BBN data with external datasets such as measurements of the baryon acoustic oscillation scale, producing~\cite{Pagano:2015hma}
    \begin{equation}
    \label{eq:N_eff-constraint}
        \Omega_\mathrm{GW}< 1.2\times 10^{-6}\qty(\frac{H_0}{100\,\mathrm{km}\,\mathrm{s}^{-1}\,\mathrm{Mpc}^{-1}})^{-2}.
    \end{equation}

Given these strong constraints on $\Omega_{\rm GW}$, it is reasonable to consider GWs as perturbations in the Universe, which do not influence the bigger picture but obey the propagation rules, the structure, and the energy content of the Universe that have been set out. 

The density parameter appearing in Equation~\eqref{eq:N_eff-constraint} encapsulates the \emph{total} energy density of GWs at all frequencies\footnote{Strictly speaking, GWs with wavelengths larger than the size of the cosmological horizon are frozen out by Hubble friction and, thus, do not contribute to the effective energy density. As a result, the density parameter appearing on the left-hand side of Equation~\eqref{eq:N_eff-constraint} should be interpreted as an integral over $\Omega_\mathrm{GW}(f)$ from a minimum frequency of $f\sim10^{-15}\,\mathrm{Hz}$, which corresponds to the size of the horizon at the epoch when the CMB photons were emitted.}, as this is what determines the Hubble expansion rate.
However, all other searches for gravitational waves focus on particular frequency bands; thus, it is convenient to define the GW energy density frequency spectrum~\cite{Christensen2018},
\begin{equation}
	\Omega_{\rm GW}(f) = \frac{1}{\rho_\mathrm{c}} \frac{d\rho_{\rm GW} (f)}{d\ln f}\,;
\label{eq:Omegarho}
\end{equation}
note that the frequency $f$ here is the \emph{measured} frequency, assuming an observer and the intervention of an instrument which mediates the true signal and the measurement.
Furthermore, beyond the isotropic value of  GWB over the entire sky it is possible to include any anisotropy by adding a directional dependence $\Omega_{\rm GW}(f) \rightarrow \Omega_{\rm GW}(f,\hat{\bm n})$ where $\hat{\bm n}$ is the unit vector of a line of sight.

It is now possible to recover the cosmological information contained in the $\Omega_{\rm GW}$ parameter, which is effectively the cumulative GW energy history weighted per redshifted frequency bin, as presented in~\cite{Phinney2001}. This may be observed by noting that homogeneity and isotropy throughout the universe imply
\begin{equation}
	\rho_{\rm GW} = \int_0^\infty \frac{df}{f} \int_0^\infty dz \frac{N(z)}{1+z} \,f_{\rm r} \,\frac{d E_{\rm GW}}{df_{\rm r}}\,,
	\label{eq:rhoGW}
\end{equation}
where $z$ is redshift, $N(z)$ is the number of GW emitters as a function of redshift, and $f_{\rm r}$ is the emission frequency of GWs in the rest frame of the emitter, $f_{\rm r} = f (1+z)$. $d E_{\rm GW}/df_{\rm r}$ is then the spectral energy density of the source population. Equation~(\ref{eq:Omegarho}) may be rewritten~as
\begin{equation}
	\rho_{\rm GW} = \int_0^\infty \frac{df}{f} \rho_{\rm c}\, \Omega_{\rm GW}(f)\,,
\end{equation}
hence, equating Equations~\eqref{eq:rhoGW} and~\eqref{eq:Omegarho} frequency by frequency yields
\begin{equation}
	\Omega_{\rm GW} (f) = \frac{1}{\rho_\mathrm{c}} \int_0^\infty dz\,\frac{N(z)}{1+z}\,\qty[f_{\rm r} \frac{d E_{\rm GW}}{df_{\rm r}}]_{f_{\rm r} = f(1+z)}\,.
	\label{eq:phinney}
\end{equation}
Equation~(\ref{eq:phinney}) implies that the fractional energy density of GWs per measured logarithmic frequency intervals at redshift zero, $\Omega_{\rm GW}$, is proportional to the integral over the cosmic history of the comoving number density of the sources multiplied by the emitted energy fraction of each source, in the appropriately redshifted frequency bin.
This directly relates $\Omega_{\rm GW}$ to the source event rate as a function of redshift, $N(z)$. In the case of different types of sources, this may be extended to a sum over all contributors,
\begin{equation}
	N(z)\,\qty[f_{\rm r} \frac{d E_{\rm GW}}{df_{\rm r}}]_{f_{\rm r} = f(1+z)} \quad\longrightarrow \qquad
\sum_i N_i(z)\,\qty[f_{\rm r} \frac{d E_{{\rm GW},\,i}}{df_{\rm r}}]_{f_{\rm r}=f(1+z)}\,.
\end{equation}

$\Omega_{\rm GW}(f)$ may also be re-written in terms of the intensity parameter $I$ defined in Equation~\eqref{stokey}~\cite{Allen1996}.
\begin{equation}
	\Omega_{\rm GW}(f) = \frac{4\pi^2f^3}{\rho_\mathrm{c} G}I(f)\,.
\label{eq:omegatoI}
\end{equation}
We can derive this relationship using the Isaacson formula for the energy density of the GWB~\cite{Isaacson:1968hbi,Isaacson:1968zza},
\begin{equation}
    \label{eq:isaacson}
    \rho_\mathrm{GW}=\frac{1}{32\pi G}\ev{\dot{h}_{ij}\dot{h}_{ij}},
\end{equation}
where the angle brackets indicate an average over many wavelengths.
Inserting the plane-wave decomposition in Equation~\eqref{expan} and rewriting the two-point statistics of each plane wave in terms of the Stokes parameters in Equation~\eqref{stokey}, we obtain
    \begin{equation}
        \rho_\mathrm{GW}=\frac{\pi}{2G}\int_0^\infty\dd{f}\int_{S^2}\dd{\vu*n}f^2I(f,\vu*n).
    \end{equation}
Note that we have assumed ergodicity here such that the spatial average in Equation~\eqref{eq:isaacson} is equivalent to the ensemble average in Equation~\eqref{stokey}.
If we treat  intensity as isotropic, $I(f,\vu*n)\to I(f)$, then we immediately recover Equation~\eqref{eq:omegatoI}.
This fundamental relation allows us to connect GWB observations to cosmological implications of the background and may easily be extended to a directional statement as
\begin{equation}
	\Omega_{\rm GW}(f,\hat{\bm n}) = \frac{4\pi^2f^3}{\rho_\mathrm{c} G}I(f,\hat{\bm n})\,.
\end{equation}
Note that this implies
\begin{equation}
    \Omega_\mathrm{GW}(f)=\frac{1}{4\pi}\int_{S^2}\dd{\vu*n}\Omega_\mathrm{GW}(f,\vu*n)\,,
\end{equation} 
however, one must take care of factors of $4\pi$ as other conventions are sometimes used. Note that we are using the conventions as in~\cite{Allen1996}, where all power spectra are one-sided.

As for the other Stokes' parameters, these do not contribute to $\Omega_{\rm GW}$ and are equal to zero in the case of an unpolarised background. This may be easily gleaned by their definitions, in Equation~(\ref{stokey}), as the second moments of the $h_+$ and $h_\times$ fields should be equal and the expectation value of any cross correlations between the two should vanish.
$\Omega_{\rm GW}$, thus, inherits the spectral properties from the GWB intensity as per Equation~(\ref{eq:omegatoI}). As will emerge in the following sections, the spectral dependence is often reduced to a power law, in which case it is common practice to write\footnote{Often in the literature, the letter $\alpha$ is used to denote a different spectral index, hereby denoted $\alpha'$, of the characteristic strain spectrum of the stochastic gravitational wave background $h(f) \equiv \sqrt{f S(f)} \propto f^{\alpha'}$, where $S(f)$ is the power spectral density of the background. This leads to the power-law index of $2\alpha' + 2$ for $\Omega_\text{GW}$ in Equation~\eqref{eq:Omegascale} instead of $\alpha$.\label{fn:alphas_general}}
\begin{equation}
    \Omega_{\rm GW}(f) = \Omega_{\rm GW}(f_{\rm ref})\left(\frac{f}{f_{\rm ref}}\right)^{\alpha}\,,
    \label{eq:Omegascale}
\end{equation}
where $\alpha$ is the spectral index of the signal, and $f_{\rm ref}$ is a reference frequency. This assumption is often extended to be valid in each direction independently, such that the spectral and directional dependence of the GW signal factor out as
\begin{equation}
    I(f,\hat{\bm n}) = I(f_{\rm ref},\hat{\bm n})\left(\frac{f}{f_{\rm ref}}\right)^{\alpha-3}\,.
    \label{eq:intef}
\end{equation}
This is particularly useful when performing multiparameter estimations such as map-making as it allows one to integrate out the spectral dependence of the reconstructed signal and solve for an intensity sky map. Note that in the ensemble, averaging is required to recover the Stokes' parameters, and all  temporal phase information in the GWs is discarded; as argued in~\cite{Renzini:2021iim}, no physical information may be recovered from the coherent phase component. 

\section{Sources of Stochastic Backgrounds}\label{sec:sources}

GWBs fall broadly into two distinct categories based on the underlying generation mechanism. The first is a superposition of signals of astrophysical origin, galactic and/or extra-galactic, that are not individually detected or resolved. This type of background will, thus, be unresolved and incoherent in that the temporal phase of the signal is expected to be random and is expected to be stochastic in the limit where the number of sources is very large. 
The second category is cosmological or primordial backgrounds. These include GWBs generated by the evolution of vacuum fluctuations during an inflationary epoch that end up as superhorizon tensor modes at the end of inflation~\cite{Grischuk1975,Maggiore1999}. It also includes GWBs generated by nonlinear phenomena at early times such as phase transitions or via emission by topological defects~\cite{Hogan1986,Battye1997,Vilenkin2000}. For the most part, primordial backgrounds are formed via stochastic processes, resulting in stochastic signals; on top of this, propagation effects contribute to isotropising a primordial signal, in most cases erasing any initial anisotropies~\cite{Bartolo:2018evs, Bartolo:2019zvb}. In particular, stochasticity here implies no initial information is retained by the temporal phase of the background~\cite{Margalit:2020sxp}. Cosmological GWBs can also be stochastic due to many overlapping signals, for example, in the case of cosmic strings.
Assuming that the orientation of individual astrophysical sources is random on the sky, these are in general expected to be unpolarised, as an equal distribution in the polarisation modes will cancel out any local asymmetry. Primordial signals are also mostly assumed to be unpolarised, as the underlying stochastic process is not expected to select a preferential frame, with some exceptions as we explain below.
What then characterises backgrounds of different origin is the spectral dependence of the source energy density in Equation~(\ref{eq:phinney}): the power spectrum of the signal depends on the specific GW emission mechanism, which is imprinted in the measured strain. Note that this measurement includes non-negligible selection effects, as \emph{qualitatively} different backgrounds contribute from different redshift shells and from different directions. 

\begin{figure}[t]
    \centering
    \includegraphics[width = 0.85\linewidth]{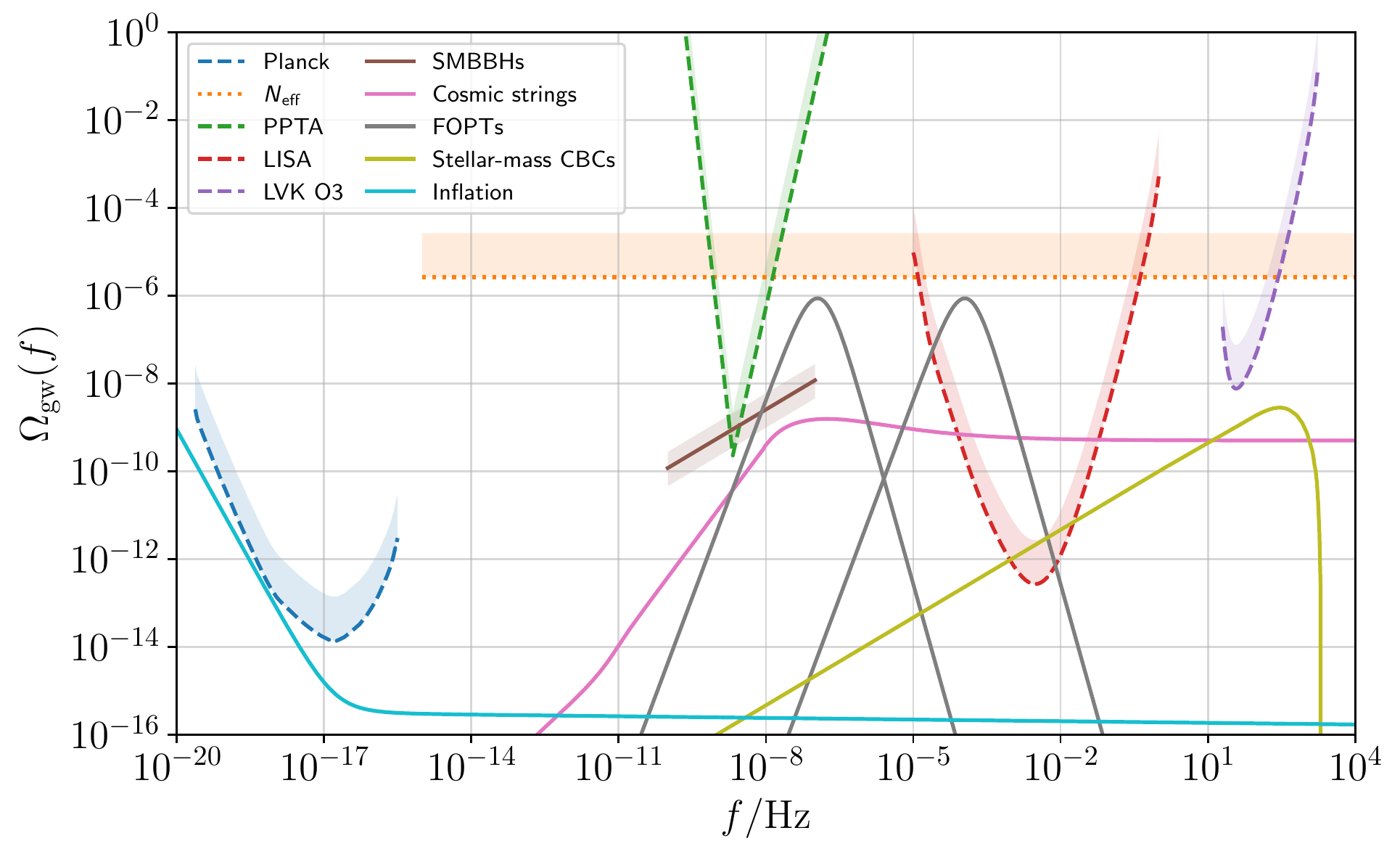}
    \caption{{An overview of potential GWB signals across the frequency spectrum. The light blue curve~shows the prediction for single-field slow-roll inflation with a canonical kinetic term, with tensor-to-scalar ratio $r_{0.002}=0.1$~\cite{Maggiore:2018sht}. The pink curve shows a GWB from Nambu--Goto cosmic strings, using ``model 2'' of the loop network, with a dimensionless string tension of $G\mu=10^{-11}$~\cite{Auclair:2019wcv}. The brown curve shows a GWB from inspiralling supermassive BBHs, with the amplitude and shaded region shown here corresponding to the common noise process in the NANOGrav 12.5-year data set~\cite{ArzoumanianBaker2020}. The two grey curves show GWBs generated by first-order phase transitions at the electroweak scale ($\sim$200 {GeV}) and the QCD scale ($\sim$200 {MeV}), respectively~\cite{Caprini:2015zlo}. The yellow curve shows a GWB generated by stellar-mass compact binaries, based on the mass distributions and local merger rates inferred by LVK detections~\cite{LVK:2021kbb}. The dashed curves show various observational constraints, as described further in Section~\ref{sec:efforts} (this includes the PPTA constraint, which intersects the possible NANOGrav SMBBH signal); 
the dotted curve shows the integrated constraint from measurements of $N_\mathrm{eff}$, which cannot be directly compared with the frequency-dependent constraint curves but is shown here for indicative purposes.} }
    \label{fig:GWBplot}
\end{figure}

In this section, we review both astrophysical and cosmological GWBs, providing the necessary background for the targeted searches discussed in Section~\ref{sec:efforts}. We also comment on the \emph{observational properties} of the signal which are essential to understand when building an optimal search method. The various sources are also summarised in Figure~\ref{fig:GWBplot}, which includes the sensitivity of several GW detection efforts for reference.


\vspace{6pt}

\subsection{Astrophysical Backgrounds}\label{ssec:astro_sources}
Astrophysical GWBs are the collection of all GWs generated by astrophysical processes which are individually unresolved by your GW detector. These can be either individual subthreshold signals, or they can be so numerous that they add up incoherently and form a continuous signal in the timestream. 

Perhaps the most studied signal in the literature is a background sourced by a collection of inspiralling and merging compact binary systems. These include black hole binaries, neutron star binaries, white dwarf binaries, and systems counting a mixed pair of these objects. Black hole binaries in particular are a vast category of sources, as the mass of each black hole in the binary ranges between 5 and 10$^{10}M_\odot$.
The lower mass of $5M_\odot$ here refers to the lack of black holes between the Tolman--Oppenheimer--Volkoff limit (which sets the maximum mass of a neutron star to $2.9M_\odot$~\cite{Kalogera1997}) and $5M_\odot$, as seen in low-mass X-ray binary observations~\cite{BailynJain1998,OzelPsaltis2010}\footnote{There is still much uncertainty around this black hole ``lower mass gap'' and it remains a main focus of research in the field; see, e.g., the discussions in~\cite{BailynJain1998,OzelPsaltis2010,Gupta:2019nwj}.}.
Specifically, a BH is considered \emph{massive} \mbox{($M_{\rm BH} \sim 10^5$--$10^{10} M_\odot$)}, \emph{intermediate} ($M_{\rm BH} \sim 10^2$--$10^{5} M_\odot$), and \emph{stellar mass} ($M_{\rm BH} \sim 5$--$10^{2}M_\odot$)~\cite{Celotti:411555}. There are binaries that have a mass ratio close to one, where the two black holes in the binary fall within the same mass category, and there are also so-called extreme mass ratio (EMR) binaries where the two objects belong to different families, including extreme mixed binaries such as massive black hole--neutron star pairs. 
There is a direct relation between  binary masses and  GW emission frequencies; however, overall, the GW power spectra have a similar shape. In particular, in the early emission stage of the binary, i.e., the \emph{inspiral} phase when the energy emission increases adiabatically as the binary's orbit shrinks very gradually, energy emission can be described by very few parameters and is well-approximated by a power law as explained below. 
The masses also impact the characteristic amplitude of a GW signal, which in turn may be linked to the detection threshold via the signal-to-noise ratio (SNR) of each signal. The latter sets the maximum expected redshift depth at which each type of signal may be observed by the detectors probing that frequency range. A good example of this is the background generated by double white dwarfs, which is expected to be observable by the LISA detector only at very low redshift~\cite{Cooray2004} such that LISA will only be sensitive to white dwarfs from the Milky Way and, perhaps, to those in the local group~\cite{Korol:2020hay}.


Other astrophysical sources of GWs include asymmetric supernovae explosions and rotating neutron stars which are not spherically symmetric and may produce triaxial emission~\cite{Regimbau2011}. These are considered minor sources of GWs compared to compact binaries as their characteristic signal is weaker. Furthermore, they are also considerably harder to model; hence, the discussion that follows is mainly focused on the former; nevertheless, many equations presented are general and may be applied to these secondary sources as well. In principle, the fact that these are harder to model also implies that they will necessarily be targets of stochastic searches, as the latter may be considered as generic, ``un-modelled'' searches of GW power.

It may seem odd to use the fractional parameter $\Omega_{\rm GW}$ to describe  GW energy from astrophysical events, as historically this is considered a cosmological quantity; however, several authors and collaborations have deemed it a useful convention allowing scientists from different fields to easily compare results. With this in mind, it may be considered as a compact measure of the GW history of the low-redshift Universe, after galaxy and star formation have taken place. To see this in more detail, following~\cite{Phinney2001}, let $N$ be the number of events of a particular signal type in a comoving volume. This may be decomposed in  event rate $\dot{N}$ multiplied by  cosmic time slice $dt_r/dz$,
\begin{equation}
	N(z) = \dot{N} \frac{dt_r}{dz} \equiv \frac{\dot{N}}{(1+z) H(z)}\,.
\end{equation}
Then, as in~\cite{Callister2020}, the fractional energy density may be rewritten as
\begin{equation}
	 \Omega_{\rm GW} = \frac{f}{\rho_{\rm c}}\,\int_{0}^{z_{\rm max}} dz \, \frac{\dot{N}(z) }{(1+z)H(z)} \,\expval{ \frac{d E_{\rm GW}}{df_{\rm r}}\eval_{f_{\rm r}= f(1+z)}} \,,
	 \label{eq:Omegaexpval}
\end{equation}
where the angle brackets indicate an averaging over  particular source population parameters $\theta$, such that
\begin{equation}
	\expval{\frac{d E_{\rm GW}}{df_{\rm r}}} = \int d\theta p(\theta)\frac{d E_{\rm GW}(\theta; f_{\rm r})}{df_{\rm r}}\,,
\end{equation}
and $p(\theta)$ is the probability distribution of  source parameters. In the case of a stochastic background, one would expect the parameter space to be fully sampled by the population; however, note that selection effects imposed by the detector may break this limit.
One would expect the event rate $\dot{N}$ to locally increase with redshift, as the volume of each redshift shell increases. In fact, the event rate for stellar mass compact binary coalescences has been found to increase with redshift when analysing the direct detections reported by the LVK~\cite{Fishbach2018}. However, this rate cannot increase monotonically out to infinity and must have a turnaround point at some peak redshift $\hat{z}$ and then decay to zero as star formation ceases in the early Universe. In general, this may be modelled phenomenologically with~\cite{MadauDickinson}
\begin{equation}
	\dot{N}(z) = {\cal C} (\alpha_\text{r}, \beta_\text{r}, \hat{z}) \frac{\dot{N}_0 (1+z)^{\alpha_\text{r}}}{1+\qty(\frac{1+z}{1+\hat{z}})^{\alpha_\text{r} + \beta_\text{r}}}\,,
\end{equation}
such that at low redshift, the rate increases as $\dot{N} \propto (1+z)^{\alpha_\text{r}}$, then reaches a maximum at $z\sim\hat{z}$, and finally decreases at high redshift as $\dot{N} \propto (1+z)^{-\beta_\text{r}}$. $\cal C$ is the normalisation constant
\begin{equation}
    {\cal C}(\alpha_\text{r}, \beta_\text{r}, \hat{z}) = 1+ \qty(1+\hat{z})^{-(\alpha_\text{r}+\beta_\text{r})}    
\end{equation}
which ensures $\dot{N}(z = 0) = \dot{N}_0$.  Constraining $\hat{z}, \alpha_\text{r}$ and $\beta_\text{r}$ provides insight on binary formation history and, consequently, star formation history as well. In practice, however, stochastic searches probe the redshift-integrated observable $\Omega_{GW}(f)$ only. The redshift history may be reconstructed with careful modelling and cross-correlation with direct measurements of individual events for which  redshift is estimated; see the approach presented in~\cite{Callister2020}. 
Note that selection effects due to the mass distribution of sources need to be taken into account in this sort of analysis. Alternatively to this phenomenological approach, the redshift-dependent event rate may also be probed by conducting a careful study of the specific astrophysical conditions which favour compact binary inspirals and mergers; for example, the evolution of galaxy star formation and chemical enrichment can be combined with prescriptions for merging of compact binaries obtained via stellar evolution simulations~\cite{Boco:2019teq,Boco:2020pgp}. 

\begin{figure}[t]
    \includegraphics[width = 0.75\textwidth]{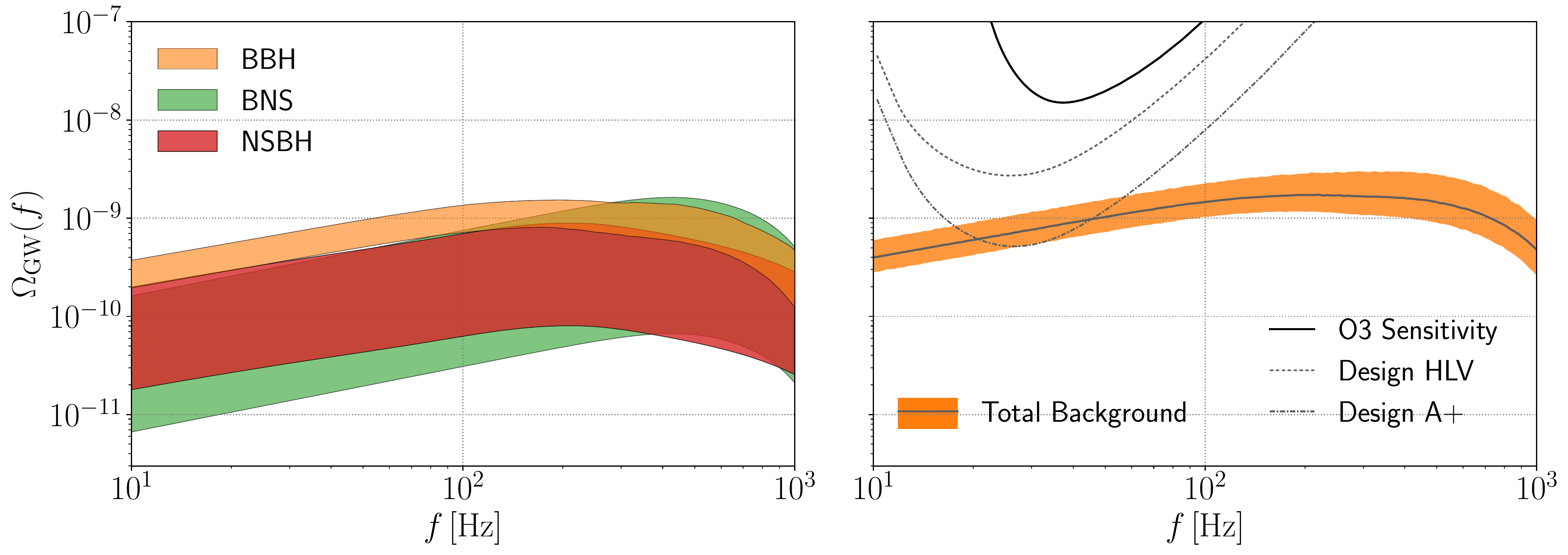}
    \caption{Projections for the CBC stochastic background using the third GW transient catalogue (GWTC-3), first presented in~\cite{LIGOScientific:2021psn}. On the right, the total expected background is compared to the integrated sensitivity of current and future configurations of the 2G detector network. This plot was obtained using  open data published in~\cite{LVK:open_data_pops}.} 
    \label{fig:LIGO_CBCBackground}
\end{figure}
To study the spectral dependence of different backgrounds, we can plug in different source models in Equation~(\ref{eq:phinney}) and calculate the spectral shape of $\Omega_{\rm GW}$.
 This is particularly useful in the case of an astrophysical background as $N$ and $E_{\rm GW}$ are directly related to the star formation history of the Universe, and astrophysical GWs from different source populations will contribute to $\Omega_{\rm GW}$ in different amounts at different times. This may result in a method that differentiates between astrophysical backgrounds and also test different stellar and galaxy evolution models~\cite{Callister:2016ewt}. 
 
An example of astrophysical background with a simple expected spectral dependence is that of compact binaries in the adiabatic inspiral phase~\cite{Sesana2008}. 
In this case, it is sufficient to model  GW radiation at leading quadrupole order~\cite{Thorne1987},
\begin{equation}
	\frac{dE_{\rm GW}}{d {\rm ln} f_{\rm r}} = \frac{\pi^{2/3}}{3} G^{2/3}{\cal M}^{5/3} f_{\rm r}^{2/3}\,,
\end{equation}
where ${\cal M}$ is the \emph{chirp mass}, given by
\begin{equation}
	{\cal M} = \frac{(m_1m_2)^{3/5}}{(m_1+m_2)^{1/5}}\,,
\end{equation}
where $m_1$ and $m_2$ are the masses of the two compact objects in the binary.
The co-moving number density will depend on  chirp masses in the binary distribution and may be rewritten as an integral over $\cal M$,
\begin{equation}
	N = \int_{{\cal M}_{\rm min}}^{{\cal M}_{\rm max}}  \frac{d N}{d{\cal M}} d{\cal M}\,,
\end{equation}
where the integration extremes will depend on  source population characteristics. Rearranging the terms in Equation~(\ref{eq:Omegaexpval}), one obtains
\begin{equation}
	 \Omega^{\rm CB}_{\rm GW} \propto f^{2/3}\,\int_{0}^{z_{\rm max}} dz \, \frac{1 }{(1+z)^{1/3}} \,\int_{{\cal M}_{\rm min}}^{{\cal M}_{\rm max}} d{\cal M} \frac{d N}{d{\cal M}}{\cal M}^{5/3} \,,
\end{equation}
where the average over the population parameters is simply reduced to the integral over all possible chirp masses. Here, the superscript CB labels compact binaries generically. This provides a simple spectral dependence for $\Omega^{\rm CB}_{\rm GW}$, which may be condensed as\footnote{As mentioned in the footnote~\ref{fn:alphas_general}, the energy density spectral index $\alpha = 2/3$ is related to the strain amplitude spectral index $\alpha'$ such that $\alpha = 2\alpha'+2$. Therefore $\alpha'=-2/3$, which is the quantity often referred to as $\alpha$ in publications. Due to the parametric choices made in the different search pipelines, $\alpha$ is most frequently used in LVK literature, while $\alpha'$ is most frequently used in the PTA literature.}
\begin{equation}
	\Omega^{\rm CB}_{\rm GW} = \Omega^{\rm CB}_{\rm GW}(f_\text{ref})\, \qty(\frac{f}{f_\text{ref}})^{2/3},
	\label{eq:astro_BG}
\end{equation}
where $\Omega^{\rm CB}_{\rm GW}(f_\text{ref})$ is the appropriately normalised energy density calculated at a reference frequency $f_\text{ref}$.

This simple model may be extended to all general inspiral-dominated GWBs as the same principles apply and has been reprised all over the literature and used in almost every GWB search with  data available. The same principle, for example, has been used in the estimation of the expected background from stellar mass black hole and neutron star binaries in the LIGO detector frequency band~\cite{LIGOimpli2018}, with some tweaks to account for the chirp signal from merging black holes which has a substantially different energy output per frequency. This result is shown in Figure~\ref{fig:LIGO_CBCBackground}.

\subsection{Primordial Backgrounds}
There are a range of potential physical processes occurring in the primordial Universe which would result in a rich variety of GW signals. We report the most noteworthy here; however, this is a nonexhaustive list of all sources which have been studied in the literature. We focus especially on sources which have corresponding search pipelines and for which constraints have been derived with existing GW data. For a more in-depth review of the subject, see~\cite{Caprini2018}.

First and foremost, all inflationary models include an irreducible GWB as a byproduct of inflation~\cite{Guzzetti:2016mkm, Caprini2018}. 
Any inflationary process results in the stretching of quantum fluctuations which are parametrically amplified into classical density perturbations, producing particularly tensor perturbations which we identify as GWs. The amplification process causes  perturbation wavelengths to expand until they reach the size of the horizon, when they exit causal contact and their evolution remains frozen until the Universe's horizon returns to their scale in the post-inflationary era and they may re-enter. At time of re-entry, the GWB is formed by standing waves, and their phases are correlated between modes with opposite momenta~\cite{Contaldi:2018akz}. 
In most models, the perturbations are adiabatic, Gaussian, and approximately scale-invariant, resulting in a similarly defined GWB. In this case, the fractional energy density will be approximately frequency independent at frequencies $f\gg H_0\sim10^{-18}\,\mathrm{Hz}$,
\begin{equation}
	\Omega^{\rm Infl.}_{\rm GW}(f) \approx \text{const.}\,,
\end{equation}
and will be primarily characterised by its amplitude.
The perturbations are also expected to imprint a pattern of B-modes in the polarization of the CMB~\cite{Kamionkowski1996}, which has not been detected yet and is the aim of numerous CMB experiments~\cite{COrE:2011bfs,Matsumura:2013aja,CMB-S4:2016ple,SimonsObservatory:2018koc}. This nondetection imposes an upper bound on the inflationary GWB amplitude, setting it well below the reach of present or planned GW detectors. The prevalent opinion in the community is that precision measurements of the CMB are still the best chance to garner decisive evidence of inflation and discriminate between models.

Beyond the irreducible background, however, the inflationary era may also source a second GWB which in certain cases dominates over the irreducible component, often presenting strongly scale-dependent features or characteristic peaks in the power spectrum~\cite{Caprini2018}. %
For example, the presence of additional fields during the inflationary epoch other than the inflaton could have an effect on GW emission, as these may have interactions resulting in strong particle production or may act as spectator fields and exhibit a sub-luminal propagation speed~\cite{Sachiko2009,InflParticles2000}. %
The presence of additional symmetries in the inflationary sector would also have an impact, as these would result in  breaking of space reparametrizations, allowing the graviton to acquire a mass. %
Alternative theories of gravity (other than general relativity) governing the inflationary era would have an impact on all aspects of primordial GWs. 
The scalar density perturbations generated during inflation may also source GWs~\cite{Domenech:2021ztg}; this may occur either at second order in perturbation theory~\cite{Ananda:2006af} or by acting as seeds for primordial black holes~\cite{Clesse:2018ogk,Garcia-Bellido:2017aan}.
Finally, GWs can also be produced during preheating~\cite{Khlebnikov:1997di,Easther2006,Garcia-Bellido:2007nns,Easther2007} due to the violent transfer of energy from the inflationary sector to  Standard Model plasma.
All these possible inflationary scenarios are effectively extremely hard to probe with electromagnetic observations; hence, GWs present a unique window into this epoch. 
See for example~\cite{Bartolo:2016ami} for a discussion of inflationary backgrounds in the LISA band and how a positive detection would revolutionise the current science of inflation. %
For more details and references to the specific models that may give rise to the effects discussed above, refer to the review~\cite{Caprini2018}.

After inflation, primordial GWs may have been generated as a consequence of phase transitions which may have occurred as the primordial Universe cooled down in its post-inflationary adiabatic expansion. First-order phase transitions in particular would proceed via the nucleation of true-vacuum bubbles, sourcing GWs through the collision of these bubbles~\cite{Grischuk1975,Hogan1986}, as well as through the resulting acoustic and/or turbulent motion of the primordial plasma~\cite{Caprini:2015zlo}.
The characteristic power spectrum of a background formed by bubble collisions presents a sharp peak at the frequency related to the energy scale at which the transition occurred, $T_*$,
\begin{equation}
    f_\mathrm{peak}\sim10^{-6}\,\mathrm{Hz}\times\frac{\beta}{H_*}\times\frac{T_*}{200\,\mathrm{GeV}},
\end{equation}
where $\beta/H_*$ is the rate of the transition in units of the Hubble rate at the transition epoch.
For typical values $\beta/H_*\sim10^3$, we see that this peak lies in the megahertz (mHz) band probed by LISA for transitions at the electroweak scale $T_*\sim200\,\mathrm{GeV}$, allowing us to test a myriad of particle physics models which predict such signals.

Phase transitions may also give rise to stable topological defects, the most prominent example of which are \emph{cosmic strings}. These line-like topological defects are generated through spontaneous symmetry breaking in the early Universe~\cite{Battye1997,Vilenkin2000} and are a generic prediction of many theories beyond the Standard Model of particle physics~\cite{Jeannerot:2003qv}.
On macroscopic scales, cosmic strings are effectively described by a single parameter, the dimensionless string tension $G\mu$, which is determined by the energy scale at which they were formed.
Once formed, dynamical evolution results in a dense network of cosmic string loops which oscillate at relativistic speeds, producing copious GWs which combine to produce a strong SGWB signal for which its amplitude and spectral shape are set by $G\mu$.

As mentioned above, a potential GW signal of cosmological origin is the SGWB from \emph{primordial black holes} (PBHs); i.e., black holes formed in the early Universe rather than through stellar evolution.
Since black holes are massive and non-baryonic and interact only through gravity, PBHs are very natural and well-motivated cold dark matter (CDM) candidates.
The cosmic mass density of PBHs is, therefore, usually expressed as a fraction of the total CDM mass density, $f_\mathrm{PBH}\equiv\rho_\mathrm{PBH}/\rho_\mathrm{CDM}$.
Similarly to stellar-origin black holes, PBHs can form binaries (either by forming in close proximity to each other or through dynamical encounters) which then inspiral and emit GWs or they can emit GWs through close hyperbolic encounters~\cite{Garcia-Bellido:2021jlq}.
The cumulative signal from many such binaries produces a SGWB spectrum which is determined by the PBH mass spectrum and  DM mass fraction~$f_\mathrm{PBH}$. 

In general, GWBs emitted in the primordial Universe are considered stochastic under the \emph{ergodic} assumption, which equates the ensemble average of an observable with its time and/or spatial average. 
Essentially, we assume that by observing large enough regions of the Universe today, or a given region for long enough time, we have access to many independent realisations of the system. This is realistic given two fundamental premises of cosmology, namely that the Universe is statistically homogeneous and isotropic, such that the initial conditions of the GW-generating process may be considered the same at every point in space and that the GW source fulfills causality, operating at a time when the typical size of a region of causal contact in the Universe was smaller than today.
It can be shown~\cite{Caprini2018} that the correlation scale today of a GW signal from the early Universe is tiny comparable to the present Hubble scale, verifying this requirement.

Stochasticity would also imply the lack of a preferred polarisation mode in the signal, as all mode configurations should be fully sampled in the observation set. While this is intuitively true for astrophysical backgrounds, the argument for primordial GWs is more delicate, and while a background may be stochastic for the reasons described above, there may be specific formation channels which will violate parity and induce an asymmetry in the polarisation modes, effectively generating circularly polarised waves. This effect may be found in alternative theories of gravity, such as Chern--Simons gravity~\cite{Alexander:2012ge} or certain quantum gravity models~\cite{Freidel:2005sn}. 
The study of parity-violating theories of gravity is inspired by grand unification, as the electro-weak sector of the standard model is chiral and, thus, maximally violates parity~\cite{Alexander:2007mt,Smolin:2007rx}.

\subsection{Anisotropies in Stochastic Backgrounds}

We can distinguish two root causes of anisotropy in stochastic signals. Firstly, there may be inhomogeneities in the GW source mechanisms, such as a particular distribution of the sources on the sky, which necessarily produces an anisotropic signal. Secondly, as  GWs propagate, they accumulate line-of-sight effects~\cite{Contaldi:2016koz}, crossing different matter density fields which are (if mildly) inhomoheneously distributed in our Universe. 
For astrophysical backgrounds, the spatial distribution of GW events is a biased tracer of the underlying light matter distribution, while 
primordial backgrounds will present anisotropies linked to the particular physical processes that occurred in the early Universe. 
In order to investigate these effects, it is useful to define the fractional GW energy density per solid angle $d\hat{\bm n}$,
\begin{equation}
	\Omega_{\rm GW}(f,\hat{\bm n}) = \frac{1}{\rho_c} \frac{d\rho_{\rm GW}}{d\ln f \, d\hat{\bm n}}\,,
\end{equation}
such that
\begin{equation}
	\overline{\Omega}_{\rm GW}(f) = \frac{1}{4\pi}\int_{S^2} d\hat{\bm n}\,\Omega_{\rm GW}(f,\hat{\bm n}) \,,
	\label{eq:Omegan}
\end{equation}
as in~\cite{Jenkins:2018nty,Jenkins:2018kxc}. $\overline{\Omega}_{\rm GW}(f)$ is the GWB monopole, where the bar is introduced to highlight the fact that it is an isotropic all-sky average over the directional fluctuations. The latter may be described by the GW density contrast,
\begin{equation}
	\delta_{\rm GW}(f,\hat{\bm n}) = \frac{\Omega_{\rm GW}(f,\hat{\bm n}) -\overline{\Omega}_{\rm GW}(f)}{\overline{\Omega}_{\rm GW}(f)}\,,
\end{equation}
in analogy with the formalism used to quantify the level of anisotropy in  CMB temperature~\cite{durrer_2008} and in galaxy clustering~\cite{Peebles2020}. This dimensionless parameter provides an unambiguous metric to compare different results.

At first order in the perturbations, the density contrast may be decomposed into three distinct contributions related to the origin of the anisotropies,
\begin{equation}
	\delta_{\rm GW} \simeq \delta^{\rm s}_{\rm GW} + \delta^{\rm los}_{\rm GW} + {\cal D} \hat{\bm n} \cdot \hat{\bm v}_{\rm obs}\,.
	\label{eq:deltaGW}
\end{equation}
Here, $\delta^{\rm s}_{\rm GW}$ is the \emph{source} term resulting from anisotropies at the GW source, for example, due to the specific distribution of events for astrophysical backgrounds, while $\delta^{\rm los}_{\rm GW}$ is the \emph{propagation} term~\cite{Contaldi:2016koz}, which encapsulates accumulated line-of-sight effects. The final term is the dipole of magnitude ${\cal D}$ induced by the observer's peculiar velocity ${\bm v}_{\rm obs}$. Note that the propagation term includes contributions similar to those encountered in CMB anisotropy calculations: the Sachs--Wolfe effect, due to the local curvature at emission, and the integrated Sachs--Wolfe effect, due to the curvature perturbations encountered along the line-of-sight~\cite{Sachs:1967er}; a Doppler term due to the source's peculiar velocity; and higher order effects such as gravitational lensing and redshift-space distortions~\cite{Bartolo:2019oiq}. 
All these terms have been the subject of intense study in recent years and both  astrophysical and primordial contributions have been modelled extensively (see for example~\cite{Cusin:2018,Jenkins:2018uac,Capurri:2021zli} for independent calculations of astrophysical background anisotropies and~\cite{Bartolo:2019zvb,Jenkins:2018kxc,Geller:2018mwu} for examples of primordial ones).
The propagation term is often assumed to be negligible in the case of astrophysical backgrounds or it is directly included as part of the source term; in~\cite{Bertacca:2019fnt,Bellomo:2021mer}, it is expressly estimated and shown to account for at most $\sim$10\% of the total anisotropy. In the case of a truly primordial background dating back as far as or before the CMB, however, this term will be considerably larger and will be comparable to the intrinsic anisotropies at large scales as shown in~\cite{Contaldi:2016koz}. 

We analyse here the general scenario where the density contrast at the source behaves as a Gaussian random field, as this is mostly assumed in the literature as well. In this case, $\delta^{\rm s}_{\rm GW}$ is characterised by its two-point correlation function,
\begin{equation}
	C_{\rm GW} (f,\cos\theta) = \expval{\delta^{\rm s}_{\rm GW}(f,\hat{\bm n})\, \delta^{\rm s}_{\rm GW}(f,\hat{\bm n}') }\,,
\end{equation}
in analogy with CMB anisotropies, as the first moment $\expval{\delta^{\rm s}_{\rm GW}}$ is zero by definition, and all higher order moments will either vanish or may be expressed in terms of $C_{\rm GW}$~\cite{Jenkins:2018kxc}. Here, $\cos\theta = \hat{\bm n}\cdot\hat{\bm n}'$ and the angular brackets represent an averaging over all $\hat{\bm n},\,\hat{\bm n}'$ direction pairs of aperture $\theta$. $C_{\rm GW} (f,\cos\theta)$ will then be different from zero where there are consistent spatial correlations on the sky associated to $\theta$, at frequency $f$. To observe this formally, we expand the two-point function in spherical harmonic space as
\begin{equation}
	C_{\rm GW} (f,\cos\theta) = \sum_{\ell = 0}^\infty \frac{2\ell + 1}{4\pi} C^{\delta}_\ell(f) P_\ell(\cos\theta)\,,
\end{equation}
where $P_\ell$ is the $\ell$th Legendre polynomial. By inverting this equation, we obtain the angular power spectrum $C^\delta_\ell(f)$ for the GW energy density contrast,
\begin{equation}
	C^{\delta}_\ell(f) = 2\pi\,\int_{-1}^{+1} d\cos\theta \, C_{\rm GW} (f,\cos\theta) \, P_\ell(\cos\theta)\,,
	\label{eq:Cell}
\end{equation}
which may be estimated from observations. Note that  quantity $\ell(\ell + 1)C_\ell/2\pi$ is roughly the contribution per logarithmic $\ell$ bin to the variance of $\delta^{\rm s}$; hence, this will often be the quantity shown in plots quantifying the $C^{\delta}_\ell$ power spectrum from different signal models and in GW data sets. 

Alternatively, we can also calculate the angular power spectrum of the $\Omega_{\rm GW}$ parameter itself, first by decomposing it into spherical harmonic components,
\begin{equation}
	\Omega_{\rm GW} (f,\hat{\bm n})= \sum_{\ell=0}^\infty \, \sum_{m=-\ell}^{+\ell}\,\Omega^{\rm GW}_{\ell m}(f) \, Y_{\ell m}(\hat{\bm n})\,,
\end{equation}
and then by defining
\begin{equation}
	C_\ell^\Omega = \frac{1}{2\ell + 1} \sum_{m=-\ell}^{+\ell}\expval{\Omega^{\rm GW}_{\ell m}(f)\, \Omega^{\rm GW}_{\ell m}(f)}\,,
    \label{eq:Cell_omega}
\end{equation}
where  angle brackets indicate an ensemble average over random realizations of $\Omega_{\rm GW}$. The information encapsulated in $C_\ell^\Omega$ and $C_\ell^\delta$ is the same, and they are both dimensionless quantities; however, note that the numerical scale of $C_\ell^\Omega$ is roughly that of $\Omega^2_{\rm GW}$, whereas $C_\ell^\delta$ shows the fractional contribution of each mode to the power spectrum with respect to the monopole $\overline{\Omega}_{\rm GW}$.

To quote the expected levels of anisotropy for astrophysical backgrounds from calculations in the references mentioned above, one may compare Figure 2 in~\cite{Cusin:2018} and Figure~4 in~\cite{Jenkins:2018uac}. It has been pointed out in the literature~\cite{Jenkins:2018kxc,Cusin:2018ump,Jenkins:2019cau,Pitrou:2019rjz} that there is a tension between these two predictions; the difference in the estimates highlighted here stems probably from different treatments of the source effects at very low redshift, where the clustering at small scales induces non-linear effects in the field which the overall anisotropy level is very sensitive to. Cosmological backgrounds are expected to present CMB-like anisotropies of $\delta^{\rm s}_{\rm GW}\sim10^{-5}$~\cite{Contaldi:2016koz}. A comparison between these estimates and the expected sensitivity-per-multipole of various GW detectors may be found in~\cite{Alonso2020}.

Finally, note that to properly assess the sourced GW anisotropies from direct measurements, it is necessary to subtract out the kinematic dipole term induced by the observer's peculiar velocity as seen in Equation~(\ref{eq:deltaGW}). This may be performed similarly to CMB analysis, potentially using the independent estimate of $\bm v_{\rm obs}$ from CMB data themselves.

\subsection{Observational Properties of Stochastic Backgrounds}\label{ssec:observational_props}


The observational properties of the SGWB such as time-domain characteristics and selection effects imposed by the observation method merit special attention as these are crucial for developing effective search methods. For the sake of simplicity, this discussion is focused on astrophysical backgrounds where individual events are clearly defined and are not extremely model dependent. However, most of the following considerations generally apply to primordial backgrounds as well and have intuitive extension to backgrounds from topological defects and primordial black holes for example.

\begin{figure}[t]
\centering
\includegraphics[width= .7\textwidth]{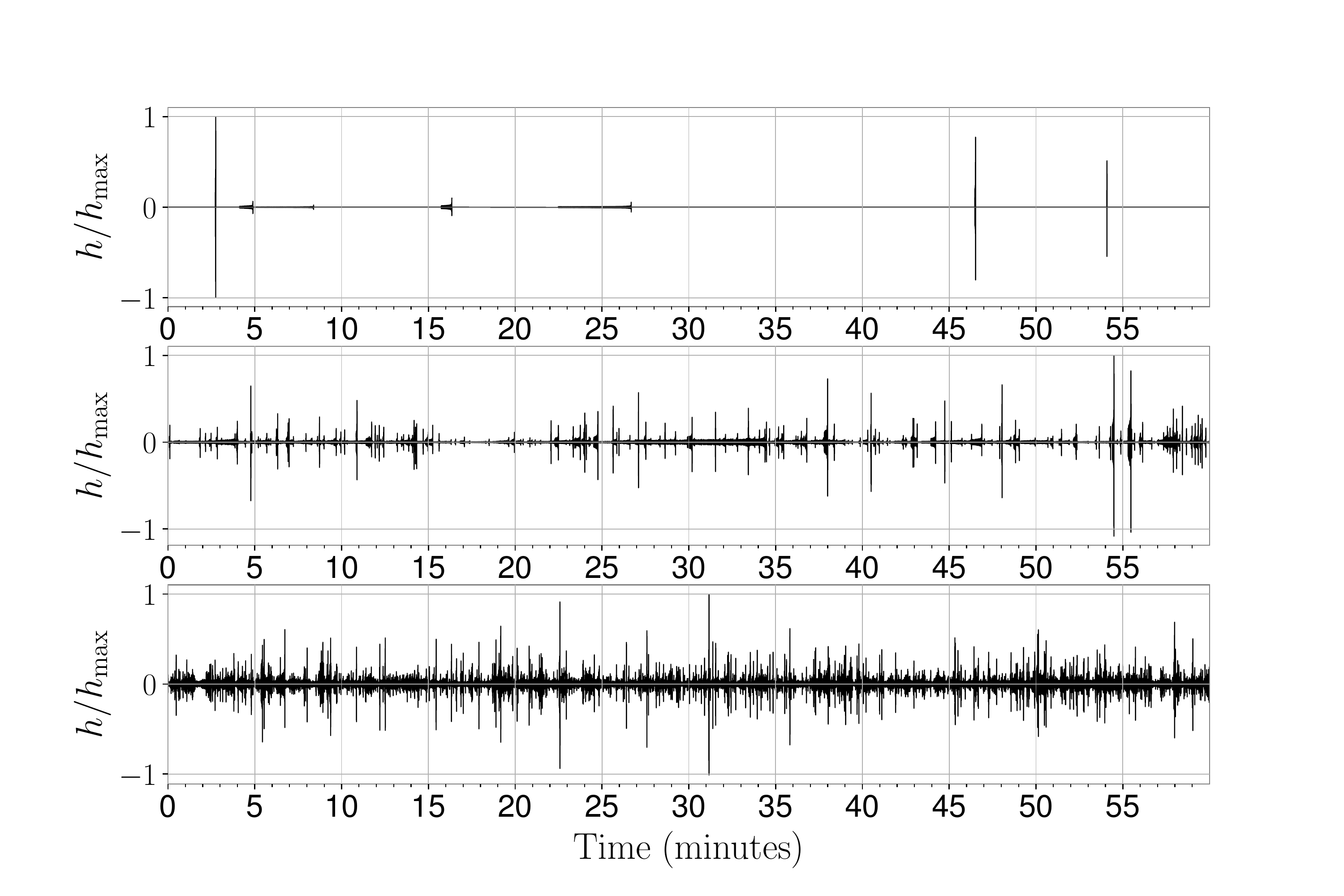}
\caption[Time-domain properties of gravitational-wave backgrounds.]{Examples of interferometer timestreams populated by GWBs with different time-domain properties, inspired by~\cite{Regimbau2011}: Poisson noise (top panel), popcorn (middle panel), and Gaussian (bottom panel). The strain on the $y$-axis is normalised to show the fractional contribution of each burst/wavelet to the signal, while the $\Delta$ parameter is the \emph{duty cycle}, i.e.,  the mean ratio between event duration and the time interval between consecutive events. }  
\label{fig:3types}
\end{figure}
 Three main regimes are identified in the literature based on the frequency of occurrence of the events in time. These are referred to as \emph{Poisson noise} regime, where  GWs appear as isolated bursts in the timestream, \emph{Gaussian} regime, where the timestream is packed with numerous overlapping signals and the central limit theorem applies, and  \emph{popcorn} or \emph{intermittent} regime, which lies somewhere in between. 
 A visualisation of these is provided in Figure~\ref{fig:3types}, generated using  mock-data generation open source package presented in~\cite{MDC1, MDC2}. The discriminating factor between regimes is encapsulated in the {\it duty cycle}, which is the mean ratio between event duration and the time interval between consecutive events. This will also be equal to the average number of events present at the detector at a given observation time.
While it is intuitively clear that the Gaussian background in Figure~\ref{fig:3types} is stochastic, this is not obvious of the other regimes. Operationally, a signal may be considered stochastic if it is more effective, in the Bayesian sense of the word, to search for it in the data using a stochastic model for the waveform as opposed to a deterministic one~\cite{Cornish:2015pda}. Furthermore, a signal is defined as \emph{resolvable} if it can be decomposed into non-overlapping and individually detectable signals, in either the time or frequency domain. With this in mind, the only way to know whether a signal is stochastic, is to search for it and see what method yields the best result or more stringent upper limit.
In the case of astrophysical GW signals, there is a continuous distribution of events where the highest SNRs are expected to be resolved, and beneath a certain detector-imposed cutoff, the events accumulate to form the background signal. The boundary between the resolved and unresolved population is not obvious, and a few bright signals stand out above the confusion background. When these are resolvable, deterministic signals, they should be subtracted from  data, leaving a residual background that may be analysed with stochastic tools. 
In~\cite{Cornish:2015pda}, the authors present simulations designed to establish whether certain signals are best searched for with stochastic or deterministic methods. They find, similarly to~\cite{Rosado:2015epa} and somewhat unsurprisingly, that deterministic signal models are favoured whenever the background is made up of a small number of sources, and stochastic models are preferred for larger source numbers. The boundary between regimes however is not well defined, and hybrid models comprised of a deterministic signal model for individual bright sources and a Gaussian--stochastic signal model for the confusion background are generally most effective to extract the population parameters.

These analyses highlight the distinction between \emph{event rate} and \emph{detection rate}. The event rate is the actual rate at which binaries are merging in the visible Universe, while the detection rate is the rate with which we can detect them, which is plagued by selection effects depending on both the binary's intrinsic and extrinsic parameters and the detector's characteristics. 

There is a further degree of complication caused by these unresolved, bright signals in the timestream which behave as \emph{shot noise} and may hinder the correct interpretation of a detection. These signals represent  random, Poissonian fluctuations in the finite population of sources and will induce a white noise component in the angular power spectrum. The observational effect that arises is induced by the fact that nearby, bright sources approximately randomly distributed around the observer will mask the underlying structure of the Universe. This is analysed and quantified in~\cite{Meacher2014} and~\cite{Jenkins:2019uzp,Jenkins:2019nks} for the monopole and anisotropies of an astrophysical background, respectively. Here,  Poisson white noise is modelled with a free parameter representing the cutoff distance from Earth beyond which the GW signal may be considered a stochastic background by the definition given above. The cutoff naturally depends on the sensitivity of the detector array under consideration. This is analysed in~\cite{Jenkins:2019uzp,Jenkins:2019nks} for the anisotropic background, and it is found that the white noise component dominates current LIGO and Virgo detectors; however, the planned upgrades of these detectors and the future 3G detectors may be sufficient to curb this effect.
As discussed in~\cite{Alonso:2020mva,Canas-Herrera:2019npr}, another strategy to mitigate the shot noise is to cross-correlate the GW signal with a galaxy catalogue, as this cross-correlation spectrum has a much lower level of shot noise.
Furthermore, as GW sources are finite in time, then Poisson statistics dictate that this noise decays as the inverse of the observation time; hence, long observing runs will progressively mitigate this effect.
Quantifying this distinction is an evergrowing effort. The current event rate estimates based on astrophysical models informed by the LVK detections show that there is approximately one binary neutron star merger every 13 s and one binary black hole merger every 223 s~\cite{LIGOimpli2018}. Clearly, the vast majority of these lie well beneath the threshold for detection. This is quantified for example in~\cite{Dvorkin2018} where the detection rate per year is encoded as a function of the merging binary masses. It is shown that there are several astrophysical models for compact binaries which fit the detection rate of $\sim$10 events per year from the aLIGO O1 and O2 data runs. It is clear that the rate of ``intermediate'' subthreshold events from an astrophysical population following these models is low and remains far from the stochastic limit as defined above.

It is worth pointing out that these observational features of astrophysical backgrounds discussed here may themselves provide insight in compact binary population parameters such as  binary coalescence rates for different binary species, as detailed in~\cite{Mukherjee:2019oma}. Furthermore, the time dependence of astrophysical backgrounds in general may be a powerful discriminating factor between these and backgrounds of primordial origin.

\section{Detection Approaches and Methodologies}\label{sec:approaches}

Given its stochastic nature, the main challenge in measuring an SGWB is confidently distinguishing signal from the (stochastic) noise in GW detectors. The interplay between signal and noise bears a central role in stochastic search methods, and in particular the validity of a given method may depend on the expected \emph{signal-to-noise regime} or \emph{ratio} (SNR): whether the noise dominates the measurement, i.e., the detector operates in the \emph{low signal regime}, or whether the signal is comparable to or stronger than the detector noise, i.e., \emph{high signal regime}. Here, we classify search methods based on assumptions made primarily on the signal, such as Gaussianity and isotropy, starting from minimal assumptions on SNR, and we highlight how the latter then impacts the formulation of the optimal estimators. 

Cross-correlation searches have cast a wide net, targeting a variety of stochastic backgrounds within the same style of search and also refining methods to extract specific signals from the data---see, e.g., the dark photon search~(e.g., \cite{LIGOScientific:2021odm}) or the cosmic string search~(e.g., \cite{LIGOScientific:2021nrg}). 
We divide  searches into three broad categories. We present the methodology to search for an isotropic stochastic background, which relies on the isotropic assumption to simplify the estimator considerably. Then, we show what happens when this assumption is relaxed in searches for anisotropic backgrounds. Finally, we review special searches which target elusive signatures which may imply deviation from general relativity and/or the standard model. However, first, let us review the common denominators in stochastic searches, namely the \emph{cross-correlation} statistic and the correlated detector response.

\subsection{Interdetector and Spatial Correlations}\label{ssec:correlations}

The first step in most stochastic searches is to consider the auto- and cross-correlation of datastreams from one or more detectors, as this operation discards all phase information and targets the signal power directly. To then deconvolve the response of the detectors to the signal, many searches rely on an \emph{overlap function}, which determines the cross-power response of a pair of detectors to a stochastic background. 
The most common approach is to estimate the significance of excess correlated power in multiple detectors, integrated over all available observing time, often assuming noise in different detectors
not to be correlated\footnote{There are, in fact, sources of spatially correlated noise. They can be distinguished from SGWBs by either a deterministic component or by an overlap reduction function different to those of SGWBs. We provide more details in Section~\ref{sec:efforts}.}. Throughout this section, these methodologies will be reviewed in detail; first, however, let us formally introduce the cross-correlation statistic and derive the overlap functions for different detector setups as these will be the building blocks for the rest of this review.

To start, consider a Gaussian, stationary GWB observed through two distinct detectors, 1 and 2. Each detector timestream may be decomposed into signal and noise components~as
\begin{equation}
    d_1(t) = R_1(t)* h(t) +n_1(t),\qquad d_2(t) = R_2(t) * h(t) +n_2(t)\,,
    \label{eq:data_model}
\end{equation}
where $R_1$ and $R_2$ represent the individual detector response functions to GW signals, and ``$*$'' represents a convolution. These are characterised by functional dependencies on intrinsic and extrinsic parameters of the GWs such as polarisation and incoming direction; however, we omit these now as they may be easily included at a later stage.

We consider both $h$ and $n$ fields in Equation~\eqref{eq:data_model} as mean zero Gaussian variates; hence, these are fully described by their second-order moments. Let us work in the Fourier domain, where the assumption of stationarity implies that the noise covariance is diagonal, i.e., the noise in each frequency bin is uncorrelated\footnote{Time segments with non-stationary noise are removed from analyses of real data~\cite{LVK:2021kbb}, the stationarity is usually determined empirically. Note that when applying a non-trivial windowing function to  time-domain data, e.g., for computing a discrete Fourier transform, we introduce correlations between frequency bins. This effect and the methods to mitigate it are described in~\cite{Talbot:2021igi}.},
\begin{equation}
    \ev{n_i(f)n_i(f')} = \delta_{ff'} P_{f,i}\,,
    \label{eq:noise_PSD}
\end{equation}
with $P_{i}$ as the noise power spectrum in detector $i$. We define the discrete Fourier transforms of the data streams starting at timestamp $\tau$ as $\tilde{d}^\tau_1$, $\tilde{d}^\tau_2$, and drop the time label immediately for the sake of conciseness. We then take the cross-correlation of these quantities as
\begin{equation}
    C_{12}(f) = \tilde{d}_1^{}(f) \tilde{d}_2^\star(f)\,,
\end{equation}
which, assuming uncorrelated noise between detectors, has expectation value of
\begin{equation}
    \langle C_{12}(f) \rangle = R^{}_1(f) R^\star_2(f) \langle \tilde{h}_1^{}(f) \tilde{h}_2^\star(f) \rangle  = T_{\rm obs} \Gamma_{12}(f) I_{\rm GW}(f)\,.
    \label{eq:crosscorr}
\end{equation}
$\Gamma_{12}$ is the \emph{overlap reduction function}, where the \emph{reduction} comes from the fact that it is a sky-integrated quantity (more details to follow), and $I_{\rm GW}(f)$ is the sky-integrated power spectrum of the background. Note the normalisation $T_{\rm obs}$ as we assume the observation lasts a finite amount of time.
The covariance of the correlated measurement is as
\begin{equation}
    {\rm Cov}[C^{}_{12\,f,f'}] = \langle C_{12}(f)C_{12}^\star(f')\rangle - \langle C^{}_{12}(f)\rangle \langle C_{12}^\star(f')\rangle \approx T_{\rm obs}\delta_{ff'} P_{1, f} P_{2, f}\,,
    \label{eq:crosscorr_cov}
\end{equation}
taking the weak signal limit in the final equality. In this case, the cross-correlation statistic has been shown to be a \emph{sufficient statistic}~\cite{Matas:2020roi}, making it a good choice for stochastic searches. This is because there is little information to be drawn out from the auto-correlated measurements, as these are effectively measurements of the detector noise power; conversely, to use auto-correlations to estimate the signal, one would need an independent and accurate measurement of the noise power to model and subtract out the noise. This approach would require an iterative estimator where both signal and noise are included in the fit~\cite{Littenberg:2020bxy}.
Note that using the cross-correlation component only of the data and, thus, ignoring the auto-correlations allows for considerable simplification of the optimal estimators and computations necessary in data analysis.

The functional form of the overlap function of a detector network can be derived from Equation~\eqref{eq:crosscorr} by appropriately combining the individual response functions of each detector. In the case of laser interferometers, the basic response function is the \emph{arm response function}, and arms are then combined to form the detector. This may be performed physically, by using laser interferometry as is performed in LIGO/Virgo~\cite{ALIGO2015}, or in post-processing, with \emph{time-delayed interferometry} (TDI) as in the case of LISA~\cite{TDI1,TDI2}. Considering arm ${\bm u}_{PS}$ spanning from point $P$ to point $S$, the one-way arm response $\overrightarrow u: P\rightarrow S$ in the frequency domain is described as~\cite{Romano2017}
\begin{equation}
    R^A_{\overrightarrow{u}} (f, \hat{\bm n}) = \frac{1}{2} {\bm u}\otimes {\bm u} : {\bm\epsilon}^A {\cal T}_{\overrightarrow{u}}(f, \hat{\bm n}\cdot {\bm u}) e^{i \frac{2\pi f}{c} \hat{\bm n}\cdot {\bm x}_S}\,,
    \label{eq:oneway_arm_response}
\end{equation}
with the one-way \emph{timing transfer function} relative to a photon travelling along $\bm u_{PS}$, ${\cal T}_{\bm u}$, as
\begin{equation}
    {\cal T}_{\overrightarrow{u}}(f, \hat{\bm n}\cdot {\bm u}) = \frac{L}{c} e^{-\frac{\pi f L}{c}(1+ \hat{\bm n}\cdot \hat{\bm u})} \text{sinc}\qty(\frac{\pi f L}{c}(1+ \hat{\bm n}\cdot \hat{\bm u}))\,.
    \label{eq:one_way_timing}
\end{equation}

\begin{figure}[t]
     \subfigure[Ground-based network]{
        \label{fig:overlap_reduction_function_LVK}
        \includegraphics[width=0.48\linewidth]{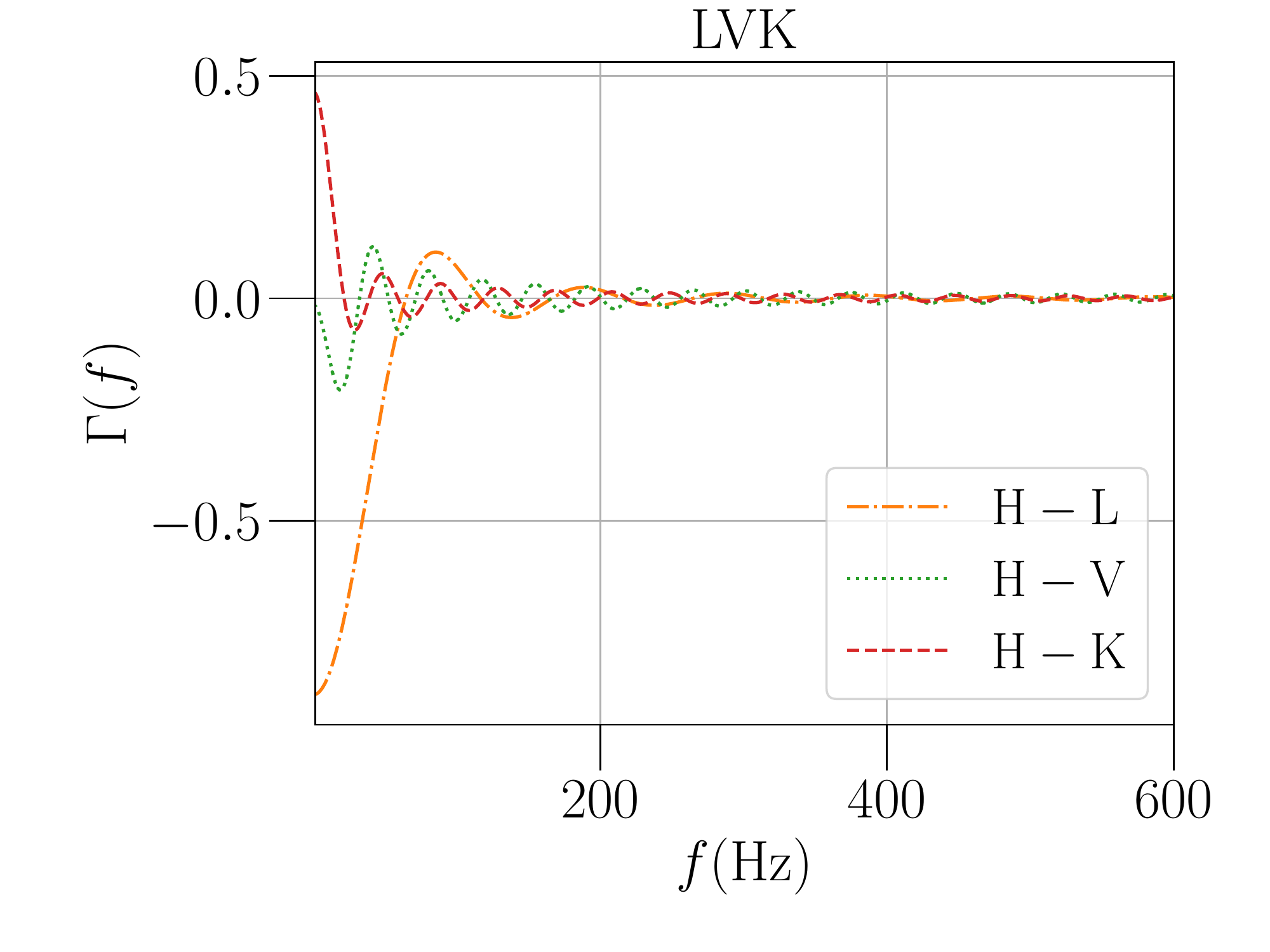}
    }
    \subfigure[LISA TDI channels]{
        \label{fig:overlap_reduction_function_LISA}
        \includegraphics[width=0.48\linewidth]{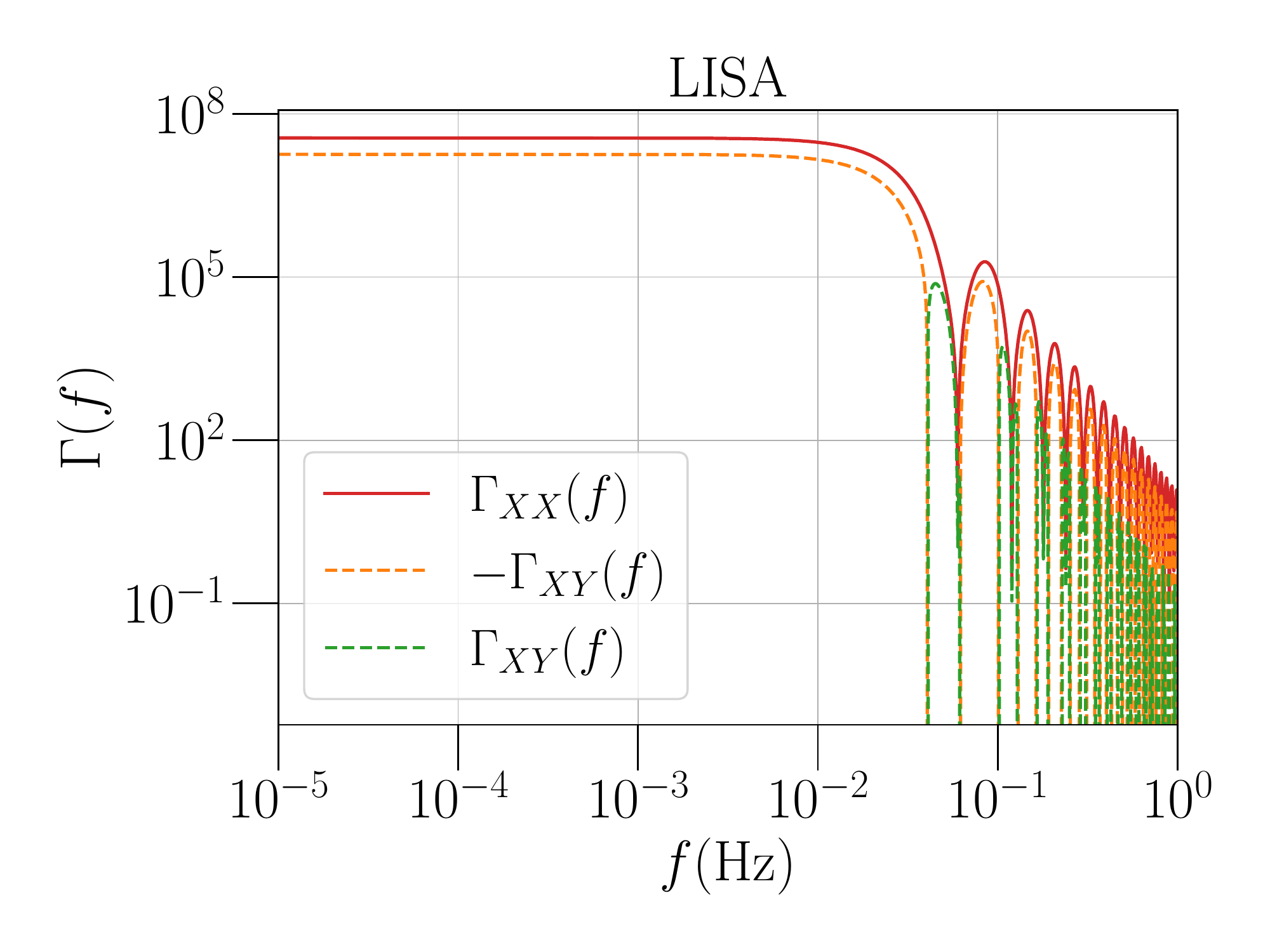}
    }
    \caption{Overlap reduction functions for different GW interferometer detectors: ground-based (left) and space-based (right). 
    Note that, in panel (\textbf{b}), the positive and negative parts of the LISA TDI overlap function $\Gamma_{XY}$ are plotted separately, as the y-axis is log-scaled.}
    \label{fig:overlap_reduction_function}
\end{figure}

\begin{figure}[t]%
    \includegraphics[width=0.99\linewidth]{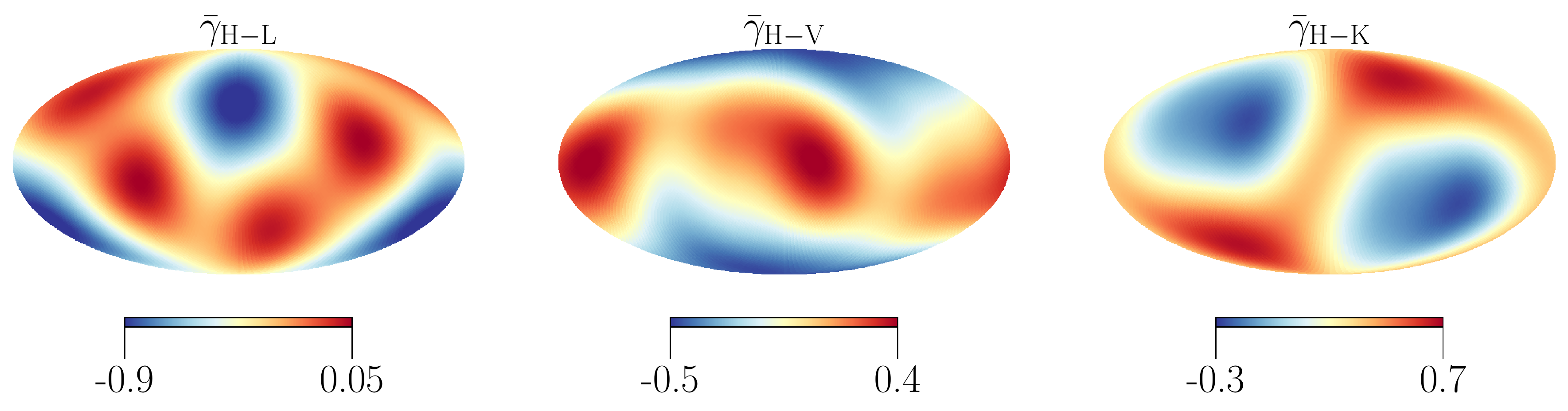}
    \caption{Geometric components of the overlap functions for different ground-based detector combinations mapped on the sky in geocentric celestial coordinates at a common time.}  
    \label{fig:overlap_function_maps}
\end{figure}

In the case of LIGO-like detectors, which are $90^\circ$-arm, Michelson--Morley style interferometers, a laser photon performs a round trip along one arm before interfering with the beam in the other arm; hence, the relevant response function is the \emph{two-way} arm response function along $\overleftrightarrow u: P\leftrightarrow S$, which we can write in terms of the one-way timing transfer function as
\begin{equation}
    R^A_{\overleftrightarrow u} (f, \hat{\bm n}) = \frac{1}{2} {\bm u}\otimes {\bm u} : {\bm\epsilon}^A \qty({\cal T}_{\overrightarrow u}(f, \hat{\bm n}\cdot {\bm u}) e^{i \frac{2\pi f}{c} (\hat{\bm n}\cdot {\bm x}_S - L)} + {\cal T}_{\overleftarrow u}(f, \hat{\bm n}\cdot {\bm u}) e^{i \frac{2\pi f}{c} \hat{\bm n}\cdot {\bm x}_P})\,;
\end{equation}
note the extra $-L$ term which accounts for the extra phaseshift incurred when travelling over the same distance twice. Then, assuming that the detector is not moving during the measurement, we can substitute in ${\bm x}_S = {\bm x}_P + {\bm u}$, and rewrite the above as
\begin{equation}
    R^A_{\overleftrightarrow u} (f, \hat{\bm n}) = \frac{1}{2} {\bm u}\otimes {\bm u} : {\bm\epsilon}^A {\cal T}_{\overleftrightarrow u}(f, \hat{\bm n}\cdot {\bm u}) e^{i \frac{2\pi f}{c} \hat{\bm n}\cdot {\bm x}_P}\,,
    \label{eq:two_way_response}
\end{equation}
where the two-way timing transfer function along the same arm is derived as
\begin{align}
    {\cal T}_{\overleftrightarrow u}(f, \hat{\bm n}\cdot {\bm u})&= \frac{L}{c} e^{-i\frac{2\pi f L}{c} }\bigg[ e^{-i\frac{\pi f L}{c}(1-\hat{\bm n}\cdot \hat{\bm u})} {\rm sinc}\qty(\frac{\pi f L}{c}(1+ \hat{\bm n}\cdot \hat{\bm u}))\notag\\
    & + e^{i\frac{2\pi f L}{c}(1+\hat{\bm n}\cdot \hat{\bm u})} {\rm sinc}\qty(\frac{\pi f L}{c}(1 - \hat{\bm n}\cdot \hat{\bm u}))\bigg]\,.
\end{align}
Note that in the \emph{small antenna limit}, i.e., when the scale of the detector is much smaller than the wavelength probed, $L \ll c/f_{\rm GW}$, we have that $|{\cal T}_{\overleftrightarrow u}| \longrightarrow 1$, implying that all the frequency dependence of the detector response is encoded in the phase term of Equation~\eqref{eq:two_way_response}. Consider now detector 1 which is made up of arms $\bm u$ and $\bm v$, joined at point $P$; its polarisation response in frequency space, in the small antenna limit, will simply be
\begin{equation}
    F^{A}_1(f, \hat{\bm n}) = R^A_{\overleftrightarrow u} (f, \hat{\bm n}) - R^A_{\overleftrightarrow v} (f, \hat{\bm n}) = \, \frac{1}{2} \qty({\bm u}_1\otimes {\bm u}_1 - {\bm v}_1\otimes {\bm v}_1): {\bm\epsilon}^A e^{i \frac{2\pi f}{c} \hat{\bm n}\cdot {\bm x}_1}\,,
    \label{eq:MM_response}
\end{equation}
where we have substituted in ${\bm x}_1$ for ${\bm x}_P$ as we consider it to be the location of the detector itself.
Finally, the unpolarised overlap function\footnote{Note that this is result is derived from the detector response to a fully unpolarised stochastic background; in the case of a polarised signal, extra terms need to be considered. More details on this can be found, e.g., in~\cite{Seto2007, Seto2008}.} of a detector pair is given by combining the response functions of the two detectors~\cite{Romano2017},
\begin{equation}
    \gamma_{12} (f, \hat{\bm n}) \, = F^{+}_1(f, \hat{\bm n})F^{+}_2(f, \hat{\bm n}) + F^{\times}_1(f, \hat{\bm n}) F^{\times}_2(f, \hat{\bm n})\,,
    \label{eq:liorf}
\end{equation}
while the overlap reduction function $\Gamma$ is obtained by sky-integrating the above,
\begin{equation}
    \Gamma_{12}(f) = \int_{S^2} d \hat{\bm n}\, \gamma_{12} (f, \hat{\bm n})\,.
    \label{eq:skyintegralorf}
\end{equation}
Note that  polarisation $F$ functions here defined specifically for ground-based interferometers correspond to the individual detector response functions seen in Equation~\eqref{eq:crosscorr}. We have chosen to incorporate the frequency-dependent phase terms into the response functions directly; however, in the literature, these are sometimes left out of  $F$ functions to underline the fact that, in the small antenna limit, there is no interaction between  GW frequency and  detector geometry~\cite{Romano2017}. In the overlap function, the phase terms add, giving rise to a time delay term representing the frequency-dependent phase shift between the two detectors; explicitly,
\begin{equation}
    \gamma_{12} (f, \hat{\bm n}) = \bar{\gamma}_{12} (\hat{\bm n}) e^{i \frac{2\pi f}{c} \hat{\bm n}\cdot {\bm b}}\,,
\end{equation}
where $\bar{\gamma}$ is the geometric component of the overlap function, and $\bm b$ is the \emph{baseline vector} that points from detector 1 to detector 2. The numerical values of $\bar{\gamma}$ as a function of direction then depend specifically on the coordinate basis chosen to represent the function; often, the choice is to represent it in an Earth-centered celestial frame, where the function will be time-dependent, as the arms sweep round and round on the sky following the Earth's rotation on its axis. This behaviour gives rise to the specific \emph{scan strategy} of the detector array, which is essential for directional stochastic searches. On the other hand, the frequency-dependent phase term highlights the role of the baseline length in a stochastic search: the ratio between $\abs{\bm b}$ and GW wavelength will determine what frequencies will be suppressed in the search; the longer the baseline, the higher the frequencies accessible will be. The overlap reduction functions for the LIGO Hanford and Livingston detector pair (L--H), LIGO Hanford and Virgo (H--V), and LIGO Hanford and Kagra (L--K) are shown in Figure~\ref{fig:overlap_reduction_function}a, while the respective (instantaneous) overlap functions are plotted on the sky in Figure~\ref{fig:overlap_function_maps} at a particular common~time. 

\unskip
As mentioned above, long arm length, spaceborne detectors such as the future LISA three-satellite constellation, oriented in roughly an equilateral triangle, will measure GWs by combining the one-way arm responses~\eqref{eq:oneway_arm_response} of several arms at different times composing TDI configurations or \emph{channels}\footnote{Note that directly interfering laser beams in the case of LISA is impossible due to energy dispersion along the arm~\cite{Amaro2017}.}. The most basic configuration can be thought of as a TDI--Michelson--Morley interferometer, where one constructs a response such as that in Equation~\eqref{eq:MM_response}. However, the great advantage of TDI is that it is possible to construct channels that suppress  detector noise and enhance the signal~\cite{Bayle:2021mue}; this is a very active area of study~\cite{BayleLilley2019,Vallisneri:2020otf,BayleVallisneri2021,TintoDhurandhar2021b,PageLittenberg2021,BaghiThorpe2021}.
For the purposes of this review, let us cite the TDI $\{X,Y,Z\}$ set of channels, which are based on the Michelson--Morley style measurement in Equation~\eqref{eq:MM_response}, where nominally $X$ is centered around spacecraft 1, referred to hereafter as $S_1$; $Y$ around spacecraft 2, $S_2$; and $Z$ around spacecraft 3, $S_3$. However, to minimise  noise, each light path is ``reflected'' back and forth between spacecrafts multiple times such that the noise interferes destructively. The total number of reflections is then linked to a version number for the TDI channel; e.g., ``TDI 1 $X$'' refers to a channel constructed by ``interfering'' the following light paths:
\begin{equation}
    \qty( S_1 \rightarrow S_2 \rightarrow S_1 \rightarrow S_3 \rightarrow S_1) - \qty( S_1 \rightarrow S_3 \rightarrow S_1 \rightarrow S_2 \rightarrow S_1 ).
\end{equation}
For more details on this calculation, see also official LISA documentation~\cite{LDC_manual}. Another useful set of channels is the orthonormal set which approximately diagonalise the $X, Y, Z$ noise-correlation matrix, $A, E, T$~\cite{Tinto:2004wu}. This set is often used for simplicity as one may consider the auto-correlations of $A$ and $E$ only, treating them as separate measurements, and use $T$ as a \emph{Sagnac} channel which is mostly sensitive to detector noise and blind to the signal~\cite{adamscornish1, adamscornish2}. However, in practice diagonalizing $X, Y, Z$ is not a simple task as the detectors continuously move with respect to the measurement. The auto- and cross-correlated overlap functions for LISA are constructed as in Equation~\eqref{eq:liorf}, where now the polarisation responses are referred to TDI channels. Note that LISA does not operate in the small antenna limit, hence the timing transfer function remains frequency dependent, inducing oscillations in  TDI responses at high frequency, as observed in Figure~\ref{fig:overlap_reduction_function}b. 

\begin{figure}[t]
    \centering
    \includegraphics[width = 0.8\linewidth]{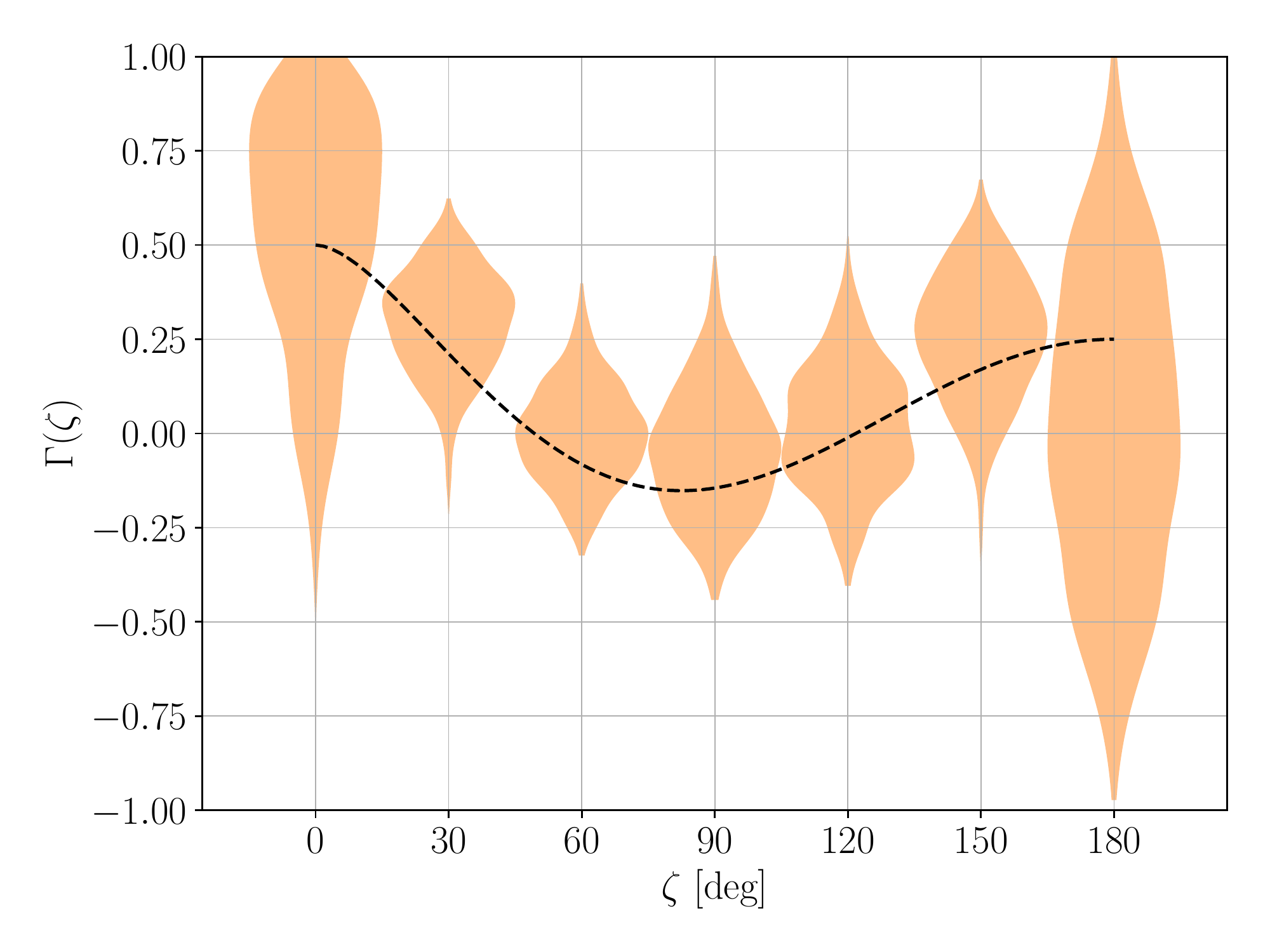}
    \caption{The dashed line is the Hellings--Downs curve described by Equation~\eqref{eq:orfhd}. $\zeta$ is the angle between Earth--pulsars baselines, and $\Gamma$ measures the spatial correlation. At $\zeta=0$, the function assumes the pulsars located at different distances. When distances are the same, $\Gamma(0) = 1$. Shaded regions correspond to constraints on spatial correlations obtained with the second data release of the Parkes Pulsar Timing Array except PSR~J0437--4715~\cite{goncharov2021cpgwb}.}
\label{fig:overlap_reduction_function_PTA} 
\end{figure}

In the case of pulsar timing arrays, we can use the one-way transfer function defined in Equation~\eqref{eq:one_way_timing}, as in this case the arm vector $\hat{\bm u}$, points from the solar system barycenter to the pulsar, and the arm length $L$ is the distance from the solar system barycenter to the pulsar. One can expand the sinc function using Euler's identity and isolate the exponentials with arguments that depend on $fL/c$. Those exponential terms vary rapidly with sky direction for $fL/c \gg 1$ averaging down to zero in Equation~\eqref{eq:skyintegralorf} for Earth--pulsar baselines greater than $100$ parsec in the nanohertz band~(e.g., Figure 1 in \cite{AnholmBallmer2009}). 
From here, it is straightforward to calculate the (now frequency-independent) overlap reduction function by considering the cross-correlation of data from two pulsars, which serve as one-way timing beams (see, e.g.,~\cite{Romano2017,HellingsDowns1983,AnholmBallmer2009} for a more complete discussion).
The overlap reduction function for a pair of pulsars $a$ and $b$ now depends on the angle between Earth--pulsar baselines, $\zeta_{ab}$~\cite{HellingsDowns1983}:
\begin{equation}
    \Gamma_{\text{PTA}}(\zeta_{ab}) = \frac{1}{2} - \frac{1}{4} \bigg( \frac{1 - \cos{\zeta_{ab}}}{2} \bigg) + 
     \frac{3}{2} \bigg( \frac{1 - \cos{\zeta_{ab}}}{2} \bigg) \ln{\bigg(\frac{1 - \cos{\zeta_{ab}}}{2}\bigg)}.
\label{eq:orfhd}
\end{equation}
where $\Gamma_{\text{PTA}}$ is commonly referred to as the \emph{Hellings--Downs} curve and may be observed in Figure~\ref{fig:overlap_reduction_function_PTA}.

\subsection{Isotropic Background Search Methods}
\label{ssec:detection:isotropic}
In this section, we review the case of isotropic backgrounds searches. We start by presenting the approach for ground-based interferometers, which may be considered the simplest case as these operate in the low-signal regime. We then lay out the detection strategy of PTAs. The application of these methods on real data and corresponding search results are reviewed in Section~\ref{sec:efforts}.

\subsubsection{Ground-Based Detectors}
\label{sssec:detection:isotropic:LVK}

The flagship LVK stochastic search targets a Gaussian, isotropic, stationary, and unpolarised background~\cite{LVK:2021kbb}. This set of assumptions results in a simple, ready-to-use estimator which results in a relatively rapid analysis of LIGO-Virgo data. All other collaboration searches use, as a starting point, the breakdown of one of these assumptions, and most extend the estimator to a particular case. 

As detailed in Section~\ref{sec:sources}, a stochastic signal which satisfies all assumptions listed above is fully described by its spectral energy density, which may be expressed in terms of the fractional energy density $\Omega_{\rm GW}(f)$. Under the assumption the latter presents a power-law spectrum as in Equation~\eqref{eq:Omegascale}, we can use the cross-correlation statistic presented in Equations~(\ref{eq:crosscorr}) and (\ref{eq:crosscorr_cov}) to write down an \emph{optimal filter} to apply to the data and draw out an estimate for the amplitude $\Omega_{\rm GW}(f_{\rm ref})$ at reference frequency $f_{\rm ref}$. To start, we remind the reader of Equation~\eqref{eq:omegatoI} relating the measured GWB intensity to the energy density, which we may rewritten as
\begin{equation}
    \Omega_{\rm GW}(f_{\rm ref}) = {\cal H}(f, f_{\rm ref}) I(f)\,,
\end{equation}
with
\begin{equation}\label{eq:intensity_spec}
    {\cal H}(f, f_{\rm ref}) = \frac{4\pi^2}{\rho_c G} f_{\rm ref}^3 \qty( \frac{f}{f_{\rm ref}} )^{3-\alpha}\,.
\end{equation}
By inspection, we already oberve that ${\cal H}(f, f_{\rm ref})$ is an unbiased filter to use on the data to derive an estimate of the background, in the case where the GWB obeys Equations~(\ref{eq:omegatoI}) and (\ref{eq:crosscorr}). In general, there are several methods to show that this filter is optimal and write an estimator for $\Omega_{\rm GW}$, and different assumptions on the signal and noise components result in distinct solutions. We report the derivation shown in~\cite{Allen1999}. The approach here is to find the function which maximises the SNR of the \emph{estimated} GWB signal, defined as
\begin{equation}
    {\rm SNR} = \frac{\hat{\Omega}_{\rm GW}|_{f_{\rm ref}}}{\sigma^2_\Omega}\,,
\end{equation}
where
\begin{equation}
    \sigma^2_\Omega = {\rm Cov}[\hat{\Omega}_{\rm GW}|_{f_{\rm ref}}] \,.
    \label{eq:sigmasqu}
\end{equation}
Re-writing the expectation value for the data correlation $C(f)$ in terms of $\Omega_{\rm GW}(f_{\rm ref})$ using Equations~(\ref{eq:omegatoI}) and (\ref{eq:crosscorr}),
\begin{equation}
    \langle C(f) \rangle = T_{\rm obs} \Gamma_{12}(f){\cal H}^{-1}(f,f_{\rm ref})\Omega_{\rm GW}(f_{\rm ref})\,,
\end{equation}
and assuming the low-signal limit following Allen and Romano~\cite{Allen1999}, one obtains
\begin{equation}
    \hat{\Omega}_{\rm GW}|_{f_{\rm ref}} = F_{f_{\rm ref}}^{-1}\int_0^\infty df \frac{\Gamma_{12}(f){\cal H}(f, f_{\rm ref})}{P_1(f)P_2(f)} C(f)\,,
\end{equation}
where $F_{f_{\rm ref}}$ may be thought of as a normalisation which depends on the chosen reference frequency,
\begin{equation}
    F_{f_{\rm ref}} = T_{\rm obs}\int_0^\infty df \frac{\Gamma^2_{12}(f) {\cal H}^2(f, f_{\rm ref})}{P_1(f)P_2(f)}\,.
\end{equation}
In fact, in deriving the variance $\sigma^2_\Omega$ as formulated in Equation~(\ref{eq:sigmasqu}), it turns out that $F_{f_{\rm ref}}$ is equal to the inverse variance of the estimator, up to a factor of 2,
\begin{equation}
    \sigma^2_\Omega = \frac{1}{2} \qty[\int_0^\infty df \frac{\Gamma^2_{12}(f) {\cal H}^2(f, f_{\rm ref})}{P_1(f)P_2(f)}]^{-1} \equiv  \frac{F_{f_{\rm ref}} }{2}\,.
\end{equation}
where $P_1$ and $P_2$ are the noise power spectra in detectors 1 and 2, respectively, as defined in Equation~\eqref{eq:noise_PSD}. In principle, these are not known and need to be estimated carefully so as not to bias the search. When operating in the weak signal, it is reasonable to calculate the noise power spectra from the data directly; however, one should be cautious not to use these estimates on \emph{the very same} data stretch they have been estimated from. In practice, for stochastic searches using ground-based interferometers, the data are segmented into short time segments and the full estimator takes the average of $\Omega_{\rm GW}$ over the entire dataset. To estimate $\Omega_{\rm GW}$ in a single data segment $\tau$, one may use noise power spectra estimated from \emph{other} segments within a time window over which  detector noise is stationary; for example, in the LVK stochastic searches, the average of $P_i$ in the adjacent segments is taken. Otherwise, it is also possible to estimate the noise power spectrum using a parametric fit, as in~\cite{Renzini2018,Renzini2019a,Renzini2019b}, or by using a wavelet expansion using methods as shown for example in~\cite{Chatziioannou2019}.

Another approach to construct an estimator for the SGWB is to construct a likelihood function for the signal given the data, and maximise it to obtain a maximum likelihood detection statistic. Let us switch to a compact vector notation, where the data vector ${\bm d} = (d_1, d_2, ..., d_N)$ spans the detector space comprising $N$ detectors, and  data covariance is the generalised $N\times N$ correlation matrix $\bm C$, where each entry is $C_{ij} = d_i d_j$.
Starting with a zero-mean Gaussian assumption for both the signal $h$ and noise $\bm n$ terms in the data, we can write
\begin{equation}
    {\cal L}(\bm d) \propto \prod_{f,\tau} \frac{1}{|\bm C|^{1/2}} e^{\frac{1}{2} \bm d^\dagger \bm C^{-1} \bm d}\,,
    \label{eq:likelihood_data}
\end{equation}
where all observed time segments $\tau$ and frequency modes $f$ are considered independent. A method based on this likelihood employs both auto-correlations and cross-correlations of the datastreams to reconstruct  SGWB and is discussed in the context of ground-based detectors in~\cite{Cornish:2013nma}. Using auto-correlations is not recommended in scenarios where the detector noise terms are not independently measured, as these will be inextricable from the auto-power spectra. This likelihood is in fact proposed for future LISA data analysis in~\cite{adamscornish1,adamscornish2}, making full use of the independent estimate of the noise from GW-insensitive channels (see details on Sagnac channels here~\cite{Cornish:2001bb}). A more simple, alternative route can be taken by noting that, in the low-signal approximation, the Gaussian assumption can actually be applied to the average of the two-detector residual $C_{ij} - \langle C_{ij} \rangle$, leading to a different formulation,
\begin{equation}
    {\cal L}(C_{ij}) \propto \prod_{f,\tau} \frac{1}{\sigma^2} e^{\frac{1}{2} (C_{ij} - \langle C_{ij}\rangle)\sigma^{-2} (C_{ij} - \langle C_{ij}\rangle)^\star}\,.
    \label{eq:likelihood_residuals}
\end{equation}
This holds in the limit where  data are averaged over many segments such that the distribution of the residuals approximates a Gaussian; otherwise, a single residual will not follow this distribution~\cite{Coughlin:2014}.
In~\cite{Matas:2020roi}, Matas and Romano show that likelihood~\eqref{eq:likelihood_data} is well approximated by~\eqref{eq:likelihood_residuals} in the low-signal limit.
In this case, maximising either likelihood with respect to $\Omega_{\rm GW}(f_{\rm ref})$ yields the same solution as above. The extension to include multiple baselines is straightforward, assuming each $C_{ij}$ may be considered as an independent measurement. The full likelihood ${\cal L} (C)$ becomes the product over the individual baseline likelihoods as
\begin{equation}
    {\cal L}({\bm C}) \propto \prod_{b} {\cal L}({\bm C}_b) ,
\end{equation}
where the baseline index $b$ cycles over the detector pairs $ij$ in the set, without double counting baselines. 

The approach can be taken one step further to formulate a hybrid frequentist--Bayesian analysis as proposed in~\cite{PhysRevLett.109.171102} by re-expressing the likelihood in terms of $\Omega$ and the model parameters we would like to constrain,
\begin{equation}
    {\cal L}(\hat{\Omega}_{\rm GW}(f) | \Theta) \propto {\rm exp}\qty[{-\frac{1}{2} \sum_{b, f} \qty(\frac{\hat{\Omega}^b_{\rm GW}(f) - \Omega_{\rm M}(f|\Theta)}{\sigma_{\Omega, b}^2(f)})^{2}}]\,,
\label{eq:Omega_like}
\end{equation}
where $\Omega_M$ is the fractional energy density of the \emph{model} $M$ which is being tested, $\Theta$ are the parameters on which the model depends, and $\hat{\Omega}^b_{\rm GW}(f)$ is an estimator for the strength of the SGWB in a single frequency bin~(see, e.g., \cite{Callister:2017ocg}).

Equation~(\ref{eq:Omega_like}) is used as the basis for parameter estimation of models for the GWB, and Bayesian model selection~\cite{PhysRevLett.109.171102}. In the case of parameter estimation, one estimates the posterior distribution of the parameters of the model, $\Theta$,
\begin{align}
    p(\Theta | \hat\Omega_{GW}(f), M) = \frac{\mathcal{L}(\hat{\Omega}_{\rm GW}(f) | \Theta) p(\Theta)}{p[\hat{\Omega}_{\rm GW}(f) | M]},
    \label{eq:omgw_posterior_ground_based}
\end{align}
where now we have specified  model $M$ explicitly, $p(\Theta)$ is the prior on the model parameters, and
\begin{align}
p(\hat{\Omega}_{\rm GW}(f)) =\int \mathcal{L}(\hat{\Omega}_{\rm GW}(f) | \Theta) p(\Theta)\,d\Theta
\end{align} 
is the model evidence or ``marginal likelihood''. The posterior is typically estimated either by brute force estimation or, in high dimensional cases, by MCMC methods. 

In addition to estimating the posterior in Equation~(\ref{eq:omgw_posterior_ground_based}),
one can perform Bayesian model selection by comparing the marginal likelihood for two separate models.
This is used to construct an odds ratio between two models $\mathcal A$ and $\mathcal B$,
\begin{align}
    \mathcal O^{\mathcal A}_{\mathcal B} = \frac{\pi(\mathcal A)}{\pi(\mathcal B)}\frac{p[\hat{\Omega}_{\rm GW}(f) | \mathcal A]}{p[\hat{\Omega}_{\rm GW}(f) | \mathcal B]}.
\end{align}
The first term on the right-hand side is the \emph{prior odds}, where one specifies any prior information about the preference for one model over the other. The second term is simply the ratio of  evidences. Odds ratios, or Bayes factors (simply the second term on the right-hand side, i.e., equal prior odds), can be used to distinguish between models for the GWB. A general heuristic guide for their use is discussed in Chapter 3 of~\cite{Romano2017}. Bayes factors have been used recently as the detection statistic in searches that seek to measure alternative polarizations of GWs~\cite{Callister:2017ocg,LIGOScientific:2019vic,LVK:2021kbb}, in searching for a background from superradiant instabilities around black holes~\cite{Tsukada:2018mbp,Tsukada:2020lgt}, as well as searching for models of cosmological backgrounds~\cite{Romero:2021kby,Romero-Rodriguez:2021aws,Martinovic:2021hzy}. This hybrid-Bayesian methodology has also been proposed in attempting to distinguish between a GWB and globally correlated magnetic noise~\cite{MeyersMartinovic2020}, and simultaneously search for multiple GWB models contributing at once~\cite{Martinovic:2020hru}. We discuss the problem of globally correlated magnetic noise for current and future detectors in more detail in Section~\ref{sec:efforts}. In addition, a Fisher matrix formalism has also been proposed to simultaneously search for multiple GWB contributions~\cite{ParidaSuresh2019}.


The detection approaches outlined above have been validated both through mock data challenges~\cite{Meacher:2015iua} and hardware injection campaigns~\cite{Biwer:2016oyg}. The latter consisted in generating an isotropic Gaussian stochastic signal by manipulating the test masses at both LIGO sites and successfully recovering the injection via a cross-correlation stochastic search. However, the mock data challenge in~\cite{Meacher:2015iua} considers the case of a semirealistic astrophysical background of binary black holes and neutron stars. Here, the authors prove that the cross-correlation statistical approach assuming a Gaussian background remains valid even in the case of an intermittent astrophysical signal, although it is suboptimal. The fully optimal search methods for this sort of signal are described in Section~\ref{ssec:TBS}.

\subsubsection{Pulsar Timing Arrays}
\label{sssec:detection:isotropic:PTAs}

In describing the typical Bayesian analysis that is currently performed by most PTAs to search for an isotropic GWB, we follow the methods laid out in, e.g.,~\cite{vanHaasterenLevin2009,vanHaasterenLevin2011,LentatiAlexander2013,vanHaasterenVallisneri2014,LentatiAlexander2014}, and refer to more specific studies throughout our discussion.  Unlike for ground-based interferometers where the detector strain time series is uniformly sampled with equivalent time stamps at each detector, the data collected by PTAs are the arrival times of pulses from an ensemble of pulsars on the sky. Hence, it is preferable to work in the time domain directly, and the typical starting point is the \textit{timing residuals} of all pulsars, $\timeresids$, which are left over after subtracting off the best-fit timing model constructed using deterministic parameters of the pulsars (e.g., rotation frequency, spin-down, binary parameters, sky position, etc.).  The likelihood of obtaining $n$ timing residuals given model parameters $\bm{\theta}$ for a given pulsar is a multivariate Gaussian as
\begin{equation}
\mathcal{L}(\delta\bm{t}| \bm{\theta}) \propto \frac{1}{\abs{\bm{C}}^{1/2}} e^{-\frac{1}{2} (\delta\bm{t} - \bm{s})^{\text{T}} \bm{C}^{-1} (\delta\bm{t} - \bm{s})} ,
\label{eq:pta_likelihood}
\end{equation}
where $\bm{s}$ includes contributions from \textit{deterministic} signals with explicitly modelled time dependence, including the evolution of arrival times as described by the timing model. 
The covariance matrix $\bm{C}$ represents contributions from \textit{stochastic} processes. Note that while we have used ``$\propto$'', the specific normalization does matter in this analysis in general, as the spectra of some of the stochastic processes that contribute to $\bm{C}$ are themselves described by a set of free parameters.

The diagonal elements of $\bm{C}$ yield ``white'' noise that is uncorrelated in time whereas off-diagonal elements of $\bm{C}$ manifest as time-correlated ``red'' processes, including the GWB.
Red processes are modeled in the frequency domain with a power-law for the power spectral density of timing residuals
\begin{equation}
    P(f|A,\gamma) = \frac{A^2}{12\pi^2} \text{yr}^3 \left(f~\text{yr}\right)^{-\gamma},
    \label{eq:PTA_PSD_residuals}
\end{equation}
where $f$ is the GW or noise frequency, and $A$ and $-\gamma$ are the power-law amplitude and spectral index, respectively. We discuss sources of red noise that affect pulsars individually in more detail in Section~\ref{sssec:efforts_prospects:ptas:challenges}.
For the GW case, amplitude $A$ corresponds to the GW strain amplitude at the reference frequency $f_\text{ref} = \text{yr}^{-1}$ and $\gamma$ corresponds to $\alpha$ in Equation~\eqref{eq:Omegascale} and $\alpha'$ explained in footnote~\ref{fn:alphas_general}: $\gamma = 3 - 2\alpha' = 5 - \alpha$ = 13/3.
The lowest frequency and the size of the frequency bin $\Delta f$ are usually selected to be equal to the inverse of the total observation time $T_\text{obs}$, which is typically more than a decade.
The data are collected every few weeks.
The observed power spectral density of timing residuals $P(f)$ [s$^3$] is derived from the GW strain power spectral density $S(f)$ (s) in Section A4 in~\cite{HobbsJenet2009}.
Neglecting frequencies $f$ other than multiples of $T_\text{obs}^{-1}$, the densities and strain are related as
\begin{equation}
    P(f) = \frac{S(f)}{12\pi^2 f^2} = \frac{h(f)^2}{12\pi^2 f^3}.
\label{eq:pta_psd_strain}
\end{equation}

With each pulsar modelled by tens of timing model parameters and the necessity to invert the covariance matrix that models both temporal and spatial correlations for every likelihood evaluation, in practice it is necessary to marginalise the likelihood over nuisance parameters, which include timing model parameters and Gaussian coefficients that yield a specific realization of red noise.
Below, we describe the marginalization procedure~\cite{ArzoumanianBrazier2016,TaylorLentati2017} that is employed by recent searches~(e.g., \cite{ArzoumanianBaker2018,ArzoumanianBaker2020,goncharov2021cpgwb}). First, the likelihood is rewritten as
\begin{align}
    \mathcal{L}(\timeresids | \boldsymbol{\epsilon}, \boldsymbol{a}, \boldsymbol{\eta}) &\propto \frac{1}{\abs{\bm{N}}^{1/2}}
    e^{-\frac{1}{2}(\timeresids - \boldsymbol{s})^T\boldsymbol{N}^{-1}(\timeresids - \boldsymbol{s})},\\
    \textrm{where } \boldsymbol{s} &= \boldsymbol{M}\boldsymbol{\epsilon} + \boldsymbol{F}\boldsymbol{a} + \boldsymbol{s}'.
\end{align}
The covariance matrix $\boldsymbol N$ is now diagonal and, thus, represents the white noise associated with each observation; $\boldsymbol{\epsilon}$ represents deterministic timing model parameters and the ``design matrix'' $\boldsymbol{M}$ maps them to the timing residuals; $\boldsymbol{F}\boldsymbol{a}$ represents low frequency ``red'' noise. The latter is modelled by a Fourier series with a Fourier amplitudes vector $\boldsymbol{a}$ and a matrix $\boldsymbol{F}$ of unit-norm frequency Fourier modes.
Typically, the frequencies of the Fourier series are multiples of $1/T_{\rm obs}$~\cite{LentatiAlexander2013}. $\boldsymbol \eta$ represents \emph{hyperparameters} that govern the shape of the red noise spectrum defined by $\boldsymbol F$ and $\boldsymbol a$, and $\boldsymbol{s}'$ denotes deterministic terms that are not marginalised over. 
To be precise, the matrix $\boldsymbol N$ is block-diagonal, consisting of the nominal error in  pulse arrival time, but parameterised by ``EFAC,'' which multiplies  arrival time errors; ``EQUAD'',  which is added to the arrival time errors in quadrature; and ``ECORR'', which accounts for correlated measurement uncertainties in contemporaneous, multi-band observations and contributions to the off-diagonal terms~\cite{TaylorLentati2017,ArzoumanianBaker2018,ArzoumanianBaker2020}.

Further compactifying the equations, we express 
\begin{align}
\boldsymbol{T} &=\begin{pmatrix}
\boldsymbol{M}& \boldsymbol{F} 
\end{pmatrix},\\
\boldsymbol{b} &= \begin{pmatrix}\boldsymbol{\epsilon}\\\boldsymbol{a} 
\end{pmatrix}.
\end{align}
Gaussian priors are then placed on $\boldsymbol{b}$,
\begin{align}
\pi(\boldsymbol{b}|\boldsymbol{\eta}) = \frac{\exp\big(-\frac{1}{2}\boldsymbol{b}^\mathrm{T} \boldsymbol{B}^{-1} \boldsymbol{b}\big)}{\sqrt{2\pi\det{\boldsymbol{B}}}},
\label{eq:pta_coeff_prior}
\end{align}
with covariance $\boldsymbol B$, for which its parameters have previously been referred to as $\boldsymbol{\eta}$,
\begin{align}
    \boldsymbol{B} = \begin{pmatrix}
    \infty & 0 \\ 
    0 & \boldsymbol\varphi \\ 
    \end{pmatrix},
    \label{eq:pta:bmatrix_prior}
\end{align} where we have left the prior on the timing model parameters unconstrained (the upper left corner indicates a block with diagonal entries of infinity).
The posterior is then proportional to the original multivariate likelihood described by Equation~\eqref{eq:pta_likelihood} times the prior described by Equation~\eqref{eq:pta_coeff_prior},
\begin{align}
    \mathcal{P}(\boldsymbol{b},\boldsymbol{\eta}| \timeresids) \propto \mathcal{L}(\timeresids | \boldsymbol{b}) \times \pi(\boldsymbol{b}|\boldsymbol{\eta}) \times \pi(\boldsymbol{\eta}).
    \label{eq:pta:posterior_unmarg}
\end{align}

The initial timing model parameters and the design matrix are obtained with least-squares fitting.
The matrix $\boldsymbol{\varphi}$ contains information about the spectrum of low frequency noise, including intrinsic red noise for each pulsar and the GWB. If we label pulsars as $a$ and $b$ and frequency bins (multiples of $1/T_{\rm obs}$) as $i$ and $j$, then the elements of this matrix~are
\begin{align}
[\boldsymbol\varphi]_{(ai),(bj)} = \Gamma_{ab}\rho_{i}\delta_{ij} + \eta_{ai}\delta_{ab}\delta_{ij},
\label{eq:red_noise_phi}
\end{align} where $\rho_{i}$ is the spectrum of the GWB (and is, therefore, present in all cross-correlations and auto-correlations), and $\eta_{ai}$ is the spectrum of intrinsic red noise of pulsar $a$ in frequency bin $i$ (and is, therefore, only present on the diagonals).
The vectors of maximum-likelihood values of $\boldsymbol{b}$, $\hat{\boldsymbol{b}}$, are obtained by solving
\begin{align}
    \hat{\boldsymbol{b}} &= \boldsymbol{\Sigma}^{-1} \boldsymbol{d},\\
    \boldsymbol{\Sigma} &= \boldsymbol{T}^\text{T} \boldsymbol{N}^{-1} \boldsymbol{T} + \boldsymbol{\varphi}^{-1},\\
    \boldsymbol{d} &= \boldsymbol{T}^\text{T} \boldsymbol{N}^{-1} \delta \boldsymbol{t}.
\label{eq:pta_bmatrix}
\end{align}
Alternatively, one can explicitly marginalise over $\bm b$ as
\begin{align}
\mathcal{P}(\boldsymbol{\eta} | \timeresids ) &= \int d \bm{b}\, \mathcal{P}( \boldsymbol{b},\boldsymbol{\eta} | \timeresids)\nonumber\\
&= \frac{\exp(-\frac{1}{2}\timeresids^T \boldsymbol{C}^{-1} \timeresids)}{\sqrt{2\pi \det C}},
\end{align}
where $\boldsymbol{C}=\boldsymbol{N} + \boldsymbol{T}\boldsymbol{B}\boldsymbol{T}^T$. $\boldsymbol{C}$ can be efficiently inverted using the Woodbury formula to increase computational efficiency.

In principle, $\boldsymbol\varphi$ is block diagonal with independent frequency bins but correlated noise between pulsars when a GWB is present. However, a GWB signal is expected to reveal itself as a strong common red noise process before cross-correlations are evident.
Recent analyses have, therefore, first evaluated the term in Equation~\eqref{eq:red_noise_phi} with $\Gamma_{ab}=\delta_{ab}$~\cite{ArzoumanianBrazier2016} as the null hypothesis and $\Gamma_{ab}$ given by Equation \eqref{eq:orfhd} as the signal hypothesis.
As we discuss in Section~\ref{ssec:pta_results}, there is strong evidence for a common spectrum process ($\Gamma_{ab}=\delta_{ab}$) compared to the intrinsic red-noise only hypothesis ($\Gamma_{ab}=0$), but little compelling evidence for cross-correlations compared to a common spectrum process.

The $\eta_{ai}$ and $\rho_i$ spectra are modeled as a power law,
\begin{align}
\eta_{a, i} &= P_{\rm red}(f_i|A_{a, {\rm red}},\gamma_{a, {\rm red}}) \Delta f,\\
\rho_{i} &= P_{\rm GW}(f_i|A_{\rm GW},\gamma_{\rm GW}) \Delta f,
\end{align} 
and so instead of having $N_{\rm freq}$ parameters per pulsar (plus $N_{\rm freq}$ more parameters for the GWB), one has $2(N_{\rm pulsar}+1)$ free hyper-parameters that specify the shape of  noise processes.
White noise parameters, such as the typical EFAC, EQUAD, and ECORR parameters are measured on a per-pulsar (and per-observatory or per-observing back-end) basis.
Typical pulsar datasets are modelled by tens of such parameters; thus,  white noise parameters are often fixed to their maximum likelihood values that were calculated by analyzing  pulsars individually.

Bayesian inference with the marginalised likelihood typically yields  posterior samples for the $2(N_{\rm pulsar}+1)$ hyperparameters, and the Bayesian evidence, which is the integral of the likelihood over the prior. The evidence is used to select a model that best describes the data. In some recent analyses, explicit Bayes factors (the ratio of evidences between models) are calculated using a ``product-space'' approach that foregoes individual evidence calculations in favor of sampling from multiple models simultaneously~\cite{ArzoumanianBaker2018,ArzoumanianBaker2020}.

It is also common to consider an optimal statistic  which is built similar to the one presented for LVK searches~\cite{AnholmBallmer2009,ChamberlinCreighton2015,VigelandIslo2018}. We follow the conventions and methods in~\cite{VigelandIslo2018}, which were implemented in~\cite{ArzoumanianBaker2020}. We start from the auto-covariance and cross-covariance matrices calculated from the timing residuals,
\begin{align}
\boldsymbol C_a &= \langle \delta\boldsymbol{t}_a\otimes\delta\boldsymbol{t}_{a}\rangle,\\
\boldsymbol S_{ab} &= \langle \delta\boldsymbol{t}_a\otimes\delta\boldsymbol{t}_{b}\rangle|_{a\neq b}.
\end{align}
For a GWB with PSD given by $P_{\rm GW}(f)$, the cross-covariance matrix is given by
\begin{align}
    \boldsymbol S_{ab} &= P_{\rm GW}(f)\left(F_a\Gamma_{ab}F_b^T\right).
\end{align}
Specifically, for a background dominated by supermassive black hole binaries that evolve solely due to GW emission, $\gamma=13/3$. The optimal statistic, $\hat A^2$, is then given by~\cite{AnholmBallmer2009,ChamberlinCreighton2015,VigelandIslo2018}
\begin{equation}
    \hat A^2 = \frac{{\mathlarger\sum\limits_{ab}} \delta\boldsymbol t_a^T C_a^{-1} \tilde{\boldsymbol{S}}_{ab}C_b^{-1}\delta \boldsymbol t_b}{\mathlarger\sum\limits_{ab}\boldsymbol C_a^{-1} \tilde{\boldsymbol{S}}_{ab}\boldsymbol C_b^{-1}\tilde{\boldsymbol{S}}_{ba}},
\end{equation}
with 
\begin{equation}
A^2_{\rm GW}\tilde{\boldsymbol{S}}_{ab} = \boldsymbol{S}_{ab}.
\end{equation}
In the second line, we define the amplitude-independent cross-correlation matrix, which makes $\langle \hat{A^2}\rangle = A_{\rm GW}^2$. If $A_{\rm GW}=0$, then the variance of the optimal statistic is given by
\begin{align}
    \sigma_0 = \left[\mathlarger\sum_{ab}\boldsymbol C_a^{-1} \tilde{\boldsymbol{S}}_{ab}\boldsymbol C_b^{-1}\tilde{\boldsymbol{S}}_{ba}\right]^{-1/2},
\end{align}
which allows us to construct an SNR $\rho = \hat{A}^2/ \sigma_0$. The SNR significance can be found by estimating its noise-only distribution using simulations of the data with pulsar sky positions randomly assigned or random phase shifts applied to the data to remove a signal~\cite{CornishSampson2016,TaylorLentati2017}.

The implementation of the optimal statistic requires evaluating $\bm C_{a}$ and $\bm S_{ab}$ for a given choice of hyperparameters that define $\boldsymbol a$, which are amplitudes of sinusoids on which the red noise is decomposed. One typically uses  hyperparameters that correspond to $\hat{\boldsymbol{b}}$ in Equation~\eqref{eq:pta_bmatrix}. One can also use the maximum \emph{a posteriori} estimates of the red noise parameters from the Bayesian analysis, $A_{{\rm red}, a}^{\rm max}$, $\gamma_{{\rm red}, a}^{\rm max}$, $A_{{\rm GW}}^{\rm max}$, and $\gamma_{{\rm GW}}^{\rm max}$, but this can result in a large bias in the estimate of $\hat A^2$ because the individual red noise parameters and the common red noise parameters are highly correlated.  In practice, the red noise parameters are drawn from the posterior samples from the Bayesian analysis and for each draw the optimal statistic and its SNR are calculated. One can estimate a value of $\hat A^2$ from this distribution, which is significantly less biased than if one calculates the optimal statistic once from the maximum \emph{a posteriori} estimates of the noise parameters~\cite{VigelandIslo2018}.

While our focus in this section has been on methods for detecting an isotropic GWB with PTAs, note that there is also a recent paper that aims to measure both continuous sources from individual supermassive black hole binaries as well as an isotropic confusion background~\cite{BecsyCornish2020}.


%


\subsection{Anisotropic Background Detection Methods}\label{ssec:approaches_anisotropic}

Mapping an SGWB is a distinct problem from that of tracing back an individual source position.
Individual source mapping leverages the coherence of radiation arriving from the source, using the phase information of the signal to estimate both angular and frequency-dependent phase  along with other intrinsic and extrinsic source parameters. 
In either case, however, a basic requirement for directional analyses is multiple observations of the same signal.
These multiple observations can be made in different ways. 
Multiple detectors at different locations will impose a known phase shift on a coherent signal, which can be used to constrain the position on the sky. 
In the case of a persistent, coherent source, unequal-time auto-correlation of a single detector signal can yield angular phase information, and in the case of some sources of GWs, matched filtering can be used to pinpoint the location on the sky with a significantly larger effective baseline~\cite{PrixItoh2005}.
For the case of a persistent and incoherent source, such as an SGWB, one may use equal-time correlations of data from different detectors and the motion of individual detectors with respect to the frame in which the signal is stationary to reconstruct angular information. We focus on this approach~here. 


Let us start by discussing the weak signal limit, which applies to the 2G ground-based detector network (e.g., LIGO and Virgo). As we did for the isotropic case, we can assume that the residuals for the data cross-correlation spectra and our model are Gaussian, similarly to Equation~\eqref{eq:likelihood_residuals}.
The log-likelihood may be written as a function of an anisotropic signal $I$ explicitly as
\begin{equation}
    	\ln \mathcal{L}(C^\tau(f)|I(f, \hat{\bm n}))  \propto \frac{1}{2}\left[ \qty(C^\tau(f) - \langle C^\tau(f) \rangle) \bm{\sigma}^{-2} \qty( C^\tau(f)-\langle C^\tau(f) \rangle)^\star + \Tr \ln \bm{\sigma^2}\right]\,,
    \label{eq:likelihood_residuals_anisotropic}
\end{equation}
where we generically model the correlated data as
\begin{equation}
    \langle C^\tau(f) \rangle = \int_{S^2} d\hat{\bm n}A^\tau(f, \hat{\bm n}) I(f, \hat{\bm n})\,.
    \label{eq:model_for_cross_correlations_anisotropic}
\end{equation}
The operator $A$ \emph{projects} the sky signal $I$ into the datastream, and contains all the necessary correlated response information of the detectors. For ground- and space-based detectors, $A$ is given by Equation~\eqref{eq:liorf} up to appropriate normalisation factors. We have reintroduced the timestamp label $\tau$ here as it serves an important purpose: tracking the time-dependence of the detector response on the sky; the sky signal $I$ is considered absolutely stationary\footnote{Note that this is no longer the case in the presence of temporal shot noise, e.g., for the astrophysical GWB from compact binary coalescences~\cite{Jenkins:2019uzp}. In this case, each time segment will have random GWB intensity fluctuations due to the finite number of sources. The statistical independence of these fluctuations at different times can be leveraged to mitigate the impact of shot noise on measurements of the angular power spectrum~\cite{Jenkins:2019nks}.} and, thus, $\tau$-independent. For ground-based interferometric detectors,  operator $A$ depends only upon the sidereal time, which means the data stream can be folded on the sidereal day before we invert Equation~\eqref{eq:model_for_cross_correlations_anisotropic} to solve for $I$~\cite{Ain2015,AinSuresh2018,GoncharovThrane2018}. To probe the effects of the projection operator $A$ on the GW observations, it is useful to simulate a signal as a single point source on the sky and ``project'' it analytically (or numerically) into cross-correlated mock data, which produces the \emph{point-spread function} of the detector pair. This has often been studied to glean the effective angular resolution of the detectors to a stochastic source---we refer the reader to~\cite{Mitra:2007mc} for a complete presentation of the subject.

We can now derive a \emph{maximum-likelihood mapping} solution by maximising the likelihood with respect to the sky map. To best illustrate the structure of the solution, let us use a compact matrix notation where operators in bold generally span multiple dimensions, i.e., time, frequency, direction, and baseline, such as $A^\tau_b(f, \hat{\bm n}) \rightarrow \bm A$. These are contracted appropriately to achieve the desired result,
\begin{equation}
	\tilde{\bm I} = \left(\bm{A}^\dagger \bm\Sigma^{-1} \bm{A}\right)^{-1} \bm{A}^\dagger \bm{\Sigma}^{-1} \bm{C}\,.
    	\label{eq:map_soln}
\end{equation}
In the simple case of a single frequency and single time segment of data, we can think of $\bm C$ as a vector containing cross-correlations between different baselines, $\bm \Sigma$ is the noise covariance of the cross-correlation of the detector pairs, and $\bm A$ is now a matrix that contains geometric factors that project different sky directions onto correlations between different detectors. Explicitly, $\bm \Sigma$ entries are of the form
\begin{equation}
    \Sigma^\tau_b(f, f') = \text{Cov}[P^f_i P^{f'}_j]_\tau\,,
\end{equation}
where $i,j$ label the two detectors in the baseline $b$. More details on the calculation may be found in~\cite{AIR_thesis,Mitra:2007mc,Thrane2009}. 

In practice, this solution may need to be implemented iteratively as the noise covariance $\bm \Sigma$ may not be known \emph{a priori} and, hence, may need to be estimated from the data themselves. In the low signal limit however, as is the case with current ground-based networks, the noise covariance is estimated independently of the signal and Equation~\eqref{eq:map_soln} is, thus, a closed-form mapping solution. The first term on the right-hand side is the \emph{Fisher information matrix},
\begin{equation}
    \bm F = \bm{A}^\dagger \bm{\Sigma}^{-1} \bm{A}\,,
    \label{eq:fisher_map_soln}
\end{equation}
which is inverted and applied to the \emph{projection map},
\begin{equation}
    \bm z = \bm{A}^\dagger \bm{\Sigma}^{-1}\bm C\,,
    \label{eq:dirty_map_soln}
\end{equation}
to extract the maximum-likelihood solution $\tilde{\bm I}$. $z$ is often referred to as the \emph{dirty map} in the literature, as it effectively represents a first, na\"{i}ve representation of the cross-spectral density on the sky, while $\tilde{\bm I}$ is considered the \emph{clean map} as it is the result of a deconvolution. The Fisher matrix contains the pixel--pixel covariance information necessary to deconvolve the detector response from the dirty map.
In principle, this method can also be employed for anisotropic searches in a strong signal regime; in this case, if the signal is Gaussian, the map point-estimate remains accurate; however,  covariance $\bm \Sigma$ requires extra terms as the signal contribution may not be ignored.

However, in the presence of a strong signal term, it is preferable to include  auto-correlation terms in the likelihood, as these contain precious information. This is the case for, e.g., LISA and PTAs. Hence, we consider again the Gaussian likelihood~(\ref{eq:likelihood_data}), and as in~\cite{MingarelliSidery2013,TaylorGair2013a,Contaldi:2020rht,BanagiriCriswell2021}, we extend it to include directional information,
\begin{equation}
    {\rm ln}{\cal L}(\bm d^\tau(f) | I(f, \hat{\bm n})) \propto \frac{1}{2}\left[ \bm d^\tau(f) \bm{C}^{-1} \bm d^{\tau \star}(f) + \Tr \ln \bm{C}\right]\,,
    \label{eq:likelihood_data_directional}
\end{equation}
where the correlation matrix $\bm C^\tau (f)$ (re-introducing the time and frequency dependence explicitly here for clarity) may be estimated as

\begin{equation}
    \bm C(f) = \int_{S^2} d\hat{\bm n} \bm A(f, \hat{\bm n}) I(f, \hat{\bm n}) + \bm N\,,
    \label{eq:mapping_cov}
\end{equation}
with $\bm N$ as the noise covariance. Note that $\bm A$ is an $N\times N$ matrix representing the correlated response of the full network. 

The above notation implies that we are discussing LISA because we have specified the frequency domain and that $\bm A$ depends on $\tau$. In principle,  PTA analysis works analogously---note the similarity of Equation~(\ref{eq:mapping_cov}) to Equation~(\ref{eq:red_noise_phi}), with the added complication of a dependence on direction. However, it is worth briefly reiterating specific differences. First, the ``detector response'' is time-independent for PTAs, and in general we assume it is frequency independent. In addition, PTAs leave data in the time domain but expand the GW signal and intrinsic red noise in the frequency domain using a set of Fourier basis functions with an appropriate prior to impose a power-law spectrum and correlations between pulsars (see Equations~(\ref{eq:pta:bmatrix_prior})--(\ref{eq:pta_bmatrix})). It is the correlations between pulsars that are directly affected, with the anisotropy of the background causing a deviation from the Hellings--Downs curve that depends upon the specific realization of the background. This means that Equation~(\ref{eq:red_noise_phi}) changes:
\begin{align}
\Gamma_{ab}\rho_i \rightarrow \int_{S^2}d\hat{\bm n}\,\Gamma_{ab}(\hat{\bm n})\rho_i(\bm{\hat n}),
\end{align}
with $i$ once again specifying the frequency index, $a$ and $b$ specifying pulsars we are correlating, and $\rho_{i}(\hat n)$ specifying the GWB amplitude at frequency index $i$ in direction $\hat{\bm n}$. In this case, $\Gamma_{ab}(\hat{\bm n})$ can be considered a direct analogue to $\bm A(f, \hat{\bm n})$.

In~\cite{Contaldi:2020rht}, the authors show that extracting the maximum-likelihood map solution from Equation~(\ref{eq:likelihood_data_directional}) yields
\begin{align}
	\tilde{\bm I} &= \bm F^{-1} \cdot \frac{1}{2}\, \text{Tr}\left[ \bm C^{-1} \, \frac{\partial \bm C}{\partial \tilde{\bm I}}  \, \bm C^{-1} \, (\bm d\otimes \bm d^\star - \bm N) \right]\,, \label{eq:map_itersol}  \\
 	\bm F &=\frac{1}{2}\,\text{Tr}\left[ \bm C^{-1} \, \frac{\partial \bm C}{\partial \tilde{\bm I}} \, \bm C^{-1} \, \frac{\partial \bm C}{\partial \tilde{\bm I}}  \right]\,,
 \label{eq:fish_map_itersol}
 \end{align}
where $\bm F$ is the Fisher information matrix. This solution is in fact structurally identical to the Bond--Jaffe--Knox mapping procedure in CMB~\cite{Bond1998, Rocha2009} and requires an iterative implementation: first, a ``guess'' map is injected in $\bm C$ (Equation~\eqref{eq:mapping_cov}), then a first solution $\tilde{\bm I}$ is calculated using Equation~\eqref{eq:map_itersol}, which updates the estimate of $\bm C$ and so on until convergence. 
Beyond SNR considerations, this iterative method is also useful in the case of correlated noise terms as these may be included in the noise model $\bm N$, whereas solution~\eqref{eq:map_soln} rests upon the assumption that the noise is completely uncorrelated between different detectors/measurements.

Alternatively, in the strong or intermediate signal regime one can employ a fully Bayesian approach to the problem as presented in, e.g.,~\cite{TaylorGair2013a,CornishvanHaasteren2014,BanagiriCriswell2021}. In~\cite{TaylorGair2013a}, the authors develop a fully Bayesian search pipeline to use PTAs to map the GW sky using the formalism detailed in~\cite{MingarelliSidery2013}. In~\cite{BanagiriCriswell2021}, the authors present a Bayesian search for anisotropies in the GWB using LISA, where they decompose the sky onto the first three spherical harmonics, parameterize the noise covariance curve, and use MCMC sampling to map the diffuse sky and estimate the noise properties of the detector simultaneously.

Several different approaches can be taken when applying these map-making methods to real data~\cite{Cornish:2001bb,Ballmer_2006,Mitra:2007mc,Thrane2009,MingarelliSidery2013,TaylorGair2013a,CornishvanHaasteren2014,GairRomano2014,GairRomano2015,RomanoTaylor2015,TaylorMingarelli2015,Ain2015,Ali-HaimoudSmith2020,Ali-HaimoudSmith2021} based on specific necessities. The spectral dependence of the map is often factored out (although not always, see, e.g.,~\cite{TaylorMingarelli2015,ParidaSuresh2019,Renzini2019b,SureshAgarwal2021}), under the assumption in Equation~\eqref{eq:intef} and integrated over to obtain \emph{broadband} maps of the GW sky. Then, the direction-dependent component of the signal may be expanded and mapped in spherical harmonics,

\begin{align}
I(\hat{\bm n}) = \sum_{\ell=0}^{\infty}\sum_{m=-\ell}^{\ell} I_{\ell m}Y_{\ell m}(\hat{\bm n}),
\label{eq:methods:directional:sph_decomposition}
\end{align}
or discretized directly on the sky in pixel space,
\begin{align}
I(\hat{\bm n}) = \sum_{p'=1}^{N_{\rm pix}} I_{p'}\delta_{pp'},
\label{eq:methods:directional:pixel_decomposition}
\end{align}
where $Y_{\ell m}$ are the standard spherical harmonics, and we have used $p$ and $p'$ to enumerate pixels on the sky that are specified using some choice of pixelization (e.g., HealPix~\cite{Healpix})\footnote{Note that in these two cases $I_{p}$ and $I_{\ell m}$ have different units because in (\ref{eq:methods:directional:pixel_decomposition}) the basis function carry units $\rm{sr^{-1}}$.}. Other works have argued for the use of curl and gradient spherical harmonics, as in CMB analyses~\cite{GairRomano2014,GairRomano2015,RomanoTaylor2015}, or bipolar spherical harmonics~\cite{HotinliKamionkowski2019}, although to our knowledge these have not yet been employed on real data.

In both the pixel and spherical harmonic bases, the Fisher matrix $\bm F$ is usually very poorly conditioned for a single baseline, as the sky coverage is incomplete; this is the case even after time-integration, which takes into account the sky-scanning strategy of the baseline. Regularisation techniques are, therefore, needed to estimate maximum likelihood solutions such as those presented in Equations~(\ref{eq:fish_map_itersol}) and~(\ref{eq:fisher_map_soln}). Examples of regularization techniques include cutting off the sum in (\ref{eq:methods:directional:sph_decomposition}) a priori at some maximum $\ell$~\cite{Thrane2009}, singular value decomposition (SVD)~(see, e.g., \citep{ivezic2014statistics}), and the drastic approach of limiting the calculation to the diagonal of $\bm F$, ignoring all off-diagonal correlations~\cite{Ballmer_2006}.  In some cases, combinations of these approaches are used.

The spherical harmonic basis has mostly been employed when searching for a diffuse background, which dominates the low-$\ell$ modes, while pixel space has been employed when searching for point-like stochastic sources, although it has been shown that there is a one-to-one mapping between the two~\cite{SureshAin2021} (see, e.g., Figures 1 and 2 of~\cite{TaylorGair2013a} for an illustrative example of how specific anisotropic maps correspond to different spherical harmonic decompositions). There may however be subtle differences between the conditioning in spherical harmonic and pixel space; in interferometry in general, the choice has often been to obtain a spherical harmonic domain solution since the measurements directly constrain this domain (see e.g., \cite{Myers:2002tn} for an application to CMB interferometry). The reconstruction of the actual ``image'' is then left as a separate and straightforward conversion problem.

However, as discussed in~\cite{Renzini2018}, in the case of GW detectors, there is a non-trivial coupling of spherical modes in the sky response due to the non-compactness of the beam, indicating that adopting this approach may have significant drawbacks. 
In particular, there is no isotropic kernel for a primary beam describing the coupling of individual pairs of directions, hence limiting the number of spherical modes in the solution in a piece-wise fashion could make the problem more ill-conditioned than it actually is. Finally, in the case of PTAs it has been shown that, when working in the spherical harmonics basis, the non-uniformity of pulsars on the sky makes it impossible to decouple the monopole from other multipole moments, making the search less sensitive to isotropic backgrounds~\cite{Ali-HaimoudSmith2020,Ali-HaimoudSmith2021}. In those studies, the authors propose using a basis formed with principal component analysis of the direction-dependent sensitivity of the specific PTA.

In fully Bayesian examples, using too many spherical harmonics terms or pixels on the sky can make the problem intractable or, in a detection case, could
widen the parameter space beyond what is possible to resolve. Techniques that can be used to make the full Bayesian problem more tractable might be to only search over the first few spherical harmonics, as in, e.g.,~\cite{TaylorGair2013a,BanagiriCriswell2021}. Other techniques have also been proposed, one of which constructs a set of basis ``sky vectors'' from only the columns of the SVD matrices that are associated with non-zero singular values of the Fisher matrix. One then samples over their coefficients to reconstruct the map using a lower-dimensional representation of the sky~\cite{CornishvanHaasteren2014,RomanoTaylor2015}.


Although notably more challenging when attempting map-making with a singular $\bm F$, the spectral dependence may be solved for as well, resulting in  \emph{narrowband} directional searches. 
These have been carried out by the LVK collaboration targeting point-like sources which may have an interesting spectral emission (e.g., asymmetric neutron stars, supernova remnants, the galactic bulge) and have been performed in pixel space~\cite{GoncharovThrane2018,AinSuresh2018,LVK:2021mth}. As point sources are not correlated with other points on the sky, this type of search ignores the off-diagonal correlations in the Fisher matrix to avoid the inversion problem. Another approach employed on LIGO data in~\cite{Renzini2019b} is that of frequency-banding, where one solves for maps in separate frequency bands spanning several Hz, making minimal assumptions about the spectrum of the signal in each band. This allows  both inverting the full Fisher matrix and obtaining full sky maps in each band and attempt a model-free reconstruction of the frequency spectrum. In the case of pulsar timing arrays, the signal is dominated by a small ($\lesssim 10$) number of frequencies, and so in the fully Bayesian analysis in~\cite{TaylorMingarelli2015}, they estimate different spherical harmonic coefficients at different frequencies to allow for maps that are dominated by point sources in different directions at different frequencies.

\subsection{The Approach towards Non-Gaussian Backgrounds}\label{ssec:TBS}
The cross-correlation searches reviewed above have been shown to be \emph{optimal} in the case of a continuous background, which, for example, presents a timestream as in the third row of Figure~\ref{fig:3types}; however, these will be suboptimal in the case of a background which is intermittent, such as the first row of Figure~\ref{fig:3types}. As argued in~\cite{Drasco:2002yd}, an intermittent background will not satisfy the central limit theorem; thus, searching for such a signal under the Gaussian assumption will reduce the sensitivity of the search, resulting in delays until detection (although it will not bias the point estimate). Given the observations of GWs accumulated up until now~\cite{LIGOScientific:2021djp} and the lack of a detection of a Gaussian stochastic background, it is clear that a signal of sub-threshold GW bursts from compact binary coalescences is present in LIGO and Virgo data and may be the first stochastic signal that can be detected with ground-based detectors.

A few optimised search methods for non-Gaussian backgrounds have been proposed: a maximum likelihood approach~\cite{Drasco:2002yd} where each isolated GW making up the background is modelled as a ``stochastic burst'' of GW strain, correlated between detectors; a Bayesian search method which leverages deterministic parameter estimation~\cite{Thrane:2013kb,Smith:2017vfk}, fitting each burst with a compact binary waveform; and, soon, a fully Bayesian stochastic search~\cite{lawrence} which extends the formalism of~\cite{Drasco:2002yd}. This method has also been extended to estimate the anisotropic properties of the subthreshold CBC population~\cite{Banagiri:2020kqd}.

While substantially different in formulation and implementation, these approaches share the premise of re-parametrisation of the background signal based on a \emph{Gaussianity parameter}, $\xi$, which quantifies the degree of non-Gaussianity. Effectively, this is the same as the duty cycle introduced in Section~\ref{ssec:observational_props} and is linked to the duration of each individual GW burst and the frequency with which they appear in the detector. If normalised, $0\leq \xi \leq 1$ is the probability that a burst is present in the detector at any given time, and if $\xi=1$, the background is Gaussian. The detector noise, on the other hand, is usually still assumed to be Gaussian\footnote{In fact, in~\cite{Smith:2017vfk} the authors discuss the extension of their optimal search method in the presence of non-Gaussian noise as well. The implementation of the method becomes substantially more involved, however practical considerations about the nature and behaviour of the non-Gaussian noise can simplify it considerably.}. Then, that assuming we segment  data into short segments labelled by $i$, the likelihood function for the data vector $\bm d_i$ (for a set of detectors) given the data covariance $\bm C$ and signal model $h$ is
\begin{equation}
    {\cal L}(\bm d_i) \propto \prod_{f,\tau} \frac{1}{\abs{\bm C}^{1/2}} e^{\frac{1}{2} (\bm d_i - \bm h_i){\bm C}^{-1} (\bm d_i - \bm h_i)^\star}\,.
    \label{eq:likelihood_tbs}
\end{equation}
For an intermittent background,  model $\bm h_i$ in the likelihood will be either a GW burst or a CBC waveform, if the data segment $\bm d_i$ has a signal, or simply equal to zero if it does not~\cite{Smith:2017vfk}. The full likelihood ${\cal L_{\rm full}}$ of the data with an intermittent signal may then be written as the mixture of a signal-and-noise likelihood ${\cal L}_s$ and a noise-only likelihood ${\cal L}_n$~as
\begin{equation}
    {\cal L}_{\rm full}(\bm d_i) = \xi {\cal L}_s(\bm d_i|\bm h_i) + (1-\xi){\cal L}_n(\bm d_i|0)\,,
\end{equation}
featuring the duty cycle $\xi$ as a parameter. Maximising this likelihood (with any of these approaches) results in posterior distributions for both the signal parameters that construct $h_i$, and $\xi$; furthermore, it is also possible to include noise parameters here such that these also are estimated.

\section{Current Detection Efforts of SGWBs}\label{sec:efforts}

A variety of searches based on the methods described in Section~\ref{sec:approaches} have been implemented in interferometric and timing data, in the case of ground-based detectors and pulsar timing arrays (PTA), respectively, to either detect or constrain a wide range of stochastic backgrounds. Beyond these, many investigations have been carried out to assess the detection potential of future detectors on the ground and in space and their ability to distinguish different background sources.

\begin{figure}[t]
    \centering
    \includegraphics[width = 0.8\linewidth]{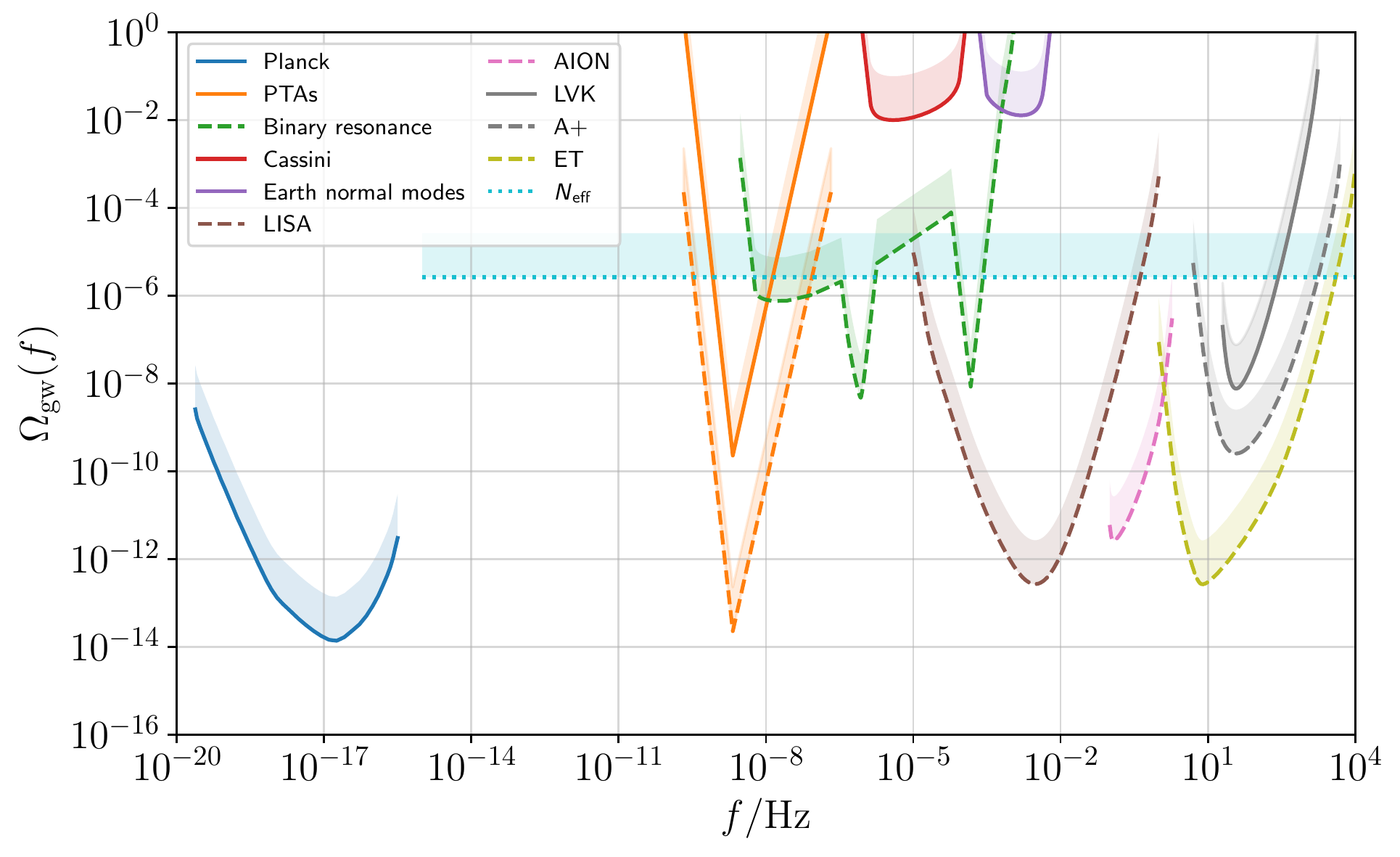}
    \caption{A survey of constraints (all at 95\% confidence) on the GWB across the frequency spectrum. Solid curves indicate existing results from LIGO/Virgo's first three observing runs~\cite{LVK:2021kbb}, monitoring of the Earth's normal modes~\cite{Coughlin:2014xua}, Doppler tracking of the Cassini satellite~\cite{Armstrong:2003ay}, pulsar timing observations by the PPTA~\cite{Lasky:2015lej}, and CMB temperature and polarisation spectra measured by Planck~\cite{Planck:2018jri,Lasky:2015lej}. Dashed curves are forecast constraints for LIGO at A+ sensitivity, Einstein Telescope~\cite{Punturo:2010zz}, AION-km~\cite{Badurina:2019hst}, LISA~\cite{Amaro2017}, binary resonance searches~\cite{Blas:2021mpc,Blas:2021mqw}, and pulsar timing with the Square Kilometre Array~\cite{Janssen:2014dka}. The dotted curve indicates the level of the integrated constraint from measurements of $N_\mathrm{eff}$~\cite{Pagano:2015hma}; note that this is a constraint on the total GW energy density over a broad frequency range and \emph{cannot} be directly compared to the other constraints. Note also that both the Planck and $N_\mathrm{eff}$ constraints apply only to primordial GWs emitted before the epoch of BBN. See Figure~\ref{fig:GWBplot} for various GWB signal predictions in relation to these constraints.} 
    \label{fig:constraints_2}
\end{figure}

In this Section, we start by discussing the most recent analysis results from the ground-based detector LIGO--Virgo--KAGRA (LVK) collaboration, obtained with the third observing run ({\tt O3}) of the 2G detectors: the \emph{Advanced} LIGO and the \emph{Advanced} Virgo. These results build on the previous data runs {\tt O1} and {\tt O2} of the 2G detectors. The overall sensitivity of the 2G detector network may be observed in Figure~\ref{fig:constraints_2}.  We then discuss future detection prospects for the 2G network, which should reach \emph{design} sensitivity with the next few upgrades of the instruments~\cite{designcurve}. When representing the sensitivity to a GWB, we use ``power-law integrated'' sensitivity curves, which show the sensitivity to a GWB at each frequency (at a chosen SNR level) with a power law spectral shape that is tangent to the curve at that frequency~\cite{Thrane2013}.

For PTAs, we provide details about existing and proposed collaborations across the world and report on their searches for nanohertz GWBs.
We review the upper limits on the amplitude of the isotropic GWB from circular supermassive black hole binaries.
Next, we briefly outline main sources of noise that limit the searches and discuss the future of timing array experiments.
We close by briefly discussing other interesting GW searches, such as resonant bar experiments and the Moon-based proposals, and highlight their bearing on stochastic searches.

\subsection{Searches with Ground-Based Laser Interferometers}
The ground-based detector communities have put in a continued, concerted effort towards detecting a stochastic GW background between $\sim$10--$10^3$ Hz.
The principal candidate for a stochastic background in the ground-based detector band is one given by inspiralling and merging binary black holes and neutron stars.
This is clear given the sensitivity of the LIGO-Virgo network, as shown for example in Figure~\ref{fig:constraints_2}, and the fact that these detectors have \emph{already detected} a large set of CBCs~\cite{LIGOScientific:2021psn}. This background has been exclusively searched for via a cross-correlation search, tailored to the astrophysical background in particular by reweighting the data in frequency according to the expected power-law spectral dependence as derived in Equation~(\ref{eq:astro_BG}).
First efforts of isotropic stochastic searches date back to 2003 and the first science run~\cite{LIGOScientific:2003jxj}, followed by several other science runs of the LIGO detectors~\cite{LIGOScientific:2005cjh, LIGOScientific:2006zmq, LIGOScientific:2011yag, LIGOScientific:2014gqo}, and the Advanced detector era since 2016~\cite{LIGOScientific:2016jlg, LIGOScientific:2019vic}, and, finally, the inclusion of Virgo~\cite{LVK:2021kbb}. 
Beyond these, several other searches have been carried out by targeting specific GWBs; notably, searches which allow for anisotropy in the signal have regularly been carried out~\cite{LIGODirectional2007,Abadie_2011, LIGOScientific:2016nwa, LIGOScientific:2019gaw, LVK:2021mth}, as well as searches for cosmic string networks~\cite{LVCStrings2014, LIGOScientific:2021nrg, LVCStrings2018}. Other targeted searches include searches for non-GR polarisation modes~\cite{LIGOScientific:2018czr,LIGOScientific:2019vic,LVK:2021kbb}. Furthermore, several stochastic search efforts have been carried out by small teams outside the LVK collaboration; let us cite here a set of directional searches complementary to the LVK ones~\cite{Renzini2019a, Renzini2019b,AgarwalSuresh2021}, a search for correlations between the anisotropic GWB and galaxy catalogues~\cite{Yang:2020usq}, searches for ultralight vector bosons~\cite{Tsukada:2018mbp,Tsukada:2020lgt}, a search for a primordial inflationary background~\cite{Kapadia:2020pnr}, and a search for parity-violating stochastic signals~\cite{Martinovic:2021hzy}.

We focus on current search results in this section, detailing the detector characterisation issues that the Advanced detectors have faced up to now. We discuss future challenges for ground-based detectors in Section~\ref{ssec:3G}, where we explore SGWB detection strategies with third generation interferometers. 

\subsubsection{Search Results for an Isotropic Background by LVK}
Applying the cross-correlation recipe described in Section~\ref{sec:approaches} to the real LIGO-Virgo datasets requires firstly identifying the valid cross-correlation times at which different detectors are simultaneously online and fully operational and, subsequently, Fourier transforming the measured timestreams to the frequency domain. 
In practice, it is convenient to divide these into smaller time segments and fast-Fourier transform (FFT) each segment, which is treated as an independent measurement. This reduces the cost of Fourier transforming, excludes frequencies that are far into the $1/f$ tail, and ensures that  noise properties are stationary within each segment. 

In the most recent isotropic stochastic analysis~\cite{LVK:2021kbb},  raw data are segmented into 192 s segments which are then Hann-windowed and FFTed. The data are downsampled from the native sampling rate of $16$ kHz to $4096$ Hz, and the full frequency band included in the analyses spans 20--1726 Hz; this choice is made so as to exclude the highest and lowest frequencies available where  noise becomes prohibitively large. 
Data quality cuts are applied based on a measure of noise stationarity between segments or when there are known detector artifacts in the data, and these cuts reduce  viable data by $\sim$20\% in {\tt O3}. In addition, frequencies where narrow spectral artifacts (``lines'') occur and are caused by known instrumental disturbances or show coherence between detectors are removed~\cite{Covas2018} and typically cut $\sim$15--20\% of the frequency band. More details on the methodology, exact values of data excised for each baseline in {\tt O3}, and implementation of the quality checks can be found in~\cite{Covas2018,LVK:2021kbb}.  Note that this is the first advanced-detector era stochastic analysis to include multiple baselines\footnote{Initial-detector era searches on LIGO Science Run 5 (S5)/Virgo Science Run 1 (VSR1)~\cite{LIGOScientific:2011yag} and LIGO Science Run 6, Virgo Science Runs 2 and 3 (VSR23)~\cite{LIGOScientific:2014gqo} used the full Hanford, Livingston, Virgo detector network.}; we highlight in this section the important implications of adding two long baselines to the network. Specifically,  {\tt O3} analysis spans Hanford--Livingston (H--L), Hanford--Virgo (H--V), and Livingston--Virgo (L--V).

Several new analysis features were implemented in~\cite{LVK:2021kbb}, as the LVK collaboration gears up for a possible SGWB detection. The results are presented by using a log-uniform prior, assigning equal weight to each order of magnitude in $\Omega_{\rm GW}$, in line with our current state of knowledge. Furthermore, a correlated noise analysis targeting magnetic noise resonances was carried out, using the framework set up by the likelihood~\eqref{eq:Omega_like}. The analysis confirmed the lack of correlated magnetic noise across the interferometer network and laid out a clear methodology, which will prove essential for similar analyses with future, more sensitive detectors (e.g., 3G). 
For the first time, short \emph{gates} ($\sim$1 s) were employed to remove frequent, loud glitches which populated the LIGO detector {\tt O3} data~\cite{gating1, gating2}. These were not as frequent nor problematic in previous data runs, and investigations are still under way to determine the sources of these noise artifacts. Without their removal, loud glitches would bias the PSD of a 192-second segment, which would then not pass the stationarity cut, reducing the effective live time of the LIGO detectors to less than 50\%, whereas  gated datasets are reduced by about 1\% of the original total length.

\begin{figure}[t]
    \centering
    \includegraphics[width=0.75\textwidth]{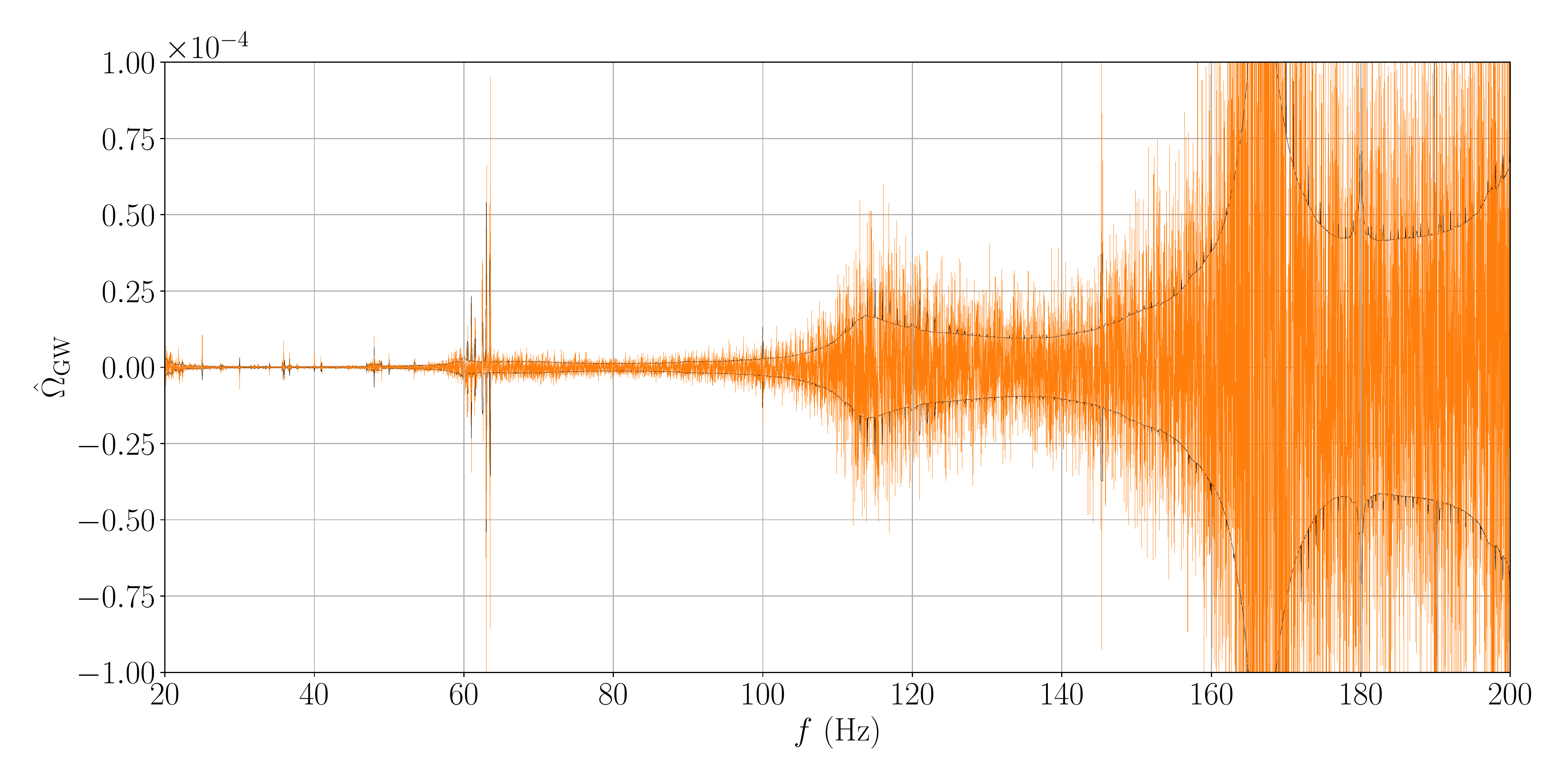}
    \caption{Combined cross-correlation spectra for the three O3 baselines. The grey lines here mark  $1\sigma$ contours. The data, by eye, appears to follow the Gaussian distribution. Note that adding the Virgo baselines has filled in the gap in the H--L data present around the 60 Hz power line (this broad line is present in both USA-based detectors, but not in Italy). The original version of this plot is presented in~\cite{LVK:2021kbb}; the one shown here was obtained using  open data published in~\cite{LVK:open_data_iso}.}
    \label{fig:C_O123}
\end{figure}

The H--L baseline contributes the most to  overall network sensitivity, while  Virgo baselines remain important for higher frequencies and, hence, especially for searches employing higher values of the spectral index $\alpha$. The combined H--L--V spectrum for all three data runs shown in Figure~\ref{fig:C_O123} is consistent with Gaussian noise, with symmetric fluctuations around zero. Note that the Virgo baselines contribute information around 64 Hz, where the overlap reduction function of H--L is null. In general, the three baseline combinations are more sensitive given that each detector has different consistent monochromatic noise lines (e.g., the harmonics of the suspension mechanism) which are excluded from analyses, inducing gaps in the frequency spectra.

\begin{figure}[t]
    \centering
    \includegraphics[width=.7\textwidth]{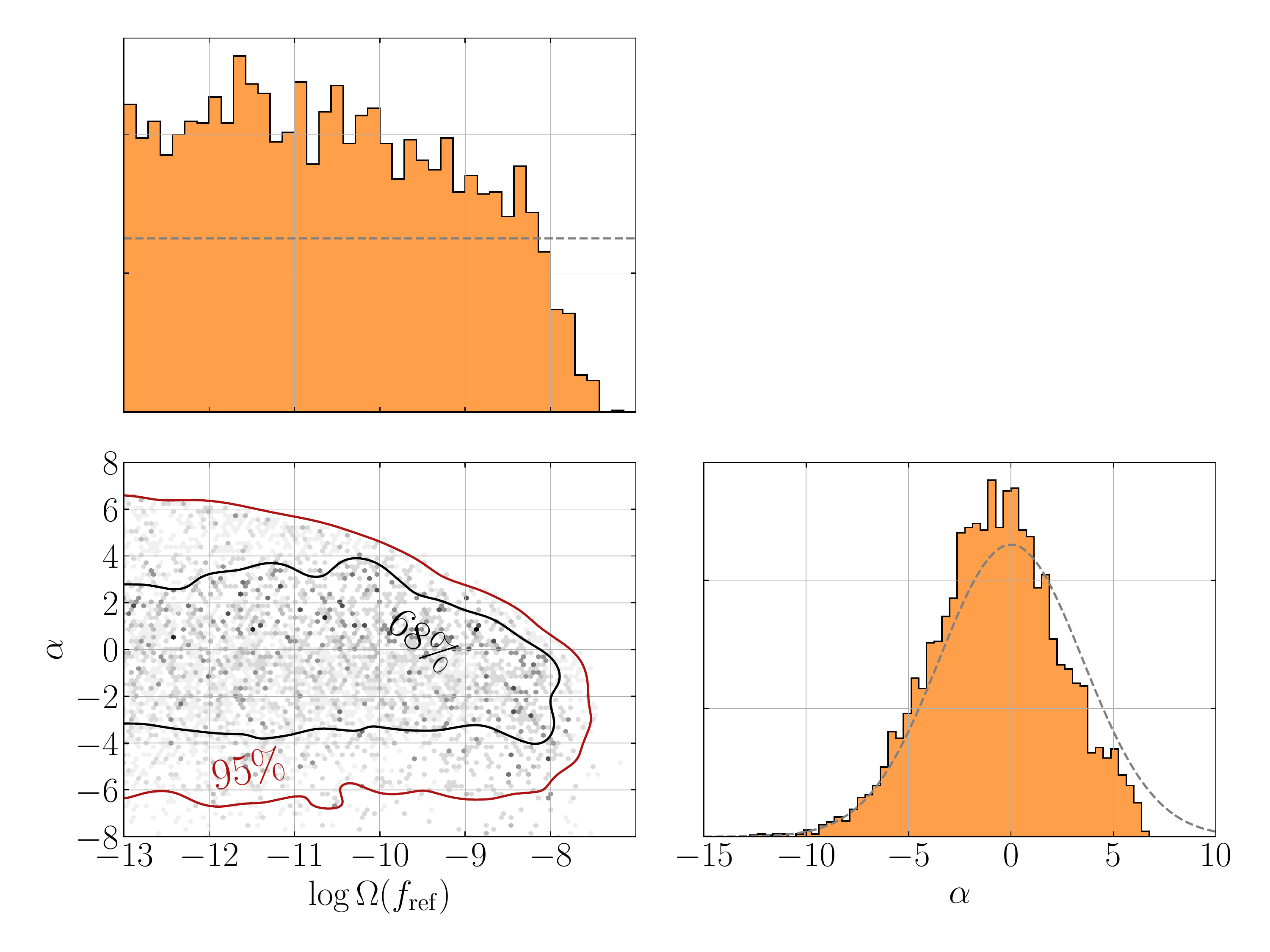}
    \caption{Posterior distributions for the GWB energy density $\Omega_{\rm GW}$ at reference frequency $f=25$ Hz and spectral index $\alpha$. The dashed grey lines show the priors used in parameter estimation. Note that a log-uniform prior was chosen here on $\Omega_{\rm GW}$, and a mean--zero Gaussian was chosen for $\alpha$ as the detector noise PSD is approximately flat. The original version of this plot is presented in~\cite{LVK:2021kbb}; the one shown here was obtained using the open data published in~{\cite{LVK:open_data_iso}}.} 
    \label{fig:OmegaAlpha_posteriors}
\end{figure}
As there is no clear evidence of a GWB, as  LVK places upper limits on the three fixed power law models: $\alpha=0$, typically associated with a scale-invariant, cosmological model~\cite{Caprini2018}; $\alpha=2/3$, which describes a population of inspiralling compact binaries~\cite{Regimbau2011}; and $\alpha=3$, which corresponds to a flat strain power~\cite{Allen1999}. These are astrophysically motivated further in Section~\ref{sec:sources}. With 95\% confidence, it was found that $\Omega_{\rm GW}|_{25 \rm Hz}<0.6\pm1.7 \times 10^{-8} \,\,\,(\alpha = 0)$, $0.3\pm1.2 \times 10^{-8} \,\,\, (\alpha = 2/3)$, and $0.4\pm1.3 \times 10^{-9} \,\,\, (\alpha = 3)$. In recent analyses, $\alpha$ is also allowed to vary as a parameter of the GWB model in likelihood~(\ref{eq:Omega_like}). In this case, a mean-zero Gaussian prior is assumed, with standard deviation $3.5$. Marginalising over $\alpha$ with this method yields a 95\% confidence upper limit of $\Omega_{\rm GW}|_{25 \rm Hz}<0.7\pm2.7 \times 10^{-8}$. When comparing the hypotheses of signal and noise to noise-only, the $\rm log_{10}$ Bayes Factor is found to be $-0.3$. The posteriors for the joint fit of $\alpha$ and $\Omega_{\rm GW}|_{25 \rm Hz}$ are shown here in Figure~\ref{fig:OmegaAlpha_posteriors}.

The results discussed so far assume that the stochastic GW signal is entirely composed of the two transverse-traceless (TT) polarisation modes predicted by GR.
However, modified gravity theories generally contain additional polarisations, up to a maximum of six: the two TT modes of GR, one scalar ``breathing'' mode, and three longitudinal modes---one scalar and two vector~\cite{Will:2014kxa}.
A detection of non-TT polarisation content would provide compelling evidence for the breakdown of GR.
Using the method first developed in Ref.~\cite{Callister:2017ocg},  LVKs have conducted searches for non-GR polarisations in  GWB, using the distinct overlap reduction functions associated with each different set of modes (scalar, vector, and tensor) and marginalising over the power-law slope $\alpha$ as described above~\cite{LIGOScientific:2018czr,LIGOScientific:2019vic,LVK:2021kbb}.
The resulting upper limits on the GWB intensity are broadly similar in each case, giving $\Omega_\mathrm{T}<6.4\times10^{-9}$ for tensors, $\Omega_\mathrm{V}<7.9\times10^{-9}$ for vectors, and $\Omega_\mathrm{S}<2.1\times10^{-8}$ for scalars with the O3 dataset~\cite{LVK:2021kbb}.

 LVK assessed future detection prospects by projecting the expected astrophysical background signal, given the current GW detections, onto the expected sensitivities of upcoming detector networks. This comparison is present in Figure~\ref{fig:LIGO_CBCBackground} in Section~\ref{ssec:astro_sources}, where we discuss the different contributions to the stochastic background. The expected signal includes contributions from BBHs, BNSs, and BHNSs, all of which have  representative direct detections in the advanced LIGO-Virgo runs.
We discuss details of the astrophysical background model in Section~\ref{ssec:astro_sources}; in essence, the total GWB at the selected reference frequency is expected to be $\Omega^{\rm CBC}_{\rm GW}|_{25 \rm Hz}<1.9\times10^{-9}$, which is about an order of magnitude away from the upper limit set with $\alpha=2/3$. Taking $\Omega^{\rm CBC}_{\rm GW}$ as an upper limit, it is compared with the 2$\sigma$ integrated sensitivity curve for {\tt O3},
2-year projections for the Advanced LIGO-Virgo network operating at design sensitivity~\cite{designcurve}, and a 2-year, 50\% duty cycle A+ network at design sensitivity~\cite{LivingRev2020}. Figure~\ref{fig:LIGO_CBCBackground} shown in Section~\ref{sec:sources} suggests that by the time design sensitivity is reached for  Advanced LIGO-Virgo detectors, these will be marginally sensitive to the background. By the A+ phase, the vast majority of the background signal should be within reach of detection. 

\begin{figure}[t]
    \centering
    \includegraphics[width=.7\textwidth]{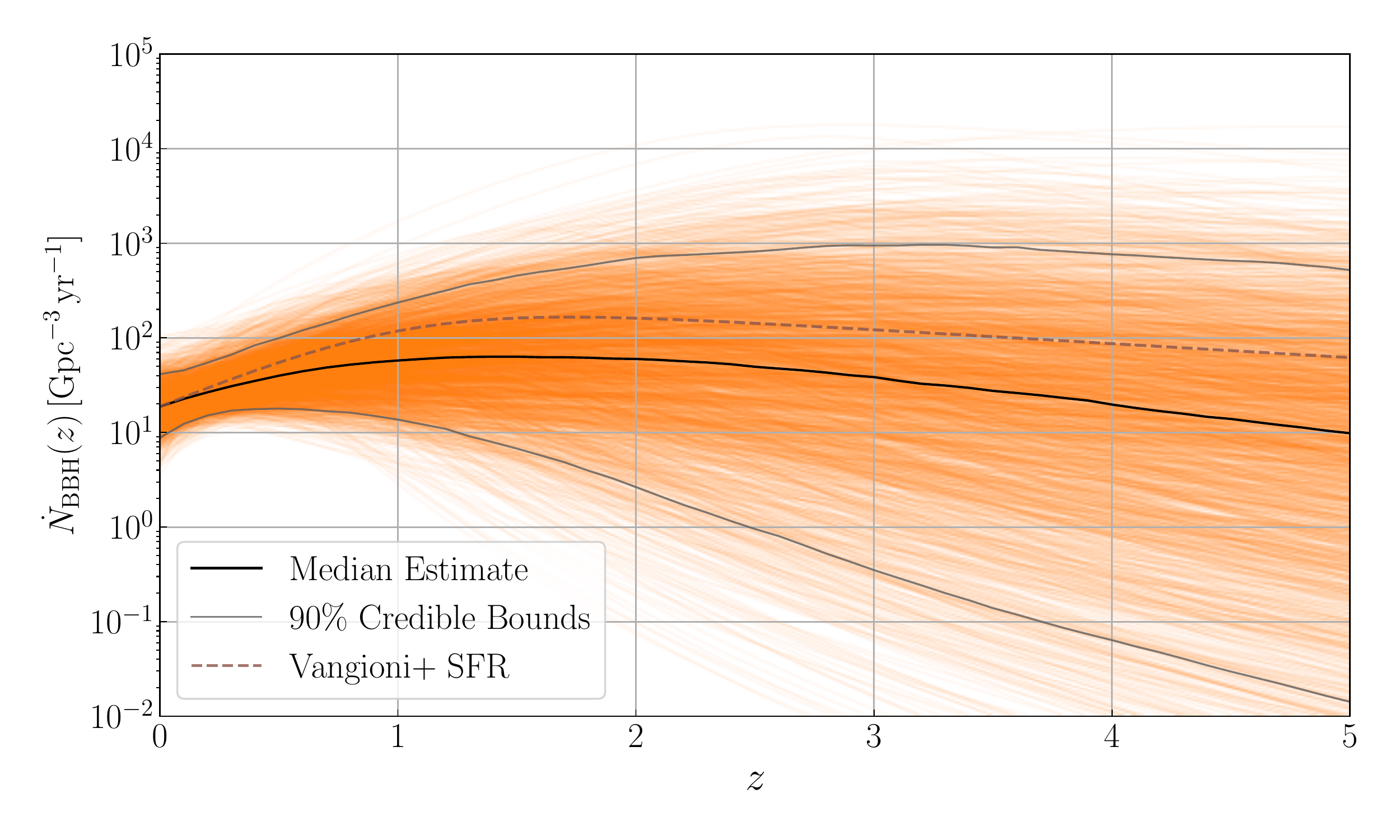}
    \caption{Collection of all posterior draws for the merger rate as a function of redshift $\dot{N}_{\rm BBH}(z)$ inferred from the combination of the resolved BBH detections in the GWTC-2 catalog~\cite{GWTC-2} and the upper limits on the stochastic background obtained with LIGO/Virgo O3 data. The original version of this plot is presented in~\cite{LVK:2021kbb}; the one shown here was obtained using  open data published in~\cite{LVK:open_data_iso}.} 
    \label{fig:bbh-rate-redshift}
\end{figure}

Both the astrophysical GWB spectrum and the set of individual BBH signals resolved by LIGO/Virgo are determined by the cosmic BBH merger rate history $\dot{N}(z)$.
The non-observation of the GWB from this category of sources sets an upper bound on this function, while the set of resolved signals provides a lower bound at low redshift.
By combining both pieces of information, we can, therefore, infer something about this merger history~\cite{Callister2020,LVK:2021kbb}, potentially revealing interesting astrophysical information about BBH abundances and formation channels, such as whether BBHs are predominantly formed through isolated binary evolution or through dynamical assembly in dense stellar environments.
As was recently pointed out in Refs.~\cite{Mukherjee:2020tvr,Buscicchio:2020cij,Buscicchio:2020bdq}, these inferences about the BBH merger rate at high redshift can also be used to set an upper bound on the fraction of individual detections that are expected to undergo strong gravitational lensing.
At present (i.e., post-O3), the uncertainty on $\dot{N}(z)$ is still large, spanning several orders of magnitude at redshifts \mbox{$z\gtrsim1$}~\cite{LVK:2021kbb}, but we can look forward to significant improvements with future LVK observing runs.
The inferred redshift shape of $\dot{N}(z)$ is currently consistent with the BBH merger rate directly tracking the cosmic star formation rate (see Figure~\ref{fig:bbh-rate-redshift}), but there are some tentative hints of a shallower growth at low redshift, as one would expect if there is a non-zero delay time between star formation and the eventual merger event. Note that the star formation rate reported here is arbitrarily normalised.

LVK stochastic searches have also been used to search for various cosmological sources of GWs that have been proposed in the literature, which we reviewed in Section~\ref{sec:sources}. Among the most promising of these sources are cosmic strings, and LVK stochastic searches can be used in particular to place upper limits on the cosmic string tension $G\mu$. The precise constraint depends on the model adopted for the string loop network, but the strongest constraint from the first three observing runs is $G\mu<4\times10^{-15}$~\cite{LIGOScientific:2021nrg}.
This is the strongest existing constraint on a cosmic string model from any experiment; for comparison, CMB constraints are typically at the level of $G\mu\lesssim10^{-7}$~\cite{Planck:2013mgr}.
Another potential signal of cosmological origin is  SGWB from PBHs. The non-observation of this signal by LIGO/Virgo implies that PBHs in the mass range $2\,M_\odot\text{--}200\,M_\odot$ cannot make up the entirety of the cosmic CDM budget~\cite{Vaskonen:2019jpv,Carr:2020gox}, $f_\mathrm{PBH}\lesssim1$; by the time LIGO/Virgo reach design sensitivity, this constraint is expected to improve to $f_\mathrm{PBH}\lesssim10^{-3}$.
Searches have also been conducted for the ``scalar-induced'' SGWB generated by the collapsing overdensities during the process of PBH formation~\cite{Romero-Rodriguez:2021aws}.
For GWs in the LVK frequency band, this corresponds to the formation of very light PBHs, with masses $\lesssim10^{13}\,\mathrm{kg}$.
With future \emph{third-generation} interferometers, these searches are expected to dominate over existing constraints coming from electromagnetic signatures of PBH evaporation.

We have aimed here to provide only a taster of some of the numerous cosmological GW sources that have been probed with LVK stochastic searches.
Many other examples exist, including first-order phase transitions~\cite{Romero:2021kby}, superradiant bosonic condensates around black holes~\cite{Tsukada:2018mbp,Tsukada:2020lgt}, and even sources which are not associated with GWs, but which couple directly to the LIGO/Virgo test masses to mimic a SGWB signal, such as dark photons~\cite{LIGOScientific:2021odm}.

\subsubsection{Search Results for an Anisotropic Background by LVK}
The anisotropic search methods outlined in Section~\ref{ssec:approaches_anisotropic} were most recently applied to LIGO and Virgo detector data in~\cite{LVK:2021mth}. For the first time, data from Virgo was included in the search, leading to a three-baseline network, similarly to the isotropic search presented in Section~\ref{sssec:detection:isotropic:LVK}. The data were pre-processed as in the latter, starting from the \emph{gated} LIGO data-sets. To increase the efficiency of the search, the data were \emph{folded} to one sidereal day before analysis~\cite{Ain2015}; this effectively reduces the cost of the search by a factor equal to the number of days in the dataset, with extremely little loss of information. 

Three different, complementary approaches were employed in the search: a broadband, pixel-based analysis targeting point sources, referenced as the broad-band radiometer (BBR) analysis; a broadband, spherical harmonic decomposition (SHD) analysis, sensitive to extended stochastic sources~\cite{Thrane2009}; and, finally, a narrowband analysis targetting three specific sky locations where a GWB signal may be expected, dubbed the narrowband radiometer (NBR) search. Notably, the so-called ``radiometer'' searches are both carried out in pixel space and ignore  off-diagonal correlations present in the Fisher information matrix (Equation~(\ref{eq:fisher_map_soln})) under the assumption that each direction is fully uncorrelated. The SHD search on the other hand is aimed at diffuse sources which span multiple directions and, thus, requires tackling the Fisher matrix inversion problem. The spherical harmonic domain is chosen here to reduce the effective number of degrees of freedom to a small number of modes on the sky, assuming the measurement is \emph{diffraction-limited}, i.e., the maximum resolvable scale on the sky is directly related to  frequency $f$ and  aperture $D$ of an interferometric measurement by the following~(see, e.g., \citep{LIGOScientific:2016nwa} for a more detailed discussion),
\begin{equation}
    \ell_{\rm max} \sim \frac{2 \pi D f}{c}\,.
\end{equation}
In a cross-correlation GW analysis with pairs of detectors, $D$ is the length of the baseline for each detector pair. To obtain an estimate for a sensible $\ell_{\rm max}$ to use in the search,  frequency $f$ is assumed to be the frequency at which the detectors are most sensitive to the stochastic signal; in practice, three different spectral indices were employed in the SHD search in~\cite{LVK:2021mth}, which approach  LIGO/Virgo sensitivity curves at different ``peak'' frequencies. These suggest a band-limiting of $\ell_{\rm max}=3,$ 4, and 16 for $\alpha=0,$ 2/3, and 3, respectively.

\begin{figure}[H]
    \includegraphics[width=1\linewidth]{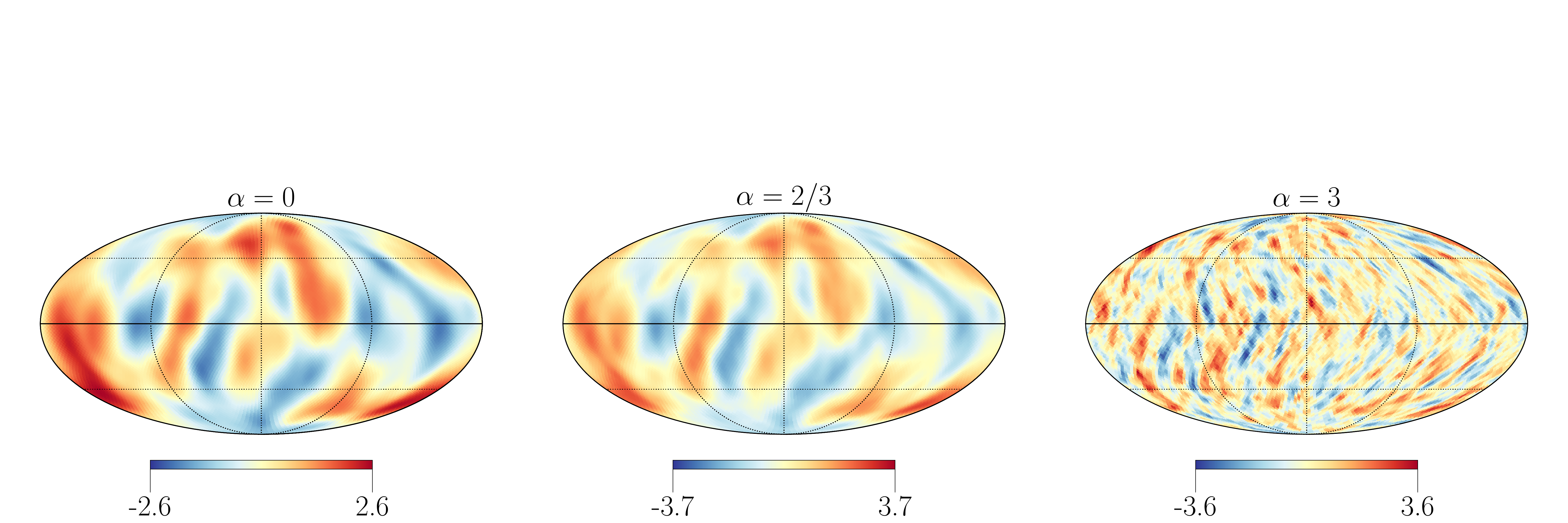}
    \caption{SNR maps obtained with the ``broad--band radiometer'' search method applied to {\tt O1}, {\tt O2}, and {\tt O3} data, for different fixed values of the spectral index $\alpha$, as explained in the text. It is clear there is no significant high SNR signal in the maps. The original version of this plot is presented in~\cite{LVK:2021mth}; the one shown here was obtained using  open data published in~\cite{LVK:open_data_aniso}. }
    \label{fig:LIGO_BBR_SNR_maps} 
\end{figure}

\begin{figure}[H]%
    \centering
    \includegraphics[width=0.65\textwidth]{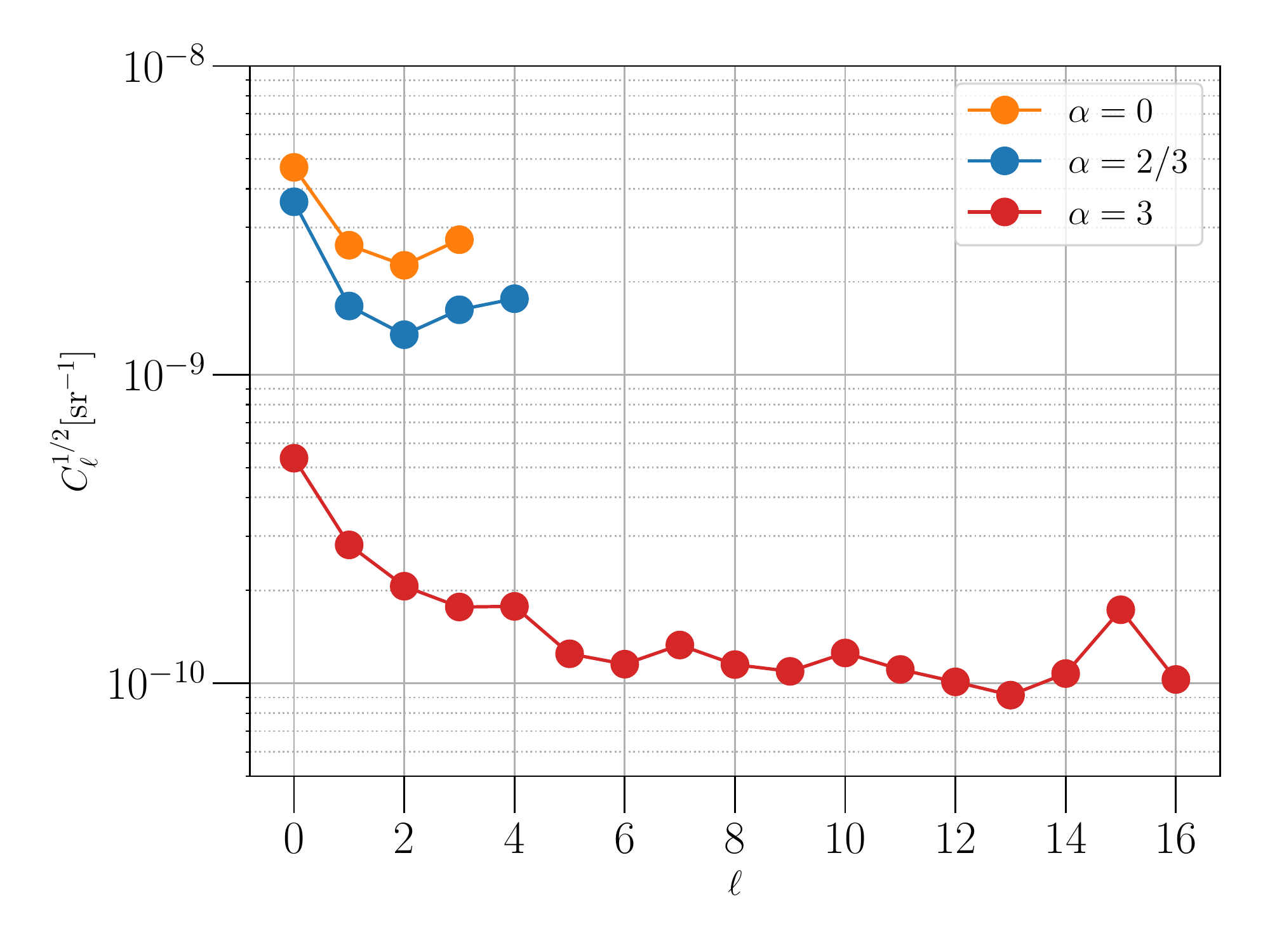}
    \caption{Angular power spectrum upper limits obtained with the ``spherical harmonic decomposition'' search method applied to {\tt O1}, {\tt O2}, and {\tt O3} data, for different fixed values of the spectral index $\alpha$, as explained in the text. The original version of this plot is presented in~\cite{LVK:2021mth}; the one shown here was obtained by using  open data published in~\cite{LVK:open_data_aniso}.} 
    \label{fig:LIGO_Cells}
\end{figure}

 The BBR analysis yields sky maps consistent with mean zero Gaussian noise and, thus, provides directional upper limits, as seen in the signal-to-noise ratio (SNR) maps presented in Figure~\ref{fig:LIGO_BBR_SNR_maps}. Note that these are obtained by assuming each direction is fully independent and, thus, should not be read as maps at all, but rather each pixel value should be interpreted as the upper limit on stochastic GWs in that particular direction in the case where that direction is the only one contributing to a stochastic signal. Similarly,  SHD analysis produces spherical harmonic coefficients for  GWB, band-limited according to the prescription above, which are then converted to the pixel domain yielding maps consistent with noise. The angular power spectrum $C_\ell^{\rm GWB}$ as defined by Equation~(\ref{eq:Cell_omega}) can be directly estimated from the coefficients; however, as the measurement is noise-dominated, the unbiased estimator is~\cite{Thrane2009, Renzini2019b}
\begin{equation}
    \hat{C'}_\ell^{\rm GWB} = \hat{C}_\ell^{\rm GWB} - {\cal N}_\ell\,,\qquad {\cal N}_\ell = \frac{1}{2\ell + 1} \sum_m F^{-1}_{\ell m, \ell m}.
\end{equation}
Here, $\hat{C}_\ell^{\rm GWB} $ is the naive angular power spectrum estimator calculated as in Equation~(\ref{eq:Cell_omega}) using the spherical harmonic coefficients obtained from the data. $\cal N_\ell$ is an estimate of the noise covariance of the GW measurement, containing the appropriately normalised inverse Fisher matrix $F$ calculated in spherical harmonic coordinates and is taken to be a measure of the uncertainty. Employing the unbiased estimator results in an angular power spectrum estimate fully consistent with zero; thus 95\%, upper limits are placed on the spectrum as shown in Figure~\ref{fig:LIGO_Cells}.

\unskip

 The NBR analysis described in~\cite{LVK:2021mth} sets upper limits on frequency spectra from three astrophysically relevant sky locations: the direction of the X-ray emitting binary system \emph{Scorpius X-1}~\cite{LVC_ScorpiusX2017}; the direction of the remnant of Supernova 1987A~\cite{PhysRevD.94.082004}; and the center of our Galaxy, which may contain relevant GW sources~\cite{LVC_galacticCenter2013}. No overdensity of GWs is found in any of these directions; hence, narrowband upper limits are placed in each frequency bin.
Recently,  LVK extended this type of search to an ``All--Sky, All--Frequency'' (ASAF) search~\cite{ASAF}, which produces sky maps for each frequency bin, in the pixel domain and at high pixelisation, assuming each pixel is uncorrelated. This type of search is best suited to quickly identify candidates for continuous, quasi-monochromatic GW wave sources coming from point sources on the sky, e.g., rotating neutron stars. The results of this search found a small set of outliers which may be investigated with continuous waves searches, for example, with methods proposed in~\cite{Tenorio:2021njf}.

\subsection{Stochastic Searches with Pulsar Timing Arrays}\label{ssec:pta_results}

Unlike  ground-based interferometers, the first astrophysical signal observed with PTAs is expected to be a stochastic signal~\cite{Rosado:2015epa} from supermassive black hole binaries.
For an insightful review of the astrophysics that can be performed with PTAs, see~\cite{Burke-SpolaorTaylor2019}, and for a broader overview of the field, see~\cite{Taylor2021}.
The concept of such experiments was proposed by Sazhin~\cite{sazhin1978} and further discussed by Detweiler~\cite{detweiler1979}.
To reach nanohertz frequencies, PTAs perform arrival time measurements from an ensemble of millisecond pulsars for a minimum of several years.
The first three PTAs are the EPTA~\cite{KramerChampion2013},  NANOGrav~\cite{McLaughlin2013}, and  PPTA~\cite{ManchesterHobbs2013}.
The dependence of the Hellings--Downs curve in Equation~\eqref{eq:orfhd} on the angular coordinates of pulsars implies that timing arrays benefit from observing pulsar pairs across the entire sky.
Specifically, it was shown by Siemens et al.~\cite{SiemensEllis2013} that it is important to increase the number of observed pulsar pairs to resolve  Earth-term spatial correlations of the background from the spatially uncorrelated pulsar-term component of the background, which manifests as time-correlated noise.

While EPTA and NANOGrav mostly observe pulsars from the northern hemisphere, PPTA has access to pulsars from the southern hemisphere.
Recently, two more efforts have emerged: the Indian Pulsar Timing Array~(InPTA,  \cite{JoshiArumugasamy2018}, which is the newest member of IPTA); and the MeerTime project~\cite{BailesJameson2020}, which is based on the MeerKAT~\cite{Jonas2009} facility in South Africa, a precursor of SKA~\cite{DewdneyHall2009}.
A recent proposal sees the new telescope FAST~\cite{NanLi2011}, the largest ``dish'' radio telescope in the world, laying the foundations for a PTA collaboration in China.
The IPTA~\cite{HobbsArchibald2010} is the global effort to accelerate the detection of the nanohertz GWB by combining data sets across the previously mentioned member collaborations.

\subsubsection{Search Results for an Isotropic Nanohertz Background}

Searches for  isotropic GWB constrain the power-law strain amplitude $A$ of the background at the GW frequency $f=1~\text{yr}^{-1}$.
Following Equation~\eqref{eq:astro_BG} and the relation between spectral indices explained in footnote~\ref{fn:alphas_general}, the strain spectrum of the background has the form $h(f) = A (f/{\rm 1\, yr^{-1}})^{-2/3}$.
The limits on $A$, denoting either confidence or credibility, are reportedly between $1.1\times 10^{-14}$ and $1.0 \times 10^{-15}$ at $95\%$ \citep{jenet2006gwb,vanHaasterenLevin2011,DemorestFerdman2013,shannon2013gwb,LentatiTaylor2015,NANOGravCollaborationArzoumanian2015,shannon2015gwb,ArzoumanianBaker2018}.

As discussed in Section~\ref{sssec:detection:isotropic:PTAs}, it is expected that, prior to the detection of Hellings--Downs spatial correlations, temporal correlations with the same spectra will be observed in pulsar data sets.
This phenomenon is now known as the ``common-spectrum process'', and it was first reported in the search for the background with the NANOGrav 12.5-year data set~\cite{ArzoumanianBaker2020}, where the authors for the first time did not place limits on the background amplitude but instead provided the measurement of the amplitude of the common-spectrum process.

After that, the PPTA collaboration reported~\cite{goncharov2021cpgwb}  statistical evidence for the common-spectrum process in a similar fashion, while also suggesting that current analyses do not consider a model where spectra are similar but not common.
Furthermore, the authors simulated data set  showed evidence for the common-spectrum process, whereas amplitudes of the injected noise power spectra were different by several orders of magnitude.
Thus, it was shown that the evidence for the common-spectrum process might arise from pulsar noise described by parameters in a similar range or by model mis-specification.

\begin{figure}[t]%
    \centering
    \includegraphics[width=0.6\textwidth]{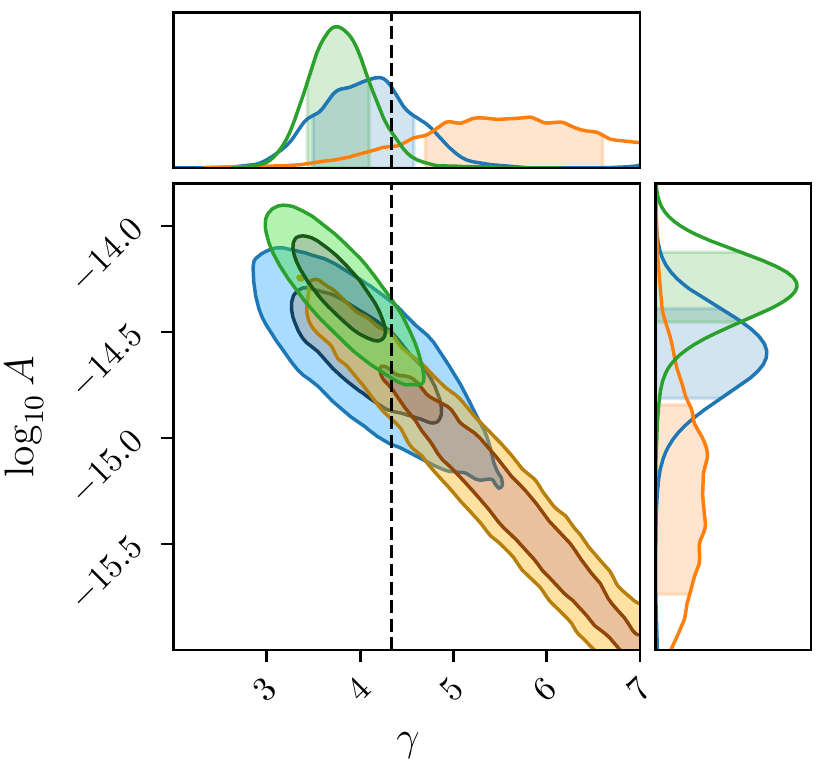}
    \caption{Power-law amplitude $A$ and the negative of the spectral index, $\gamma$ (see Equation~\eqref{eq:PTA_PSD_residuals}), of the common component of power spectral density of timing residuals $\delta\bm{t}$ as reported NANOGrav~(orange,  \cite{ArzoumanianBaker2020}), PPTA~(blue,  \cite{goncharov2021cpgwb}), and EPTA~(green,  \cite{ChenCaballero2021}).}
	\label{fig:pta_common_noise}
\end{figure}

On the other hand, compared with  NANOGrav results~\cite{ArzoumanianBaker2020}, PPTA~\cite{goncharov2021cpgwb} measured the spectral slope $-\gamma$ of the common-spectrum process to be even more consistent with the theoretical prediction for the GWB from supermassive black hole binaries, $\gamma = 13/3$.
The relation between this value and the strain spectral index $\alpha'=-2/3$ is explained in Section~\ref{sssec:detection:isotropic:PTAs}.
Furthermore, under certain analysis conditions, constraints on spatial correlations as a function of pulsar-Earth baselines obtained by the PPTA seem to be even in a better agreement with the Hellings--Down curve.
The PPTA measurement is presented here in Figure~\ref{fig:overlap_reduction_function_PTA} and the NANOGrav measurement may be seen in Figure 7 of~\cite{ArzoumanianBaker2020}.
The PPTA measurement is affected by the pulsar with strong unmodelled excess noise~\cite{goncharov2021cpgwb}.
The EPTA collaboration has also reported evidence for the common-spectrum process with a 24-year-long data set from an observation of six pulsars~\cite{ChenCaballero2021}.
Measurements of power-law parameters of the common-spectrum process by NANOGrav, PPTA, and EPTA are presented in Figure~\ref{fig:pta_common_noise}.

\begin{table}[t]
\caption{\label{tab:isotropic_results}Results of searches for the nanohertz-frequency isotropic GWB with the amplitude $A$ and the energy-density spectral index $\alpha=2/3$ in chronological order. The first three columns list details of corresponding publications. Publications that adopted alphabetical author list are denoted by asterisks. The fourth column contains 95\% upper limits on $A$ for the GWB and measurements of $A$ of the common-spectrum process (CP) with credible levels of $1\sigma$ in~\cite{goncharov2021cpgwb} and 5--95\% in~\cite{ArzoumanianBaker2020,ChenCaballero2021,AntoniadisArzoumanian2022}. The values in the last two columns are the characteristics of the data set: the total observation time $T_\text{obs}$ and the number of pulsars in the array $N_\text{psr}$.}
\begin{tabularx}{0.75\textwidth}{m{4cm}<{\raggedright}m{2cm}<{\centering}m{1.3cm}<{\centering}m{1.8cm}<{\centering}m{1.3cm}<{\centering}m{.9cm}<{\centering}}
\toprule

\textbf{Publication} & \textbf{Collaboration} & \textbf{Year} & \boldmath{$A~(\times 10^{-15})$} & \textbf{\boldmath{$T_\text{obs} (\text{yr})$}} & \textbf{\boldmath{$N_\text{psr}$}}  \\   \midrule
Jenet et al.~\cite{jenet2006gwb} & PPTA & 2006 & $<$11 & 20  & 7 \\
van Haasteren et al.~\cite{vanHaasterenLevin2011} & EPTA & 2011 & $<$6 & 11  & 5 \\
Demorest et al.~\cite{DemorestFerdman2013} & NANOGrav & 2013 &  $<$7 & 5  & 17 \\
Shannon et al.~\cite{shannon2013gwb} & PPTA & 2013 & $<$2.4 & 25  & 6 \\
Lentati et al.~\cite{LentatiTaylor2015} & EPTA & 2015 & $<$3.0 & 18  & 6 \\
 * Arzoumanian et al.~\cite{ArzoumanianBrazier2016} & NANOGrav & 2015 & $<$1.5 & 9  & 37 \\
Shannon et al.~\cite{shannon2015gwb} & PPTA & 2015 & $<$1.0 & 11 & 4 \\
Verbiest et al.~\cite{VerbiestLentati2016} & IPTA & 2016 & $<$1.7 & 21 & 4 \\
 * Arzoumanian et al.~\cite{ArzoumanianBaker2018} & NANOGrav & 2018 & $<$1.45 & 11.4  & 45 \\
 * Arzoumanian et al.~\cite{ArzoumanianBaker2020} & NANOGrav & 2020 & CP: $1.9^{+0.8}_{-0.6}$ & 12.5  & 45 \\
Goncharov et al.~\cite{goncharov2021cpgwb} & PPTA & 2021 & CP: $2.2^{+0.4}_{-0.3}$ & 15.0  & 26 \\
Chen et al.~\cite{ChenCaballero2021} & EPTA & 2021 & CP: $3.0^{+0.9}_{-0.7}$ & 24.0  & 6 \\
 * Antoniadis et al.~\cite{AntoniadisArzoumanian2022} & IPTA & 2022 & CP: $2.8^{+1.2}_{-0.8}$ & 30.2  & 53 \\
\bottomrule
\end{tabularx}

\end{table}

The results of searches for the isotropic background are summarised in Table~\ref{tab:isotropic_results}.
The observation time span $T_\text{obs}$ does not directly increase with publication year due to the inclusion or noninclusion of older data where the limited radio frequency band is not sufficient to model ``chromatic'' red noise arising from pulse propagation effects in the interstellar medium.
One also might notice that a few upper limits on the GW strain amplitude are below the amplitude of the common-spectrum process.
This is partly caused by the adoption of the older Solar System ephemerides in earlier publications, which were known to contain systematic errors.
The origin of the common-spectrum process is to be clarified in future work through better constraints on spatial and temporal correlations.
For a more thorough discussion about the noise with references please refer to the following subsections.

\subsubsection{Challenges in GWB Searches with PTAs}
\label{sssec:efforts_prospects:ptas:challenges}

PTAs, as galactic-scale gravitational wave detectors, are affected by a wide variety of noise sources. These include instrumental noise artifacts from the instruments and the hardware that processes raw telescope data; effects induced by the Solar System planets and/or the interstellar medium; and jittering of the neutron stars themselves---sources of radio waves---and their gravitationally bound companions such as other neutron stars, white dwarfs, or black holes. 
Both instrumental and environmental noise sources play an important role.
As discussed in Section~\ref{sssec:detection:isotropic:PTAs}, stochastic processes in PTAs can be uncorrelated in time, i.e., white, and time-correlated, i.e., red.
Some noise processes introduce spatial correlations between pulse arrival times in pulsars across the sky.
Below we briefly discuss prospects for future PTAs given the current knowledge of the noise.

 Environmental white noise associated with pulse shape changes is referred to as \textit{jitter}~(e.g., \cite{ShannonOslowski2014,LamMcLaughlin2019}).
Without jitter noise, measurement errors are limited by the radio pulse signal-to-noise ratio.
New scientific instruments such as FAST and SKA will significantly improve upon previous radiometer (white instrumental) noise levels, thus expanding the number of pulsars in the array and improving noise power spectra of the observed pulsars.
Therefore, the timing precision of the brightest pulsars will be limited by  environmental properties.
In particular, MPTA reported~\cite{BailesJameson2020} an extraordinary timing precision for some of the millisecond pulsars previously observed by the PPTA.
PSR~J1909--3744 showed the root-mean-square of timing residuals (rms\footnote{Often used in the literature to quantify noise in a given pulsar. For rms $\sigma^2$ of Gaussian white noise, the power spectral density of residuals $P = 2 \pi \sigma^2 \Delta t$, where $\Delta t$ is the time between observations, which is typically a couple of weeks. Note, rms residuals may increase with longer data spans due to more sources of noise becoming prominent at longer time scales. Sometimes rms residuals are provided for a particular source of noise.}) of 66 ns with 11 months of observations, whereas in the second data release of  PPTA~\cite{KerrReardon2020}, this pulsar showed rms residuals of 240 ns.
Pulse integration times may be increased in the future to further lower white noise levels.
Following Equation~\eqref{eq:pta_psd_strain}, a pulsar with an achievable timing precision of 10 ns observed biweekly, in 10 years, would be able to probe a GW strain of $5 \times 10^{-17}$ at 3 nHz.
In reality, however, the detection of a nanohertz stochastic background requires timing an ensemble of pulsars, and there are sources of red noise that limit timing arrays at low frequencies.

Red noise is resolved in most pulsars after they are observed for a couple of years.
A comprehensive study of this topic for the first data release of the IPTA can be found in~\cite{LentatiShannon2016}.
Normally, the strongest red noise in recycled pulsars appears at low radio frequencies $\nu$ from variations in the dispersion measure, electron column density towards the line of sight.
Due to  dependence $P(f) \propto \nu^{-2}$, this noise is easy to model and it is straightforward to separate its contribution from the one of the GWB.
Some red noise was found to be affecting only certain radio frequency bands or back-end observing systems.
Independent of $\nu$, ``achromatic'' red noise is also referred to as spin noise as it is associated with pulsar rotational irregularities.
Spin noise with a power spectral density which continuously increases towards low frequencies with a power-law might eventually turn over, but no such spectra were found in millisecond pulsars~\cite{GoncharovZhu2020,ChalumeauBabak2022}.
One may find the term ``timing noise'' in the literature, which is used to denote spin noise and glitches, i.e., sudden changes in pulsar spin periods followed by an exponential relaxation~(e.g., \cite{HaskellMelatos2015}).
GWBs are separated from spin noise through spatial correlations, although incorrect red noise models may result in incorrect inferences of the GWB~\cite{HazbounSimon2020}.
For the EPTA second data release, it was shown~\cite{ChalumeauBabak2022} that tailored pulsar red noise models with band- and system-related noise marginally increase evidence for Hellings--Downs correlations,
whereas  PPTA showed that the inclusion of the brightest PSR~J0437--4715 (which shows significant red noise and residual excess noise~\cite{GoncharovReardon2021}) results in an increase in the inferred correlation coefficients $\Gamma(\zeta_{ab})$~\cite{goncharov2021cpgwb}.
Before spatial correlations are resolved, both the GWB and spin noise can yield evidence for the common-spectrum process~\cite{goncharov2021cpgwb}.
Inferences of the GWB based on the common-spectrum process at this point in time may be treated as speculative, especially given that spin noise in pulsars may yield a power-law $P(f)$ with $\gamma=4$~\cite{MeyersO'Neill2021}, which is close to $\gamma=13/3=4.3(3)$ for the GWB.
Some low-frequency noise can also introduce spatial correlations which can still be distinguished from the Hellings--Downs prediction~\cite{TiburziHobbs2016}.
Time-correlated errors in terrestrial time standards, also known as clock errors, can yield monopolar spatial correlations between pulsars $a$ and $b$, $\Gamma(\zeta_{ab}) = 1$.
Pulse arrival times are referenced to the position of the Solar System barycenter; thus, systematic errors in the barycenter position can yield red noise with dipolar spatial correlations $\Gamma(\zeta_{ab}) = \cos(\zeta_{ab})$.
This noise can be mitigated byt improving Solar System ephemerides, and when improvements are not available, it can be modelled as a purely deterministic process~\mbox{\cite{GuoLi2019,VallisneriTaylor2020}}.
In an analysis with the NANOGrav 11 year data set~\cite{ArzoumanianBaker2018}, it was demonstrated that the results are affected by the choice of Solar System ephemeris, whereas deterministic modeling of systematic errors in ephemerides renders consistent results.
The good news is that the contributions of ephemeris noise are found to be rather weak for current data sets and ephemerides.
In particular, in the search for a stochastic background in the second data release of the PPTA~\cite{goncharov2021cpgwb}, the authors computed Bayesian evidence for errors in every parameter of the deterministic ephemeris model: masses and orbital Keplerian parameters of planets from Mars to Saturn, with orbital frequencies in the PTA band.
They found marginal evidence for systematic errors in older ephemerides and no evidence for errors in present-day ephemerides used for  GW analyses between 2020 and 2021.
Red noise is mostly environmental; thus, it might be worth allocating more time to fainter pulsars that do not show evidence of such noise.

\begin{figure}[t]%
    \centering
    \includegraphics[width=0.7\textwidth]{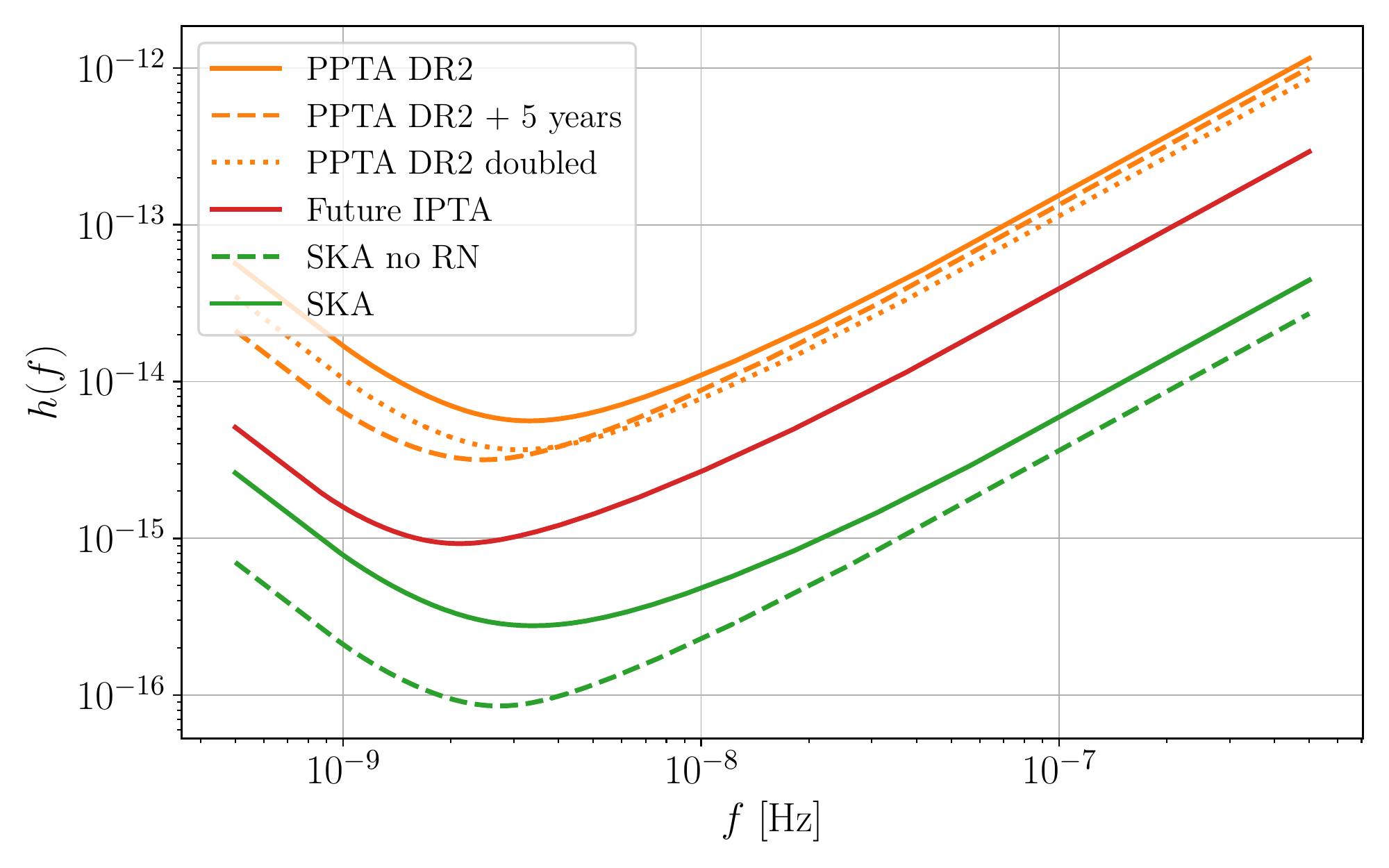}
    \caption{Power-law-integrated sensitivity~\cite{HazbounRomano2019a,HazbounRomano2019b} at the level of SNR = 3 of simulated PTAs to the SGWB strain $h(t)$. We demonstrate the effects of timing precision, red noise, and a number of array pulsars. Pulsars are distributed isotropically across the sky and observed biweekly. Orange lines represent variations of the 15-year-long PPTA DR2, with real pulsar $T_\text{obs}$, white noise levels based on the rms residuals ranging from 240 ns to 24,050 ns~\cite{KerrReardon2020} and best-fit red noise parameters from~\cite{GoncharovReardon2021}. The dashed line represents PPTA DR2 extended by 5 more years of observation, whereas the dotted line represents PPTA DR2 with a doubled number of pulsars with the same noise properties. Red line is an example of the future IPTA with 20 years of observations of 100 pulsars with white noise rms drawn uniformly between 20 and 500 ns. Red noise $\log_{10}A$ is drawn uniformly from $[-14, -20]$ and $\gamma$ drawn uniformly from $[1,7]$, a fiducial choice based on~\cite{GoncharovReardon2021}. Green lines represent SKA observing 50 pulsars with white noise rms of 30 ns for 15 years, with (solid line) and without (dashed line) red noise assumed for the IPTA. } 
    \label{fig:pta_future_sensitivity}
\end{figure}

With the above in mind, we provide sensitivity projections for a few scenarios of future PTAs in Figure~\ref{fig:pta_future_sensitivity}.
When evaluating the performance of future PTAs, several simplifications are usually made.
For example, often, red noise and effects of fitting for the timing model are neglected, as in Figures \ref{fig:GWBplot} and \ref{fig:constraints_2}.
We addressed these points in Figure~\ref{fig:pta_future_sensitivity} by adopting the formalism from~\cite{HazbounRomano2019a} and the relevant code~\cite{HazbounRomano2019b}.
The sensitivity estimates can be further refined based on more realistic noise properties and observing scenarios---a subject of future work.
For a more detailed review of measurement errors in pulsar timing arrays please refer to~\cite{VerbiestShaifullah2018}.
Challenges for GW detection with pulsar timing arrays are also discussed in~\cite{Lommen2015}.

\subsubsection{
Search Results for an Anisotropic Nanohertz Background}
To date, one search for an anisotropic GWB with PTAs has been performed~\cite{TaylorMingarelli2015}. In the latter, the authors use six pulsars from the EPTA with a timing baseline of $T_{\rm obs}=17.7\,\rm{years}$. While the results indicate that the data are unable to improve upon a physically motivated prior, the search setup, motivation, and results are useful to review.

The authors use a spherical harmonic decomposition with $\ell_{\rm max}=4$ based on analytical arguments of the PTA resolution established in~\cite{SesanaVecchio2010}. They perform three parameterizations and one independent analysis. First, they use a single set of anisotropy coefficients across the entire frequency band (lowest five harmonics of $1/T_{\rm obs}$) and chose a fixed spectral index consistent with a population of supermassive black hole binaries. Second, they allow  different anisotropy coefficients at each of the lowest five harmonics of $1/T_{\rm obs}$. Third, they allow  different spherical harmonic coefficients for each of the lowest four harmonics of $1/T_{\rm obs}$ and use a single set of spherical harmonics coefficients to describe up to the fiftieth harmonic. In all cases, they impose that the power on the sky in each pixel is positive, as negative power would be unphysical. Fourth, they independently estimate the cross-correlations between individual pulsars and then map these onto a chosen basis.

The results yielded no detection of a GWB, and limits were set on the angular power spectrum $C_\ell$ for each of the individual parameterizations (and for each individual frequency, where appropriate). The limits on strain amplitude in multipoles with $\ell > 0$ are $\lesssim 40$\% of the monopole. However, these limits do not improve upon the physically motivated prior. Current astrophysical estimates suggest that the strain amplitude in multipoles with $\ell>0$ is around 20\% of the monopole~\cite{MingarelliLazio2017}. In the case of the fourth analysis, the physical prior is not applied, and the authors observed that sensitivity to the dipole is reduced due to the distribution of the six pulsars on the sky, highlighting the desire for using a large number of pulsars that are uniformly distributed on the sky.

\subsubsection{Search Results for a Nanohertz Background Not Related to Supermassive Black Hole Binaries}
While quite a few sources of a GWB that can be measured by PTAs have been proposed, outside of the generic isotropic background, which is generally assumed to be dominated by unresolved supermassive blackhole binaries, only a few explicit searches have been carried out. 

Recently, both the PPTA and NANOGrav have placed constraints on first-order phase transitions in the early Universe~\cite{ArzoumanianBaker2021b,XueBian2021}. In~\cite{ArzoumanianBaker2021b}, the authors use 45 pulsars to search for a signal from first-order phase transitions using three different models. The common spectrum process first reported in NANOGrav data in~\cite{ArzoumanianBaker2020} is also consistent with a first-order phase transition at temperatures below the electroweak scale, with $T^\star \lesssim 10\,\rm{MeV}$ at the $1\sigma$. They calculate posterior distributions on the temperature and bubble nucleation rate when  phase transition occurs, as well as the strength of the phase transition and the friction between bubble walls and the plasma. However, it is important to note that these posteriors assume only a GWB from phase transitions is present, while the data cannot distinguish between this model and one described by a GWB from supermassive binary black holes. In addition, no spatial correlations between pulsars are found; hence, no detection of a GWB is claimed. The results of~\cite{XueBian2021} are similar---no evidence of spatial correlations between pulsars is reported, and constraints on the temperature of a potential phase transition are found to be $1\,\textrm{MeV}\lesssim T^\star \lesssim 100\,\textrm{MeV}$. For a recent discussion on separating a background from supermassive black hole binaries and first order phase transitions (or other potential sources of a nHz GWB) using a multi-messenger approach, see~\cite{MooreVecchio2021}.

Several groups have recently searched NANOGrav, PPTA, and IPTA data sets for evidence of alternative polarizations of GWs. In~\cite{ChenYuan2021,ChenWu2021}, the authors claim a Bayes factor of $\sim$100 in preference of background from spatially correlated scalar-transverse gravitational waves compared to a simple common spectrum red noise process using both NANOGrav and IPTA data. They claim similar significance when comparing a scalar transverse background to a background from tensor transverse gravitational waves. However, they found no such significant evidence when searching PPTA data~\cite{WuChen2021}. In~\cite{ArzoumanianBaker2021a}, the authors find a preference for a scalar transverse background, but with less significant evidence. However, this significance goes away when the authors account for uncertainties in the solar system ephemeris using~\cite{VallisneriTaylor2020} or when removing one pulsar, PSR J0030+0451.


\subsection{Other Stochastic Background Searches}

There are many GW experiments that we have omitted from the discussion above.
In this section, we very briefly highlight a few that are of particular interest, some of which are shown in Figure~\ref{fig:constraints_2}.
Our focus here is on the nanohertz--kilohertz frequency band; however, see the review article~\cite{Aggarwal:2020olq} and references therein for a summary of detection methods at MHz frequencies and above.

The oldest method of GW detection, pioneered by Weber~\cite{Weber:1960zz,Weber:1969bz} in the 1960s with his ``resonant bar'' experiments, is to monitor the resonant frequencies of some macroscopic test object.
When acted upon by a GW with a frequency matching one of the object's normal modes, the stretching and squeezing action they induce excites vibrations in these modes which can be amplified and detected.
Numerous such resonant-mass experiments have been operated since the 1960s~\cite{Aguiar:2010kn}, but none has had sufficient sensitivity to place meaningful bounds on the GWB below the $\Omega(f)<1$ level required to prevent GWs from over-closing the Universe.
However, the same idea has been applied with success on a much larger scale, using the Earth itself as a resonant mass.
By monitoring the Earth's normal mode frequencies using gravimeter data, Coughlin and Harms~\cite{Coughlin:2014xua} placed an upper limit on the GWB in the megahertz frequency band.
The corresponding PI curve has a minimum of $\Omega(f)\approx1.3\times10^{-2}$ at $f\approx1.6\,\mathrm{mHz}$.
Recently, a trio of similar experiments have been proposed in which gravimeters installed on the Lunar surface could be used to monitor the Moon's normal modes, providing a cleaner probe of GWs in the megahertz frequency band~\cite{LGWA:2020mma,Jani:2020gnz}.

Another GW experiment which has successfully constrained the GWB below the $\Omega(f)<1$ level is the work of Armstrong {et al.}~\cite{Armstrong:2003ay}, who carried out Doppler-tracking observations of the Cassini spacecraft.
The principle here is very similar to IFOs and PTAs in that it tracks the trajectories of photons in the presence of GWs; however, rather than measuring changes in the light-travel time, one attempts to measure changes in the photon frequency due to GW-induced Doppler shifts between the two test masses (this is essentially the time derivative of the perturbation to the light-travel time).
This analysis probed GWs in the $1\text{--}100\,\upmu\mathrm{Hz}$ band, with the corresponding PI curve reaching a minimum of $\Omega(f)\approx9.9\times10^{-3}$ at $f\approx3.9\,\upmu\mathrm{Hz}$.
This is a particularly interesting frequency range, as it straddles a significant gap between PTA constraints at nanohertz frequencies and future constraints from LISA and other space-based interferometers in the megahertz band; however, the Cassini constraint in this ``$\upmu$Hz gap'' is not strong enough to be of astrophysical or cosmological interest.
Recently, it has been pointed out that much more competitive GWB constraints could be obtained in this frequency band using high-precision observations of binary systems (in particular, binary pulsars and the Earth--Moon system), by searching for the imprints of stochastic GWs on their orbits~\cite{Blas:2021mpc,Blas:2021mqw}.
Other proposals for accessing these frequencies include an extremely long-baseline space-based interferometer such as the $\upmu$Ares concept~\cite{Sesana:2019vho}.

In addition to changing the frequency and/or time-of-arrival of photons, GWs can also cause \emph{angular deflections} in photon trajectories, such that distant astronomical objects appear displaced from their true positions on the sky.
These GW-induced deflections have a distinctive geometric pattern across the sky, allowing one to search for a GWB signal with high-precision astrometry from observatories such as GAIA and THEIA~\cite{Book:2010pf,Moore:2017ity,Garcia-Bellido:2021zgu}.
The resulting PI curve lies in roughly in the same nanohertz frequency band probed by PTAs, albeit with improved high-frequency sensitivity.
Forecasts in Ref.~\cite{Garcia-Bellido:2021zgu} indicate that next-generation astrometry with an observatory such as THEIA may be sensitive to $\Omega(f)\sim10^{-16}$ at nanohertz frequencies, surpassing even the reach of pulsar-timing observations with~SKA.

Finally, a GW detection technique which is subject to growing interest in the community is \emph{atom interferometry}, in which, rather than using interference between photons to measure the phase difference between two spatial paths, one exploits the wave-like behaviour of matter on the quantum scale to measure interference between atoms.
While first-generation atom interferometry experiments such as AION~\cite{Badurina:2019hst} and MAGIS~\cite{Graham:2017pmn,Abe:2021ksx} are in their early stages and are not expected to have significant sensitivity to GWs, proposals for future kilometer-scale experiments are forecast to provide extremely useful GWB constraints in the $0.1\text{--}1\,\mathrm{Hz}$ band.
In particular, the AION-km proposal has a forecast sensitivity of $\Omega(f)\approx2.7\times10^{-12}$ at $f\approx0.12\,\mathrm{Hz}$.


\section{Stochastic Background Detection Prospects with Future Gravitational-Wave Detectors}\label{sec:conclusions}

We close this review with an outlook on the detection capabilities of upcoming GW detectors.
We first discuss the potential of 3G detectors, which will be considerably more sensitive to GWs in the same frequency bands as  LVK instruments, as may be observed in Figure~\ref{fig:constraints_2}, and specifically prospects of performing component separation with ET and CE. 
We then review the upcoming LISA mission and lay out different stochastic search methods which may be employed with such a detector.

\subsection{Stochastic Searches with Third Generation Interferometers}~\label{ssec:3G}
By the mid-late 2030s, it is expected that ground-based gravitational wave detectors will evolve to the third generation of laser interferometers comprising ET and CE.
These observatories are expected to outperform  current 2G network sensitivity by several orders of magnitude in strain, extending the BBH detection horizon out to the earliest epochs of star formation at $z\approx20$ and beyond~\cite{Maggiore:2018sht,VitaleFarr2019}. This would allow 3G detectors to individually resolve more than $99.9\%$ of all stellar origin BBH signals in the Universe~\cite{Regimbau:2016ike}.
This eliminates the motivation to search for the background of stellar origin BBHs with standard cross-correlation methods described in Section~\ref{sec:approaches} because only very few of such signals will remain unresolved, whereas searches for the non-Gaussian background outlined in Section~\ref{ssec:TBS} might be more appropriate. In between 2G and 3G detectors, there is also the proposed ``2.5G'' Australian detector NEMO~\cite{Ackley:2020atn}, which will use very high laser power to target high frequency signals from tidal effects in the late stages of binary neutron star inspirals.

However, cross-correlation based searches might still be relevant for BNS and BHNS backgrounds because individual events will only be resolved up to redshifts $\gtrapprox 1$~\cite{Maggiore:2019uih}. The approach will be then to subtract out all the resolved sources from the timestream and run a stochastic search on the remaining data. This may be performed iteratively and/or assuming different contributions relative to the signal to perform component separation~\cite{Cutler:2005qq}.

Furthermore, cross-correlation based stochastic searches are suitable for searches for cosmological gravitational wave backgrounds, following a subtraction of individual compact binary coalescence signals~\cite{Regimbau:2016ike,Sachdev:2020bkk}.
The noise formed by residuals from the subtraction of best-fit compact binary coalescence waveforms can be reduced to negligible levels following the demonstration by~\cite{SharmaHarms2020}.

Third-generation detectors also present a unique detection opportunity for primordial black holes.
The merger rate of primordial black hole binaries per unit volume is predicted to grow continuously towards redshifts of 50 and beyond~\cite{RaidalSpethmann2019}, producing a high redshift source of stochastic GWs which could be probed with 3G cross-correlation searches.
Note that this remains subject to the theoretically uncertain rate of such events that the nature provided us with; let us cite several investigations on the matter~\cite{Mandic:2016lcn,Ali-Haimoud:2017rtz,Vaskonen:2019jpv,DeLuca:2021hde}.
Attribution of a merged black hole binary to primordial and not stellar origin based on redshift alone will not be statistically strong even with 40-kilometer Cosmic Explorer, 20-kilometer Cosmic Explorer South, and Einstein Telescope~\cite{NgChen2021}.
This means that even in the scenario where primordial black holes are abundant, studies of these events will require a combination of fitting for their ensemble properties as well as cross-correlation-based searches for the stochastic background that are sensitive to subthreshold signals. In the case of competing stochastic backgrounds, studies have been performed to carry out simultaneous Bayesian fitting of both astrophysical and primordial components~\cite{Martinovic:2020hru,Biscoveanu:2020gds}, and methods have been proposed to distinguish primordial black holes from astrophysical ones~\cite{Mukherjee:2021ags,Mukherjee:2021itf}.

Correlating anisotropies in the GWB which electromagnetic observations has also been proposed as a target for future detectors. In some cases, this means using galaxy catalogues to guide the model used for anisotropic GWB searches~\cite{Yang:2020usq}, while in other cases, the proposal is to correlate sub-threshold GW signals with EM transients to probe fundamental physics~\cite{Mukherjee:2019oma,Mukherjee:2020jxa}. 

There are quite a few technological challenges in the development of 3G detectors that are directly relevant for GWB searches, for which its sensitivity to $\Omega_{\rm gw}$ scales with $f^{-3}$ and, therefore, benefit from improved low frequency sensitivity. Below, we discuss two important and probable sources of low frequency noise that are important for GWB searches.

First, globally correlated magnetic noise from, e.g., Schumann resonances~\cite{Schumann1952,Schumann1954} will, if not subtracted or dealt with in some manner, provide a lower limit on our ability to measure a GWB in the frequency band from $\sim$8--40 Hz~\cite{Christensen1992,ThraneChristensen2013,ThraneChristensen2014,CoughlinChristensen2016,HimemotoTaruya2017,CoughlinCirone2018,HimemotoTaruya2019,MeyersMartinovic2020,HimemotoNishizawa2021,JanssensMartinovic2021}, and could even be an issue at higher frequencies~\cite{JanssensMartinovic2021}. The assumption that a GWB will look like correlated noise between multiple detectors means that correlated magnetic noise could masquerade as a GWB if we are not careful. There are several avenues we can take to avoid this issue. One of the most promising is Wiener filtering using low noise magnetometers near the detectors~\mbox{\cite{ThraneChristensen2013,ThraneChristensen2014,CoughlinChristensen2016,CoughlinCirone2018}}. Another option is spectral separation using the hybrid-Bayesian method discussed at the end of Section~\ref{ssec:detection:isotropic}, where one models the contribution of the correlated magnetic noise to the cross-correlation statistic and includes it with a model for the GWB~\cite{MeyersMartinovic2020}. Finally, reducing the coupling of magnetic fields into the strain channel of the detectors will also be vital~\cite{JanssensMartinovic2021}. In the end, some combination of all three of these ideas will likely be needed.

Second, low frequency noise due to gravity gradients caused by temperature fluctuations in the atmosphere (infrasound) or seismic waves, known as ``Newtonian noise,'' is likely a limiting noise source at frequencies up to $\sim$30\,\rm{Hz}~\cite{Saulson1984,HughesThorne1998,Punturo:2010zz,Harms2019,AmannBonsignorio2020,Hall:2020dps}. For a complete review, see~\cite{Harms2019}.  One cannot shield against Newtonian noise and, therefore, need to either subtract~\cite{DriggersHarms2012,CoughlinHarms2014,CoughlinMukund2016,CoughlinHarms2018,BadaraccoHarms2020} it or cleverly design the local environment of the detectors to reduce surface seismic waves and atmospheric perturbations~\cite{HarmsHild2014}. Estimating the specific contribution of Newtonian noise to the detectors from, e.g., seismic waves, is calculable based on a realisation of the seismic wavefield~\cite{Harms2019}, but is very difficult to achieve in practice. Proof of principle tests of using an array of seismometers or geophones to create Wiener filters or use machine learning to clean a ``target'' seismometer channel are promising, and work on optimal design of local sensor arrays is mature~\cite{DriggersHarms2012,CoughlinHarms2014,CoughlinMukund2016,CoughlinHarms2018,BadaraccoHarms2019,BadaraccoHarms2020}. However, cleaning a seismometer is quite different from cleaning Newtonian noise from a gravitational wave detector. In the end, it is likely that for 3G detectors, an array of geophones, broadband seismometers, and tiltmeters (which can be used to separate seismic tilt from seismic waves) will be needed to subtract Newtonian noise due to seismic waves~\cite{HarmsVenkateswara2016,Harms2019}. Meanwhile, an array of infrasound microphones or LIDAR detectors, along with potentially using wind fences around facilities, will be needed to reduce and cancel Newtonian noise due to atmospheric fluctuations~\cite{Creighton2008,FiorucciHarms2018}.

\subsection{Stochastic Searches with the Laser Interferometer Space Antenna}


Stochastic searches with LISA present distinct challenges from those discussed for PTAs and ground-based interferometer networks, primarily due to the fact that LISA is considered to be a \emph{single} detector; hence, it is not possible to use a cross-correlation search with a second detector which is not co-located with LISA. It is, however, possible to auto- and cross-correlate  TDI channels, as we explain here. 

The antenna itself will be comprises three spacecrafts which will be positioned on three heliocentric orbits and will specify a plane at a 60$^\circ$ angle with the ecliptic, in an equilateral triangle configuration which will be maintained throughout the entire duration of the mission (4+ years)~\cite{Amaro2017}. Each spacecraft will send a laser beam to each other spacecraft such that, in total, six laser measurements will be made, comprising the \emph{links} of the detector. The links are then combined to form TDI channels, as detailed in Section~\ref{ssec:correlations}. Basic LISA studies start from considering sets of three channels, where each channel is centered around one spacecraft; however, more sophisticated studies are already under way~\cite{Muratore2020,Vallisneri:2020otf,Bayle:2021mue}. These channels produce a set of observations, which we can treat as timestreams for the purposes of this discussion (although these are effectively combined in post-processing and are not the read-out of the detector), ${X(t), Y(t), Z(t)}$, as introduced in Section~\ref{ssec:correlations}. Taking the same approach as for real interferometer timestreams, we can decompose each of these in their signal and noise components,
\begin{equation}
    X(t) = R_X(t)*h(t) + n_X(t)\,,
\end{equation}
where $R_x$ is the specific response of the $X$ channel, and $n_X$ is its noise term. The noise terms for the three channels are not independent, as each shares (possibly multiple) links; hence, a residuals analysis such as that based on Likelihood~(\ref{eq:likelihood_residuals}) is out of the question. Hence, LISA stochastic analyses are based on Gaussian likelihoods for the data as Likelihood~(\ref{eq:likelihood_data}), where the data are now a set of TDI channels,
\begin{equation}
    {\cal L}(\bm d) \propto \prod_{f,\tau} \frac{1}{|\bm C|^{1/2}} e^{\frac{1}{2} \bm T \bm C^{-1} \bm T^\star}\,,
    \label{eq:likelihood_data_TDI}
\end{equation}
with $\bm T = (X, Y, Z)^T$, and $\bm C$ is the full covariance matrix of the data, $\bm C = \langle \bm T \otimes \bm T^\star \rangle$. The key to detecting a stochastic background with LISA is then understanding how instrument noise and GW signals manifest in different TDI channels and finding an accurate modelling approach to plug into Likelihood~(\ref{eq:likelihood_data_TDI}). The first step consists in picking the specific TDI channels and assumptions to work in---there are several options provided in the literature~\cite{TDI1,TDI2,Estabrook2000,Prince:2002hp,Sylvestre2003,Tinto:2004wu}.

Both maximum-likelihood methods and Bayesian approaches which compute the full posteriors have been proposed to extract and characterise an SGWB from LISA data. In the first case, it is possible to produce a maximum-likelihood solution by fixing a noise model and iterating over the data until the signal converges to the maximum-likelihood estimate; this is essentially the same approach as that laid out in Section~\ref{ssec:approaches_anisotropic} for anisotropic searches, based on~\cite{Contaldi:2020rht}. 
Otherwise, one may take a full Bayesian approach, parametrizing signal and noise components in the data and assuming appropriate priors for the two sets of parameters, then sampling the Likelihood to obtain informative posteriors. This is performed for example in~\cite{adamscornish1} for an isotropic stochastic background, where the data were given by the LISA Data Challenge (LDC). Here, the authors show that, using the noise orthogonal $A, E, T$ set of TDI channels, they are able to recover the stochastic signal and independently measure the relevant LISA noise parameters for each arm of the interferometer. Note, however, that this is performed in the absence of competing signals.

In fact, component separation will be a major challenge for LISA data analysis, as a variety of signals will contribute to the same frequency band and overlap in time~\cite{Crowder:2004ca}. In the case of stochastic backgrounds, as detailed in Section~\ref{sec:sources}, LISA will be sensitive to an anisotropic galactic white dwarf binary background which traces the shape of the Milky Way, a mostly isotropic background of either primordial or stellar-origin black hole and neutron star binaries~\cite{Chen2018}, and also potentially other primordial backgrounds, such as those arising from inflation~\cite{Bartolo:2016ami}, first-order phase transitions~\cite{Caprini:2015zlo,Caprini:2019egz}, and cosmic strings~\cite{Auclair:2019wcv,Boileau:2021gbr}, all of which may be either isotropic or anisotropic. Galactic binaries will by far dominate the measurement, and in fact several approaches have been proposed towards component separation, relying on characteristic, observable differences of each component such as their spectral shape~\cite{Crowder:2006eu,adamscornish2}, or their distribution on the sky, as seen in~\cite{Contaldi:2020rht, BanagiriCriswell2021}. 
In any case, it will first be necessary to disentangle the resolvable white dwarf binaries from the unresolvable white dwarf binary background in the data streams, as detailed in~\cite{Littenberg:2020bxy}.

Anisotropic stochastic searches with LISA leverage the motion of the LISA constellation around the sun to map the signal in the solar system barycenter frame. The method proposed in~\cite{Contaldi:2020rht} fixes the noise model, simulates mock data assuming the LISA response functions in~\cite{Sangria} and uses the Bond--Jaffe--Knox maximum-likelihood solution in Equation~(\ref{eq:map_itersol}) to test the mapping capabilities of the detector. It was found that the angular resolution of LISA for stochastic signals highly depends on their SNR as a function of frequency, as the detector response varies greatly as a function of the latter. The stochastic signals compete with the detector noise, which at low SNR dominates the modes of the Fisher information matrix~(\ref{eq:fish_map_itersol}). Beyond this, the angular resolution at which LISA ``sees'' the signal is diffraction-limited; hence, the higher the frequency, the better the angular resolution will be. In the best-case scenario of a high SNR signal dominating at frequencies $\sim$0.1~Hz, the angular resolution will be of the order $\ell_{\rm max}\sim 16$---see Figure 6 of~\cite{Contaldi:2020rht}. Another approach proposed in~\cite{BanagiriCriswell2021} tackles the mapping problem by setting priors on the spherical harmonic modes of the signal and solving for their amplitudes in a Bayesian framework, using a single TDI channel. This is especially useful when the sky distribution of the signal is well known, as is the case for the galactic white dwarf binary background. The technique proposed here is based on the Clebsch--Gordan expansion, which allows one to constrain the spherical harmonics with a non-negative distribution. This method has proven to be effective for the recovery of a mock galactic background and yields an informative set of posteriors on the coefficients of the signal---see Figure 3 of~\cite{BanagiriCriswell2021}. However, it has only been tested at the low resolution of $\ell_{\rm max}=4$ as the solver scales significantly in computational cost with the number of modes one tries to recover.

\vspace{6pt}
\authorcontributions{{Conceptualization}: A.I.R., B.G., A.C.J., and P.M.M. {Data curation}: A.I.R., B.G., A.C.J., and P.M.M. {Software}: A.I.R., B.G., A.C.J., and P.M.M. {Making of figures and tables}: A.I.R., B.G., and A.C.J. {Writing---original draft}: A.I.R., B.G., A.C.J., and P.M.M. {Writing---review \& editing}: A.I.R., B.G., A.C.J., and P.M.M. All authors have read and agreed to the published version of the manuscript.}

\funding{
AIR acknowledges the support of the National Science Foundation and the LIGO Laboratory.
BG is supported by the Italian Ministry of Education, University and Research within the PRIN 2017 Research Program Framework, n. 2017SYRTCN.
PMM was supported by the NANOGrav Physics Frontiers Center, National Science Foundation (NSF), award number 2020265.
Parts of this research were conducted by the Australian Research Council Centre of Excellence for Gravitational Wave Discovery (OzGrav), through project number CE170100004.
} 

\dataavailability{This manuscript shows figures created using the LVK open data~\cite{LVK:open_data_iso,LVK:open_data_aniso}, and publicly available PTA results presented in  \cite{ArzoumanianBaker2020,goncharov2021cpgwb,ChenCaballero2021,GoncharovReardon2021}.} 

\acknowledgments{The authors thank Katerina Chatziioannou and Joseph Romano for precious comments on our original draft. 
}

\conflictsofinterest{The authors declare no conflicts of interest. 
} 

%

\reftitle{References}

\end{document}